\documentclass[11pt]{article} % use larger type; default would be 10pt
\pdfoutput=1 % if your are submitting a pdflatex (i.e.\  if you have
\usepackage{jheppub} % for details on the use of the package, please
\usepackage[utf8]{inputenc}
\usepackage[normalem]{ulem}
\usepackage[english]{babel}
\usepackage{amsmath,amssymb,amsbsy,amstext,amsthm,booktabs,simplewick,exscale,relsize,slashed,graphicx,amsfonts,upgreek, xcolor,subcaption}
\usepackage{multirow,color,dcolumn,bm,enumerate}
\usepackage{hyperref}
\usepackage{newunicodechar}
\newunicodechar{，}{,}
\usepackage{dsfont}
\usepackage{ragged2e} %This was needed to resolve a red error with the bibiography, but the references still don't look right! Gah!
\DeclareUnicodeCharacter{2212}{-}
\usepackage{threeparttable}
\usepackage{makecell}
\usepackage[T1]{fontenc}

\usepackage{hyperref}

\title{Electroweak First-Order Phase Transition Triggered by Non-Gaussian Fluctuations of a $\mathbb{Z}_2$-Symmetric Spectator Scalar}

\author[a,1]{Bo-Qiang Lu}
\emailAdd{bqlu@huznu.edu.cn}

\affiliation{School of Science, Huzhou Normal University, Huzhou, Zhejiang 313000, P. R. China}

\abstract{
We propose a novel mechanism to trigger a first-order cosmological electroweak phase transition using non-Gaussian primordial fluctuations of a $\mathbb{Z}_2$-symmetric
spectator scalar field. We show that the large fluctuations of the spectator field can modify the Higgs thermal mass and enhance the thermal barrier, 
thereby enabling a strong first-order phase transition. Non-Gaussianities in the primordial fluctuation spectrum significantly increase the probability of 
large-amplitude fluctuations, allowing a substantial fraction of the Universe to undergo the transition. The spectator field also naturally serves as a cold dark 
matter candidate through its coherent oscillations, reproducing the observed relic abundance. 
The resulting stochastic gravitational-wave background peaks at $f\sim 10^{-3}-10^{-1}$ Hz, but its amplitude lies roughly ten 
orders of magnitude below the projected sensitivities of planned space-based and ground-based interferometers.
}

\begin{document}
\maketitle

\setcounter{page}{2}

\section{Introduction}
\label{sec:introduction}

The electroweak phase transition (EWPT) represents a pivotal event in the thermal history of the early Universe, during which the electroweak symmetry 
$SU(2)_L \times U(1)_Y$ is spontaneously broken to the electromagnetic $U(1)_{\text{em}}$ symmetry, endowing elementary particles with mass via the Higgs mechanism. 
The nature of this transition has profound implications for both particle physics and cosmology. The standard model (SM) of particle physics, which contains a single 
Higgs doublet with a mass of 125~GeV~\cite{ATLAS:2012yve,CMS:2012qbp}, predicts that the EWPT is a smooth crossover rather than a first-order phase 
transition~\cite{DOnofrio:2014rug}. However, a first-order EWPT is a necessary ingredient for two of the most outstanding phenomena in physics, namely the origin of the baryon 
asymmetry of the Universe~\cite{Morrissey:2012db} and the generation of a detectable stochastic gravitational wave background (SGWB)~\cite{Bian:2025ifp}.

Electroweak baryogenesis, first proposed by Kuzmin, Rubakov and Shaposhnikov \cite{Kuzmin:1985mm}, remains one of the most attractive mechanisms for explaining 
the observed matter-antimatter asymmetry~\cite{Cohen:1990py}. For EWBG to be successful, the phase transition must be strongly first-order, characterised by the condition 
$v_c/T_c \gtrsim 1$, where $v_c$ is the Higgs vacuum expectation value (VEV) at the critical temperature $T_c$. This ensures that sphaleron processes, which would 
wash out any generated baryon asymmetry, are sufficiently suppressed in the broken phase~\cite{Cline:2006ts}. Additionally, a first-order EWPT proceeds via the nucleation, expansion and 
collision of true vacuum bubbles, a violent process that generates a SGWB with a characteristic spectrum peaking in the millihertz to hertz frequency 
band~\cite{Witten:1984rs,Kamionkowski:1993fg,Grojean:2006bp,Caprini:2015zlo,Weir:2017wfa,Athron:2023xlk}. This SGWB is a primary science target for future space-based 
gravitational wave (GW) interferometers such as LISA~\cite{LISA:2017pwj,LISACosmologyWorkingGroup:2022jok}, TianQin~\cite{TianQin:2015yph}, Taiji~\cite{Luo:2019zal}, 
DECIGO~\cite{Sato:2017dkf}, and BBO~\cite{Crowder:2005nr}, providing a unique window into physics at the electroweak scale that is inaccessible to collider experiments.

Complementary to the EWPT problem, the existence of dark matter (DM) has been firmly established by cosmological observations~\cite{Bertone:2016nfn}, 
with Planck fixing its relic abundance at \(\Omega_{\text{DM}} h^2 = 0.120 \pm 0.001\)~\cite{Planck:2018vyg}. 
However, the particle nature of DM remains unknown.
Weakly interacting massive particles (WIMPs) have long been the leading candidate, motivated by the ``WIMP miracle''—the natural production of the correct relic 
density for weak-scale particles~\cite{Lee:1977ua,Hut:1977zn}.
However, after decades of extensive searches via direct detection, indirect detection, and collider experiments, 
no conclusive WIMP signal has been observed~\cite{Arcadi:2017kky}.
In particular, direct detection experiments have pushed exclusion limits down to the neutrino floor~\cite{Akerib:2022ort,XENON:2024hup,PANDA-X:2024dlo,Xia:2026bkh}.
These persistent null results across all fronts have severely constrained the traditional WIMP parameter space and motivate the exploration of alternative 
DM candidates and production mechanisms~\cite{Profumo:2019ujg,Gouttenoire:2022gwi,Baldes:2023cih,Cirelli:2024ssz}.

Given the common electroweak scale origin, it is natural to seek a unified framework that simultaneously addresses both the first-order EWPT and the DM problem. 
Over the past decades, numerous extensions of the SM have been proposed to realise a first-order EWPT. The simplest extensions introduce additional scalar fields, 
such as the real/complex singlet extension~\cite{Profumo:2007wc,Espinosa:2007qk,Barger:2007im,Espinosa:2011ax,Cline:2012hg,No:2013wsa,Profumo:2014opa,Jiang:2015cwa,
Chiang:2017nmu,Grzadkowski:2018nbc,Chiang:2019oms,Deng:2020dnf,Zhou:2020ojf,Paul:2020wbz,Bian:2021dmp,Demidov:2021lyo,Chaudhuri:2022sis,Xiao:2022oaq,Chen:2022zsh,Lu:2022zpn,
Chatterjee:2022pxf,Baldes:2022oev,Ghosh:2022fzp,Cao:2022ocg,Xu:2023lkf,Ghosh:2024ing,DEramo:2024lsk,Borah:2025ubr,Srivastava:2025oer,Liu:2025pny,Mirzaie:2025bzn,Robens:2025nev,Das:2026zuo}, 
the two-Higgs-doublet model (2HDM)~\cite{Cline:2011mm,Dorsch:2014qja,Basler:2016obg,Dorsch:2017nza,Zhang:2021alu,Ramsey-Musolf:2024zex,Aoki:2021oez,Biekotter:2021ovi,Chaudhuri:2024vrd,Lee:2025hgb,Du:2026qco},
or the $SU(2)$ dark sector~\cite{Ghosh:2020ipy,Aoki:2023xnn,Abe:2023zja}. 
In these models, a first-order phase transition is typically achieved through tree-level or loop-induced barriers in the finite-temperature effective potential. 
Supersymmetric models, such as the minimal supersymmetric standard model (MSSM) and its extensions~\cite{Katz:2015uja,Huber:2000mg,Chung:2010cd,Kozaczuk:2014kva,Huang:2014ifa,Borah:2023zsb}, can also accommodate a first-order EWPT, 
but require light stops that are increasingly disfavored by LHC searches. Furthermore, models with 
vector-like fermions~\cite{Angelescu:2018dkk,Fu:2022eun,Adhikary:2024esf}, composite Higgs sectors~\cite{Bruggisser:2018mrt,Bruggisser:2018mus,Bian:2019kmg,Banerjee:2024puz}, 
and higher-dimensional operators~\cite{Grojean:2004xa,Cai:2017tmh,Cai:2022bcf,Oikonomou:2024jms,Jahedi:2025yjz} have also been explored. Many of these models also provide natural DM candidates, 
such as the lightest neutralino in supersymmetric models, the lightest neutral inert Higgs boson in 2HDM, or the singlet scalar in xSM. However, 
most existing scenarios face challenges from experimental constraints~\cite{Huang:2015izx,Cao:2017oez,Azevedo:2018exj,Ramsey-Musolf:2019lsf,Papaefstathiou:2020iag,Papaefstathiou:2021glr,Zhang:2023mnu,Wang:2023zys,Lane:2024vur} 
and theoretical bounds~\cite{Lu:2022qjk,Balazs:2023kuk}, or the difficulty of simultaneously satisfying the requirements for a strong first-order phase transition 
and the correct DM relic abundance~\cite{Blinov:2015vma,Ghorbani:2018yfr,Chiang:2020yym}.

Cosmic inflation, a period of exponential expansion in the first $10^{-35}$ seconds after the Big Bang, provides the theoretical foundation for understanding the 
origin of structure in the Universe~\cite{Starobinsky:1980te,Guth:1980zm}. Originally proposed to resolve the horizon, flatness, and monopole problems of the standard Big Bang cosmology, inflation has 
been spectacularly confirmed by precision CMB measurements~\cite{Martin:2015dha,Planck:2018vyg}. A central prediction of inflation is that quantum fluctuations of scalar fields during the exponential 
expansion are stretched to cosmological scales, generating a nearly scale-invariant spectrum of primordial density perturbations that serve as the seeds for galaxy 
formation (see Refs.~\cite{Guth:2000ka,Bassett:2005xm,Baumann:2009ds,Achucarro:2022qrl,Odintsov:2023weg,Ellis:2023wic} for reviews). While these perturbations 
are approximately Gaussian in the simplest single-field slow-roll inflation models, most realistic inflationary scenarios predict 
a measurable level of non-Gaussianity, characterised by the nonlinear parameter $f_{\text{NL}}$. The latest Planck 2018 results give a constraint 
of $f_{\text{NL}}^{\text{local}} = -1 \pm 5.0~(68\%~{\rm CL})$~\cite{Planck:2019kim,Ellis:2023wic}, consistent with zero but leaving ample room for significant non-Gaussianity 
that will be probed by future experiments such as CMB-S4~\cite{CMB-S4:2016ple} and LiteBIRD~\cite{LiteBIRD:2022cnt}.

In this work, we propose a novel mechanism for triggering a first-order EWPT using non-Gaussian primordial fluctuations of a $\mathbb{Z}_2$-symmetric spectator scalar field. 
Traditional studies of cosmological phase transitions have almost universally assumed a perfectly homogeneous Universe, where the phase transition proceeds uniformly 
across all spatial regions. However, the primordial fluctuations generated during inflation naturally introduce small but cosmologically significant inhomogeneities 
in the values of scalar fields and physical parameters. If a light scalar field exists that couples to the SM Higgs and gauge bosons, its primordial fluctuations will modulate 
the local shape of the Higgs effective potential, leading to spatial variations in the phase transition dynamics. 
We consider a light spectator field which is a real gauge-singlet scalar field invariant under a global $\mathbb{Z}_2$ symmetry. 
The light spectator field acquires non-Gaussian primordial fluctuations generated during inflation.
In regions where the spectator field fluctuation exceeds a certain threshold, the Higgs effective potential develops a barrier between the 
symmetric and broken phases, leading to bubble nucleation and a first-order phase transition. We show that the transition can be sufficiently strong 
to suppress electroweak sphaleron processes, as required for electroweak baryogenesis.

Furthermore, the spectator field in our model naturally serves as a cold DM candidate. When the Hubble parameter falls below the scalar mass $m_s$, 
the field begins to undergo coherent oscillations around its minimum, with an energy density that scales as $a^{-3}$ during cosmic expansion (the field amplitude 
scales as $T^{3/2}$, equivalently $a^{-3/2}$). We demonstrate that the observed DM relic abundance can be reproduced over a wide range of parameters, 
thereby linking two long-standing problems of cosmology within a single framework. Our framework exhibits a ``stealth'' nature. Owing to the small portal 
coupling $\kappa$ and the high new physics scale $\Lambda$, low-energy experiments, including those at the LHC and astrophysical observations, are unable 
to impose meaningful constraints. The stochastic gravitational-wave background (SGWB) produced during the transition, which peaks in the frequency range 
targeted by future space-based interferometers, is in principle the only direct observational window. As we show in Section~\ref{sec:GW}, however, its 
amplitude lies far below the projected sensitivities of planned detectors, so that testing this scenario remains a challenge in practice.
 
We emphasise that the primary goal of this work is to establish the mechanism by which non-Gaussian spectator fluctuations trigger a strong first-order 
EWPT, which is a necessary condition for electroweak baryogenesis (EWBG). Whether this condition is met is established gauge-invariantly through the three-dimensional effective-theory criterion of Section~\ref{sec:xsm_comparison}, rather than through the gauge-dependent proxy $v_c/T_c\gtrsim1$. A complete EWBG scenario additionally requires sufficient CP violation, which the SM does not 
provide~\cite{Morrissey:2012db}. In our framework CP violation is incorporated through the UV completions. The spectator field $s$ is protected by a $\mathbb{Z}_2$ 
symmetry $s\to -s$ that enforces $\langle s\rangle=0$ throughout the EWPT; consequently the CP-violating operator must be a \emph{$\mathbb{Z}_2$-even} function of $s$, i.e.\ quadratic or higher order in $s$. 
In the vector-like fermion completion (Appendix~\ref{app:UVcompletion}) the full CP-violating Lagrangian is
\begin{equation}\label{eq:L_CP_intro}
\mathcal{L}_{\rm EWBG}\supset -\Bigl[M_0+\frac{y}{\Lambda}s^2\Bigr]\overline{\Psi}\Psi
-\Bigl[y_0+|y_s|e^{i\theta}\,\frac{s^2}{\Lambda^2}\Bigr]\frac{h(z)}{\sqrt{2}}\,\overline{\Psi}\Psi ,
\end{equation}
The $h$--$\Psi$ coupling in the second term is the unitary-gauge form of the gauge-invariant dimension-five \allowbreak operator \allowbreak $-\frac{y_{\rm CP}(s)}{\sqrt{2}\,v}\,\overline{\Psi}\Psi\,H^{\dagger}H$ 
($H$ is the SM Higgs doublet, $v=246$~GeV); after electroweak symmetry breaking $H^{\dagger}H=(v+h)^{2}/2$, whose linear part reproduces the term written above.
The first (real) term in Eq.~\eqref{eq:L_CP_intro} generates the $s$-dependent gauge kinetic function $Z(s)$ that triggers the FOPT, while the second term carries the physical CP-violating 
phase $\theta$ and is manifestly $\mathbb{Z}_2$-even. Because $s$ is spatially constant inside a given bubble, the spatial variation of the mass matrix 
$\mathcal{M}(z;s)=y_{\rm CP}(s)\,h(z)/\sqrt{2}$ comes entirely from the Higgs profile $h(\mathbf{x})$ across the wall, and the CP-violating source 
$S_{\rm CP}(z)\propto\operatorname{Im}\!\bigl[(M(s)+\mathcal{M}(z;s))^{*}\partial_{z}(M(s)+\mathcal{M}(z;s))\bigr]\propto M(s)\,\operatorname{Im}[y_{\rm CP}(s)]\,\partial_{z}h(z)$ is non-zero only at the wall. 
Its magnitude is enhanced by $M(s)\,\operatorname{Im}[y_{\rm CP}(s)]\simeq M(s)\,|y_{s}|\sin\theta\,(s^{2}/\Lambda^{2})$ to $\mathcal{O}(1)$ in precisely the high-$s$ fluctuation regions that actually undergo the FOPT. 
Thus the same large non-Gaussian $s$ fluctuation both strengthens the transition (via $Z(s)$) and amplifies the CP source; because the coupling is $\mathbb{Z}_2$-even, 
the $+s$ and $-s$ fluctuation tails contribute with the same sign and add coherently. See Appendix~\ref{app:UVcompletion} for further details.
In the supersymmetric completion, $A$-terms and gaugino masses with CP-violating 
phases play an analogous role, closely related to the MSSM electroweak baryogenesis mechanism. We also adopt the globally supersymmetric UV completion 
(Appendix~\ref{app:UVcompletion}) as the preferred framework, in which the smallness of $m_s$ is protected by the non-renormalisation theorem and exact 
boson--fermion cancellations, rendering it technically natural. 
A full EWBG calculation (baryon asymmetry prediction, experimental constraints on CP phases) is left for future work.

We emphasise that the commonly used phase-transition strength criterion $v_c/T_c\gtrsim 1$ is \emph{gauge-dependent}, since its numerical value shifts with the 
gauge-fixing prescription. It should therefore be read only as an \emph{indicative} diagnostic rather 
than a gauge-invariant statement~\cite{Patel:2011th,Katz:2015uja}. The order of the transition is instead fixed by the gauge-invariant three-dimensional effective-theory ratio $x_c\equiv\bar\lambda_3/\bar g_3^2$, whose critical value $x_c\simeq0.11$ separates a first-order transition from a crossover~\cite{Rummukainen:1998as,Kajantie:1996mn}; we apply this criterion to the present mechanism in Section~\ref{sec:xsm_comparison}. The physically meaningful, \emph{gauge-invariant} requirement for electroweak baryogenesis is the resulting suppression of the 
electroweak-sphaleron transition rate in the broken phase, $\Gamma_{\rm sph}(T_c)/T_c^4\ll 1$, which freezes out the baryon-number-violating 
processes~\cite{DOnofrio:2014rug,Quiros:2019kam}, supplemented by the percolation condition that the true-vacuum bubbles must come to fill the whole 
Universe (Section~\ref{sec:percolation}). In our setup the finite-temperature effective potential is evaluated from the exact one-loop thermal functions, 
supplemented by the Arnold--Espinosa ring term~\cite{Arnold:1993rc,Carrington:1995hz,Espinosa:1994en}, evaluated in Landau gauge. The ring term resums the longitudinal 
$W$ and $Z$ modes by shifting their masses with the leading-order (gauge-invariant) Debye mass. 
This removes what would otherwise be the longitudinal modes' contribution to the $\mathcal{O}(g^{3})$ cubic 
barrier, reducing the gauge dependence of the potential to subleading order~\cite{Patel:2011th,Katz:2015uja,Chiang:2017gwd}; the price is a lowering of the cubic 
coefficient by the factor $2/3$~\cite{Arnold:1993rc,Carrington:1995hz} and hence a somewhat weaker transition. The residual Nielsen-identity gauge variation of 
$v_c/T_c$ is quantitatively small~\cite{Patel:2011th,Katz:2015uja,Chiang:2017gwd,Wainwright:2011ki}. 
At the benchmark points the daisy-resummed four-dimensional potential gives $v_c/T_c$ of order unity, while the gauge-invariant criterion of Section~\ref{sec:xsm_comparison} shows that the transition is first order and the sphaleron rate is 
correspondingly exponentially suppressed; the gauge-dependent diagnostic and the gauge-invariant criterion are thus mutually consistent. 

The remainder of this paper is organised as follows. In Section \ref{sec:model}, we construct our model, including the tree-level potential, the field-dependent
effective gauge couplings, and the finite-temperature effective potential with the daisy-resummed full one-loop thermal functions. In Section \ref{sec:conditions}, we derive the necessary 
conditions for a first-order phase transition, analyse the parameter space, and establish perturbativity bounds on the gauge coupling enhancement. 
In Section \ref{sec:scalar_evolution}, we study the evolution of the spectator field quantum fluctuations from
inflation to the present day, discuss its properties as an oscillating DM candidate, derive the exact $\mathbb{Z}_2$-symmetric non-Gaussian probability
distribution, and analyse the isocurvature constraints and the required power spectrum shape. In Section \ref{sec:nucleation}, we calculate the thermal 
nucleation rate of true vacuum bubbles, develop a saddle-point approximation for the global
average nucleation rate, determine the nucleation temperature, compute the volume fraction of the Universe that undergoes a first-order phase transition,
and discuss the percolation condition and the validity of the local approximation.
In Section \ref{sec:GW}, we compute the SGWB produced by the phase transition, including contributions from bubble collisions,
sound waves in the plasma, and magnetohydrodynamic turbulence, with the wall velocity bounded by the model-independent local-thermal-equilibrium (LTE)
and ballistic limits.
% \blueblk{We do not fix a single wall velocity but bracket it between the local-thermal-equilibrium upper limit of 
% Ref.~\cite{Ramsey-Musolf:2024ykk} and the ballistic lower limit of Ref.~\cite{Ai:2024btx}, because the averaged transition strength 
% $\bar\alpha\sim 3\times 10^{-3}$ places the transition deep in the deflagration regime.} 
We present our conclusions and discuss future directions in Section \ref{sec:conclusion}.
Three renormalizable UV completions for the field-dependent gauge kinetic function are provided in Appendix \ref{app:UVcompletion}, 
the exact probability density function is derived in Appendix \ref{app:exact_PDF}, and the GW spectrum formulas are collected in Appendix \ref{app:GW_spectrum}.

\section{Model construction}
\label{sec:model}

\subsection{Tree-level potential}
Let us consider the simplest extension of the SM with a single real scalar field $s(x,t)$ that is invariant under a global $\mathbb{Z}_2$ symmetry:
\begin{equation}
s \mapsto -s.
\end{equation}
We assume the VEV of $s$ is zero, $\langle s \rangle = \bar{s} = 0$, so that the $\mathbb{Z}_2$ symmetry is not spontaneously broken at 
tree level and all orders in perturbation theory. The spectator field can be decomposed into background $\bar{s}$ and fluctuation $\delta s$ components:
\begin{equation}
s(x,t) = \bar{s} + \delta s(x,t) = \delta s(x,t).
\end{equation}
We therefore use $s$ to denote the quantum fluctuations of the spectator field.
At the end of inflation, \(s(x)\) is a random field on a fixed time slice. 

The total scalar potential consists of the SM Higgs potential, the spectator field potential, and the $\mathbb{Z}_2$-symmetric portal interaction is given by:
\begin{equation}
V(H,s) = -\mu_H^2 |H|^2 + \lambda_H |H|^4 - \frac{\kappa}{2}s^2 |H|^2 + \frac{1}{2}\mu_s^2 s^2 + \frac{\lambda_s}{4}s^4~,
\end{equation}
where $H$ is the SM Higgs doublet (which acquires a VEV at zero temperature $v = \sqrt{2}\langle H \rangle = 246\ \text{GeV}$), $\mu_H=\lambda_Hv^2$ is the Higgs 
mass parameter with $\lambda_H\simeq 0.13$ is the Higgs self-coupling, $\lambda_s$ is the spectator self-coupling, and $\kappa$ is the dimensionless portal coupling. 

To generate large field fluctuations during inflation, the potential of the \(s\) field must be sufficiently flat, such that its effective mass is much smaller than 
the Hubble parameter $H_{\rm inf}$. The quartic term contributes an effective mass squared $\sim 3\lambda_s s^2$ at large field values of $s$ which strongly suppresses 
the production of large-amplitude fluctuations.
We require the spectator field to be light during inflation $m_s \ll H_{\text{inf}} \sim 10^{13}\ \text{GeV}$. Furthermore,
we shall set $\lambda_s=0$ to ensure that the potential remains approximately flat within the parameter region of physical interest.

The spectator field $s$ is assumed to be minimally coupled to gravity ($\xi=0$) and to have no direct interactions with the inflaton sector. 
Its effective potential during inflation is therefore flat up to the tiny tree-level mass $\mu_s^2 \ll H_{\text{inf}}^2$.
In the supersymmetric UV completion of Appendix~\ref{app:UVcompletion} we work within global supersymmetry, i.e.\, neglecting supergravity effects; in this approximation, the dangerous Hubble-scale mass contributions from supergravity are absent. 
Consequently, the condition $m_s^{\text{eff}} \ll H_{\text{inf}}$ holds, and the quantum fluctuations of $s$ are unmodified by inflationary dynamics.

%================================================
\subsection{Field-dependent effective gauge couplings}\label{sec:gauge_coupling}

We consider the SM electroweak gauge group $SU(2)_{L}\times U(1)_{Y}$ with gauge fields $W_{\mu}^{a}$ and $B_{\mu}$.
The spectator scalar $s$ is a real gauge singlet, odd under a $\mathbb{Z}_{2}$ symmetry $s\to -s$, which guarantees $\langle s\rangle = 0$.
We modify the gauge kinetic terms as
\begin{equation}\label{eq:Lkin}
\mathcal{L}_{\text{kin}} = -\frac{1}{4}\,Z(s)\Bigl(W_{\mu\nu}^{a}W^{a\mu\nu}+B_{\mu\nu}B^{\mu\nu}\Bigr),
\end{equation}
where
\begin{equation}
W_{\mu\nu}^{a}=\partial_{\mu}W_{\nu}^{a}-\partial_{\nu}W_{\mu}^{a}+g\epsilon^{abc}W_{\mu}^{b}W_{\nu}^{c},\qquad
B_{\mu\nu}=\partial_{\mu}B_{\nu}-\partial_{\nu}B_{\mu}.
\end{equation}
Gauge invariance requires $Z(s)$ to be a positive, even function of $s$ with $Z(0)=1$.
To avoid a divergent effective gauge coupling for large $s$ while retaining a significant enhancement, we adopt the UV‑safe parametrisation
\begin{equation}\label{eq:Zparam}
Z(s)=Z_{\infty}+\frac{1-Z_{\infty}}{1+s^2/\Lambda^{2}},\qquad 0<Z_{\infty}\le 1 .
\end{equation}
Here $\Lambda$ is a high energy scale suppressing the non‑renormalisable operator, and $Z_{\infty}$ is the asymptotic value of $Z(s)$ for $s\gg\Lambda$.
For $Z_{\infty}=1$ the SM is recovered; for $Z_{\infty}<1$ the gauge kinetic term is ``compressed'' at large $s$, which, after canonical normalisation, translates into an amplification of the gauge couplings.
The form~(\ref{eq:Zparam}) interpolates smoothly between $Z(0)=1$ and $Z(\infty)=Z_{\infty}$ and is ghost‑free for any $Z_{\infty}>0$.
In Appendix~\ref{app:UVcompletion}, we provide a UV completion for the implement of parametrisation~\eqref{eq:Zparam}.

To work with canonically normalised gauge fields we perform the Weyl rescaling 
\begin{equation}
\widetilde{W}_{\mu}^{a}=\sqrt{Z(s)}\,W_{\mu}^{a},\qquad 
\widetilde{B}_{\mu}=\sqrt{Z(s)}\,B_{\mu}.
\end{equation}
The kinetic Lagrangian becomes
\begin{equation}\label{eq:gauga_kin}
\begin{aligned}
\mathcal{L}_{\text{kin}} &= 
-\frac{1}{4} \widetilde{W}_{\mu\nu}^a \widetilde{W}^{a\mu\nu} 
-\frac{1}{4} \widetilde{B}_{\mu\nu} \widetilde{B}^{\mu\nu} \\
&\quad + \frac{1}{2} \widetilde{W}^{a\mu\nu} \bigl(\widetilde{W}_\nu^a \partial_\mu\phi - \widetilde{W}_\mu^a \partial_\nu\phi\bigr)
   + \frac{1}{2} \widetilde{B}^{\mu\nu} \bigl(\widetilde{B}_\nu \partial_\mu\phi - \widetilde{B}_\mu \partial_\nu\phi\bigr) \\
&\quad - \frac{1}{4} \bigl(\widetilde{W}_\nu^a \partial_\mu\phi - \widetilde{W}_\mu^a \partial_\nu\phi\bigr)^2
   - \frac{1}{4} \bigl(\widetilde{B}_\nu \partial_\mu\phi - \widetilde{B}_\mu \partial_\nu\phi\bigr)^2 .
\end{aligned}
\end{equation}
where
\begin{equation}
\phi(s) = \frac{1}{2} \ln\!Z(s).
\end{equation}
The derivative interactions involving $\phi(s)$ in Eq.~\eqref{eq:gauga_kin} are higher-order in derivatives and are suppressed by $1/\Lambda$. For the nearly homogeneous spectator 
field configurations that dominate the nucleation process, these terms do not alter the effective gauge couplings or the finite-temperature potential at leading order. 
Hence, we omit them in the following analysis of the phase transition.

The Higgs covariant derivative takes the form
\begin{equation}
D_{\mu}H = \partial_{\mu}H - i\frac{g}{\sqrt{Z(s)}}\frac{\sigma^{a}}{2}\widetilde{W}_{\mu}^{a}H - i\frac{g^{\prime}}{\sqrt{Z(s)}}\frac{1}{2}\widetilde{B}_{\mu}H.
\end{equation}
Consequently, the physical, field‑dependent gauge couplings are
\begin{equation}\label{eq:geff}
g_{\rm eff}(s)=\frac{g}{\sqrt{Z(s)}},\qquad 
        g^{\prime}_{\rm eff}(s)=\frac{g^{\prime}}{\sqrt{Z(s)}} .
\end{equation}
A crucial observation is that the Weinberg angle $\tan\theta_{W}=g^{\prime}/g$ remains unchanged.
The operator~\eqref{eq:Lkin} therefore respects the full electroweak symmetry structure; it merely amplifies the overall gauge interaction strength in regions where $s$ is large in the early Universe.

%================================================
\subsection{Finite-temperature effective potential}

We first calculate the finite-temperature effective potential in the high-temperature approximation, retaining the leading thermal-mass and cubic-barrier terms, 
which is standard for analytical EWPT estimates~\cite{Schicho:2022wty,Athron:2022jyi}.
At finite temperature, thermal loop corrections generate temperature-dependent terms in the Higgs potential at leading-order is:
\begin{equation}
V_T(h,T) = \frac{D_{\rm eff}}{2}T^2 h^2 - E_{\rm eff} T h^3 + \frac{\lambda_H}{4}h^4,
\end{equation}
where $h = \sqrt{2}|H|$ is the physical Higgs scalar component.

The coefficients receive contributions from all light degrees of freedom.
We split $D_{\rm eff}$ into a constant non‑gauge part and an $s$‑dependent gauge part:
\begin{equation}
D_{\rm eff}(s)=D_{\rm non}+D_{\rm gauge}^{\rm eff}(s),\qquad
D_{\rm non}\equiv D_{\rm SM}-D_{\rm gauge}\approx 0.313,
\end{equation}
where the SM gauge-boson part is
\begin{equation}\label{eq:Dgauge}
D_{\text{gauge}} = \frac{1}{16}\left(3g^2 + g'^{\,2}\right) \approx 0.087 ,
\end{equation}
with $g \simeq 0.65$ is the SU(2) gauge coupling and $g' \simeq 0.35$ is the U(1) gauge coupling.
The SM thermal mass coefficient is
\begin{equation}
D_{\text{SM}} = \frac{1}{16}\left(3g^2 + g'^2 + 4y_t^2 + 8\lambda_H + \kappa\right) \approx 0.40,
\end{equation}
with $y_t \simeq 1.0$ is the top quark Yukawa coupling. $D_{\text{SM}}$ includes the thermal contributions from Higgs boson and top quark.
These coefficients refer to the normalisation $V_T\supset\tfrac{1}{2}D_{\rm eff}T^2h^2$ adopted in Eq.~\eqref{eq:loc_Veff}, so that $D_{\rm eff}$ is the thermal 
Higgs mass squared divided by $T^2$.
In this work, we consider a small coupling $\kappa \ll 0.1$, and therefore its correction to the thermal mass coefficient can be safely neglected.
The effective gauge contribution is
\begin{equation}
D_{\rm gauge}^{\rm eff}(s)\equiv \frac{1}{16}\Bigl[3\,g_{\rm eff}^{2}(s)+g_{\rm eff}^{\prime\,2}(s)\Bigr]=\frac{D_{\rm gauge}}{Z(s)} .
\end{equation}
Thus
\begin{equation}\label{eq:Deff}
D_{\rm eff}(s)=D_{\rm non}+\frac{D_{\rm gauge}}{Z(s)} .
\end{equation}

The cubic term arises from the thermal functions of the gauge bosons in the high-temperature limit $m_W \ll T$, and is therefore directly determined by the 
effective gauge couplings. In the daisy-resummed (Arnold--Espinosa ring) treatment adopted in this work the longitudinal gauge modes are Debye-screened, so the $h^3$ 
barrier is generated by the two transverse polarisations only:
\begin{equation}\label{eq:Eeff}
    E_{\rm eff}(s) \equiv \frac{1}{48\pi}\Bigl[2\,g_{\rm eff}^{3}(s)+\bigl(g_{\rm eff}^{2}(s)+g_{\rm eff}^{\prime\,2}(s)\bigr)^{3/2}\Bigr]
    =\frac{E_{\rm SM}^{\rm res}}{[Z(s)]^{3/2}},\qquad E_{\rm SM}^{\rm res}\approx 0.0063 ,
\end{equation}
The cubic coefficient $E_{\rm eff}$ in Eq.~\eqref{eq:Eeff} is a high-temperature truncation of the full thermal functions and therefore inherits a residual gauge 
dependence.  The ring (daisy) improvement screens only the longitudinal gauge modes and \emph{reduces}, though it does not eliminate, this dependence, which survives at 
subleading order from the truncation and the partial resummation.  The gauge-invariance statements in this work therefore concern the \emph{physical} nucleation 
observables ($f_{\rm FOPT}$, $\bar\alpha$, $\overline{\beta/H}$), evaluated from the Landau-gauge complete one-loop potential, rather than the truncated cubic coefficient itself.
The prefactor $1/(48\pi)=\tfrac{2}{3}\times 1/(32\pi)$ implements the restriction to transverse polarisations: the naive high-$T$ cubic term counts all $6$ 
degrees of freedom of the $W^\pm$ bosons (two charges $\times$ three polarisations) and all $3$ polarisations of the $Z$ boson, whereas in the daisy-resummed 
potential the longitudinal modes are Debye-screened and contribute only $h$-even powers, so that only the $4$ transverse $W$ and $2$ transverse $Z$ modes 
generate the $h^3$ barrier. This gives the resummed coefficient $E_{\rm SM}^{\rm res}\approx 0.0063$ (Eq.~\eqref{eq:Eeff}), i.e.\ $2/3$ of the naive SM value 
$E_{\rm SM}\approx 0.0095$. 
The full one-loop thermal functions are evaluated with this resummation, and the benchmark observables of Table~\ref{tab:ewpt_parameters} are computed directly from 
the resummed potential.

\begin{figure}
    \centering
    \includegraphics[width=0.6\linewidth]{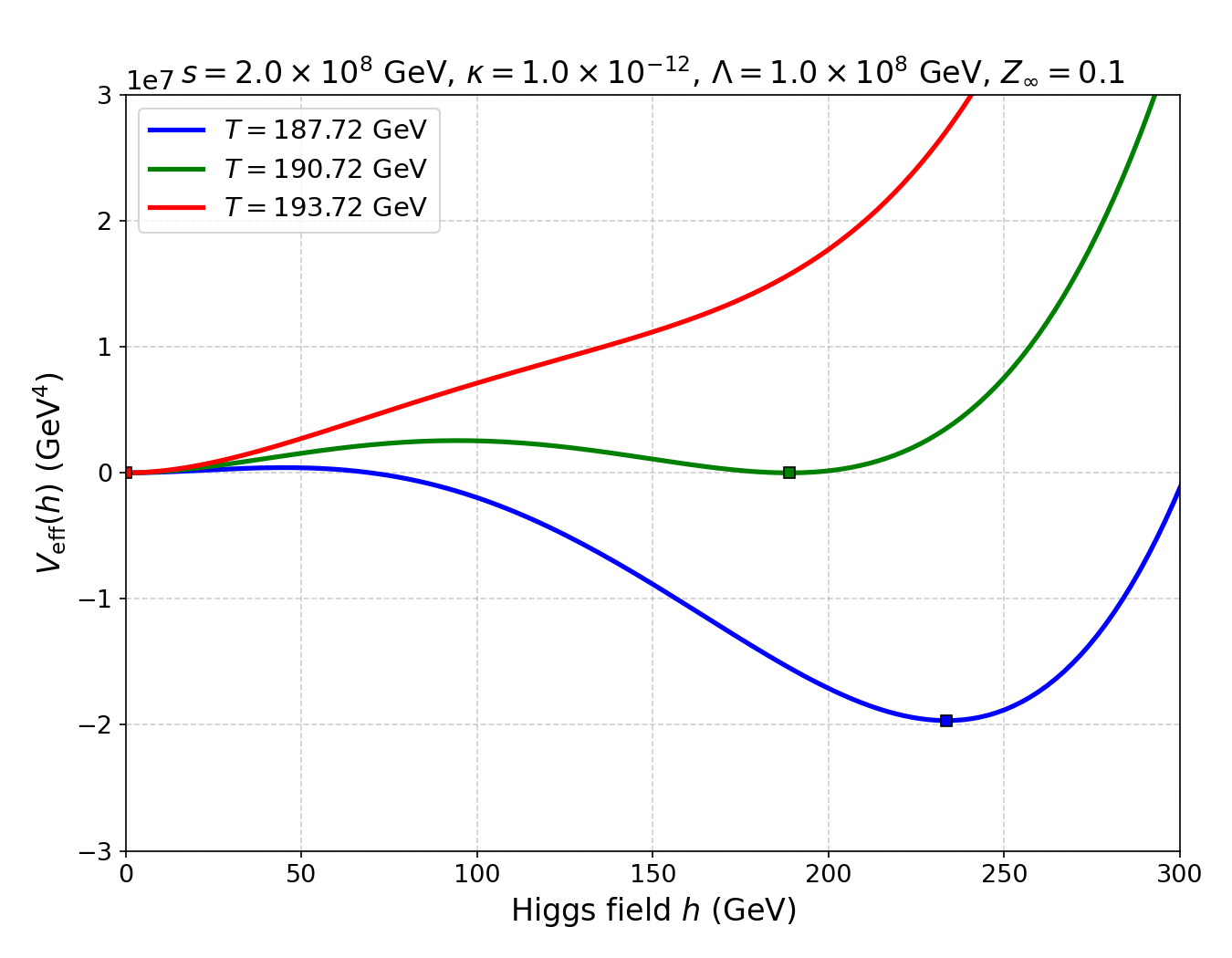}
    \caption{The effective potential as a function of the Higgs field value. We take $s=2\times 10^{8}$~GeV, $\kappa=10^{-12}$, $\Lambda=10^8$~GeV, and $Z_{\infty}=0.1$.}
    \label{fig:potential}
\end{figure}

Substituting into the Higgs portal interaction gives the space-dependent local effective potential:
\begin{equation}\label{eq:loc_Veff}
V_{\text{eff}}(h,T,s) = \frac{1}{2}m_{\text{loc}}^2(T,s)h^2 - E_{\rm eff}(s) T h^3 + \frac{\lambda_H}{4}h^4.
\end{equation}
where the local effective Higgs mass squared is:
\begin{equation}
m_{\text{loc}}^2(T,s) = -\mu^2 + D_{\rm eff}(s) T^2 - \frac{\kappa}{2}s^2.
\end{equation}
Equation~\eqref{eq:loc_Veff} is the leading-order high-temperature approximation of the full one-loop potential~\eqref{eq:Vfull}. It retains only the gauge-boson 
cubic barrier and the thermal-mass term, and omits the Goldstone, Higgs, top-quark, and daisy/ring contributions as well as the Boltzmann-suppressed regime 
$m_i\gtrsim T$. The split $D_{\rm eff}(s)=D_{\rm non}+D_{\rm gauge}/Z(s)$ isolates the gauge-boson piece (which scales as $1/Z(s)$); the Goldstone contribution is 
$s$-independent and absorbed into $D_{\rm non}$. This analytic form is used below to derive transparent estimates of the phase-transition conditions 
(Section~\ref{sec:conditions}); the extensive parameter-space scans (Figs.~\ref{fig:fFOPT}--\ref{fig:ab}) are performed in this cubic approximation 
for computational efficiency, and the complete daisy-resummed thermal functions~\eqref{eq:Vfull} are reserved for the benchmark-point evaluations of 
Table~\ref{tab:ewpt_parameters}.
We observe from Eq.~\eqref{eq:loc_Veff} that fluctuations in the spectator field can affect the EWPT by altering both the effective 
thermal Higgs mass and the tree-level thermal potential. 
The thermal mass is enhanced by a factor $1/Z(s)$, whereas the thermal barrier coefficient is thus universally amplified by the factor $1/Z(s)^{3/2}$.
In the ``bulk'' of the distribution where $s/\Lambda\ll 1$ and $Z(s)=1$, one recovers the SM value.
In the non-Gaussian tail where $s\gg\Lambda$ and $Z(s)=Z_{\infty}$, the barrier is largely enhanced by $E_{\text{eff}} \;=\; E_{\text{SM}}/Z_{\infty}^{3/2}$.
We illustrate the effective potential around the critical temperature in Fig.~\ref{fig:potential}.

\paragraph{Full one-loop thermal functions.}
The cubic potential~\eqref{eq:loc_Veff} is derived from the high-temperature expansion of the one-loop bosonic thermal function $J_B(y)$, valid when $m_i \ll T$. 
However, as we shall see in Section~\ref{sec:conditions}, within the cubic approximation the transition can be very strong 
($v_c/T_c \sim \mathcal{O}(5)$; the full one-loop resummed potential instead gives the considerably more modest values collected in 
Table~\ref{tab:ewpt_parameters}), and the effective gauge coupling is 
enhanced ($g_{\rm eff} \gtrsim 1.5$ for $Z_\infty = 0.1$), so that $m_W^{\rm eff}/T_c = (g_{\rm eff}/2)(v_c/T_c) \sim \mathcal{O}(1\text{--}5)$ is \emph{not} small. 
In this regime, the truncated cubic expansion is unreliable, and the full one-loop thermal functions must be employed for quantitative results.

The complete one-loop effective potential at finite temperature is the sum of the tree-level piece, the zero-temperature one-loop (Coleman--Weinberg) contribution, 
and the purely thermal loop:
\begin{equation}\label{eq:Vfull}
V_{\rm eff}(h, T, s) = V_0(h, s) + V_{\rm CW}(h,s) + V_T(h,T,s),
\end{equation}
where
\begin{equation}\label{eq:V0}
    V_0(h,s) = -\frac12\mu^2 h^2 +\frac14\lambda_H h^4 - \frac14\kappa s^2 h^2 + \frac12\mu_s^2 s^2
\end{equation}
is the tree-level potential, the zero-temperature Coleman--Weinberg loop is~\cite{Coleman:1973jx}
\begin{equation}\label{eq:VCW}
V_{\rm CW}(h,s) = \frac{1}{64\pi^2}\sum_i (-1)^{F_i}\,n_i\, m_i^4(h,s)\left[\ln\!\left(\frac{m_i^2(h,s)}{Q^2}\right) - C_i\right],
\end{equation}
where $n_i=\{1,1,3,6,3,6\}$ are the numbers of degrees of freedom for $\{ s,h,G,W^{\pm},Z,t\}$, $(-1)^{F_i}=+1$ for bosons and $-1$ for the fermionic top quark, and $C_i=3/2$ in the 4D~$\overline{\rm MS}$ scheme. The masses $m_i^2(h,s)$ are field dependent 
and include the $Z(s)$ modification.
The field-dependent gauge boson masses are $m_W^2(h,s) = g_{\rm eff}^2(s)\,h^2/4$ and $m_Z^2(h,s) = (g_{\rm eff}^2(s) + g_{\rm eff}^{\prime\,2}(s))\,h^2/4$, 
with $g_{\rm eff}(s) = g/\sqrt{Z(s)}$ as in Eq.~\eqref{eq:geff}. The top quark mass is $m_t(h) = y_t h/\sqrt{2}$, and the Higgs mass is 
$m_h^2(h,s) = -\mu^2 + 3\lambda_H h^2 - \kappa s^2/2$.
The $\overline{\rm MS}$ renormalisation scale $Q$ is fixed at the top quark mass $Q\simeq m_t\simeq 173\,{\rm GeV}$; 
the renormalisation condition is imposed on $V_0+V_{\rm CW}$ so that its minimum reproduces the physical 
Higgs VEV and mass, and the residual field- and $s$-dependence of the logarithms supplies the curvature that feeds the thermal barrier at $T_c$.

The purely thermal part of the potential is~\cite{Dolan:1973qd}
\begin{equation}\label{eq:VT}
V_T(h,T,s) = \sum_i \frac{n_i\, T^4}{2\pi^2}\,J_B\!\left(\frac{m_i^2(h,s)}{T^2}\right) - \sum_f \frac{n_f\, T^4}{2\pi^2}\,J_F\!\left(\frac{m_f^2(h,s)}{T^2}\right).
\end{equation}
where the bosonic and fermionic thermal functions are
\begin{align}
J_B(y) &= \int_0^\infty dx\, x^2 \ln\!\left(1 - e^{-\sqrt{x^2 + y}}\right), \\
J_F(y) &= \int_0^\infty dx\, x^2 \ln\!\left(1 + e^{-\sqrt{x^2 + y}}\right).
\end{align}
The daisy/ring-improved effective potential that we actually evaluate for the benchmark
points of Table~\ref{tab:ewpt_parameters} is obtained by adding the Arnold--Espinosa ring
correction~\cite{Arnold:1993rc,Espinosa:1994en} to the full one-loop potential~\eqref{eq:Vfull},
\begin{equation}\label{eq:Vdaisy}
V_{\rm eff}^{\rm daisy}(h,T,s) = V_{\rm eff}(h,T,s) + \delta V_{\rm ring}(h,T,s),
\end{equation}
where the ring term resums the leading-order Debye-screened longitudinal $W$ and $Z$ modes,
\begin{equation}\label{eq:Vring}
\delta V_{\rm ring}(h,T,s)
= -\frac{T}{12\pi}\sum_{i=W,Z} N_i^{(\rm L)}
\Bigl[\bigl(m_i^2(h,s)+\Pi_i(s,T)\bigr)^{3/2}
-\Pi_i(s,T)^{3/2}
-\bigl(m_i^2(h,s)\bigr)^{3/2}\Bigr],
\end{equation}
with $N_W^{(\rm L)}=2$, $N_Z^{(\rm L)}=1$ (longitudinal polarisations only) and
$\Pi_W=\tfrac{11}{6}g_{\rm eff}^2(s)T^2$,
$\Pi_Z=\tfrac{11}{6}\bigl(g_{\rm eff}^2(s)+g_{\rm eff}'^{\,2}(s)\bigr)T^2$.
The $-\Pi_i^{3/2}$ subtraction guarantees $\delta V_{\rm ring}(h=0,T,s)=0$, so the $T=0$ vacuum structure is untouched.

In the high-temperature limit $y \ll 1$, the bosonic thermal function expands as
\begin{equation}\label{eq:JBexpansion}
J_B(y) \to -\frac{\pi^4}{45} + \frac{\pi^2}{12}\,y - \frac{\pi}{6}\,y^{3/2} - \frac{1}{32}\,y^2\ln y + \cdots,
\end{equation}
where the first term gives the Stefan--Boltzmann energy density, the second gives the thermal mass correction $\propto D_{\rm eff} T^2 h^2$, and the third generates 
the cubic barrier $\propto -E_{\rm eff} T h^3$. The fermionic function contributes only to the thermal mass at leading order.

The cubic potential~\eqref{eq:loc_Veff} is therefore recovered as the leading-order high-$T$ approximation of the full expression~\eqref{eq:Vfull}. 
It remains a useful \emph{analytical} tool for deriving the phase-transition conditions (Section~\ref{sec:conditions}). For the extensive parameter-space scans 
presented in Figs.~\ref{fig:fFOPT}--\ref{fig:ab} we evaluate the phase-transition observables ($\bar f_{\rm FOPT}$, $\bar T_n$, $\bar\alpha$, $\overline{\beta/H}$) 
in the daisy-resummed high-$T$ cubic approximation, because evaluating the full one-loop thermal functions~\eqref{eq:Vfull} on the large multi-dimensional grid, 
with the fine temperature resolution required by the nucleation search, is computationally prohibitive. The daisy-resummed full one-loop thermal functions~\eqref{eq:Vdaisy} 
are instead reserved for the benchmark points of Table~\ref{tab:ewpt_parameters}, where we quote the accurate values of $f_{\rm FOPT}$, $\bar\alpha$, $\overline{\beta/H}$ 
and $\bar T_n$. We have verified that the two treatments agree to better than $5\%$ when $m_W^{\rm eff}/T \lesssim 0.5$, but diverge significantly for 
$m_W^{\rm eff}/T \gtrsim 1$, where the Boltzmann suppression $J_B(y)\to -(T/m)^{3/2}e^{-m/T}$ weakens the barrier and reduces $v_c/T_c$ relative to the cubic prediction.

\paragraph{Daisy-resummed cubic coefficient.}
In the standard resummed treatment~\cite{Arnold:1993rc,Carrington:1995hz},
the longitudinal gauge masses are screened by the Debye masses $\Pi_i(s,T)$, while the transverse modes remain unscreened.  
In the high-$T$ cubic limit this is the longitudinal-only truncation of Eq.~\eqref{eq:Vring}, 
the familiar resummed cubic barrier then arises from the transverse modes.  
In the formal $h\to 0$ expansion the longitudinal (ring) modes contribute only $h$-even powers, so the
pure $h^3$ cubic piece is set entirely by the transverse modes. 

More importantly, at the electroweak phase transition the Debye mass is \emph{not} negligible. At $v_c/T_c\simeq 1.0$ one finds
$\Pi_W/m_W^2=\tfrac{22}{3}(T_c/v_c)^2\simeq 7.3$, so the resummed longitudinal modes are dominated by the $T^2$ Debye term.  The leading-order Debye mass,
$m_D^2=\Pi_{00}(k_0{=}0)$, is gauge independent; this is precisely why the daisy-resummed potential exhibits a \emph{reduced} gauge dependence, as we now quantify.  
We therefore adopt the daisy-resummed result throughout, evaluated in Landau gauge. Although the resummed coefficient 
$E_{\rm eff}^{\rm res}=\tfrac{2}{3}E_{\rm eff}^{\rm unres}$ is smaller and yields a somewhat weaker barrier than the naive (unscreened) cubic value, 
it is the standard ring-improved result.  Screening the longitudinal modes replaces their gauge-fixed tree-level mass in the naive cubic term 
with the leading, gauge-invariant Debye mass, thereby \emph{reducing}, though not eliminating, the gauge dependence; residual gauge dependence from the 
high-temperature truncation of the thermal functions and the partial (longitudinal-only) resummation persists at subleading 
order~\cite{Arnold:1993rc,Carrington:1995hz,Patel:2011th,Katz:2015uja,Chiang:2017gwd}.  Our conclusion therefore does \emph{not} rely on the potential being gauge 
invariant; the phase-transition strength is anchored in the gauge-invariant sphaleron-rate criterion (Sec.~\ref{sec:introduction}), and the 
Table~\ref{tab:ewpt_parameters} observables ($f_{\rm FOPT}$, $\bar\alpha$, $\overline{\beta/H}$) are evaluated from the Landau-gauge complete one-loop potential and 
are gauge invariant as physical nucleation rates and energy densities.

To establish a consistent power counting in the presence of the $Z(s)$ modification, we note that in the bulk of the distribution where $Z(s) \simeq 1$, 
the standard hierarchy $g^2 \sim \lambda_H \sim \kappa$ is recovered and the daisy/ring resummation proceeds as in the SM. In the non-Gaussian tail where 
$Z(s) \to Z_\infty \ll 1$, the effective gauge coupling $g_{\rm eff} = g/\sqrt{Z_\infty}$ grows, and the loop expansion parameter $\epsilon_{\rm gauge} = g_{\rm eff}^2/(4\pi)$ increases. 
We establish the perturbativity bound on $Z_\infty$ in Section~\ref{sec:conditions} (see Eq.~\eqref{eq:pertbound} below). Within this bound, the ring-resummed 
treatment remains controlled, in the sense that the transverse-mode contribution to the cubic barrier, $E_{\rm eff}(s) \propto Z(s)^{-3/2}$, is robust under the resummation scheme, 
while higher-order corrections scale as $\mathcal{O}(g_{\rm eff}^4)$.

\paragraph{Accuracy of the cubic approximation.}
We stress that the cubic high-temperature approximation \emph{overestimates} the transition strength. The full one-loop potential~\eqref{eq:Vfull} used for the
Table~\ref{tab:ewpt_parameters} benchmark points evaluates the complete bosonic and fermionic
thermal functions $J_B$ and $J_F$ \emph{beyond} the truncated cubic expansion. In the
strong-transition regime of interest, $m_W^{\rm eff}/T_c\gtrsim 1$, the full functions capture
the Boltzmann-suppressed tail $J_B(y)\to -(T/m)^{3/2}e^{-m/T}$ that the cubic form
$J_B\simeq -\pi y^{3/2}/6$ misses, which lowers the barrier and reduces $v_c/T_c$ relative to
the cubic prediction. The benchmark quantities $f_{\rm FOPT}$, $\bar\alpha$ and
$\overline{\beta/H}$ are gauge invariant by construction (they are physical nucleation rates
and released-energy densities); their Table~\ref{tab:ewpt_parameters} values are therefore the
conservative ones relative to the cubic-scan estimates of Figs.~\ref{fig:fFOPT}--\ref{fig:ab}.
Accordingly, the cubic-approximation scans of Figs.~\ref{fig:fFOPT}--\ref{fig:ab} should be
read as an \emph{upper bound} on the FOPT volume, with the Table~\ref{tab:ewpt_parameters}
benchmarks providing the quantitative, full-potential results.

\paragraph{Gauge dependence and modern criteria.}
The ratio $v_c/T_c$, widely used as a diagnostic for the phase transition strength, is gauge-dependent (it is typically computed in Landau gauge). 
We employ it as an \emph{indicative} parameter throughout this work. The order of the transition is more rigorously fixed by the gauge-invariant three-dimensional effective-theory ratio $x_c=\bar\lambda_3/\bar g_3^2$, with the critical value $x_c\simeq0.11$ separating a first-order transition from a crossover~\cite{Rummukainen:1998as,Kajantie:1996mn}; we apply this criterion to the present mechanism in Section~\ref{sec:xsm_comparison}. The physically relevant, gauge-invariant requirement for electroweak baryogenesis is the resulting suppression of the sphaleron transition 
rate in the broken phase, $\Gamma_{\rm sph}(T_c)/T_c^4 \ll 1$, which ensures that any generated baryon asymmetry is preserved~\cite{DOnofrio:2014rug,Quiros:2019kam}. 
We also verify that the percolation criterion (Section~\ref{sec:nucleation}) is satisfied, providing an additional physically transparent diagnostic for the 
completion of the transition.

%=====================================================

\section{Conditions for first-order phase transition}
\label{sec:conditions}
\subsection{Criterion}
\label{sec:critical_condition}
Using the extremum condition $\partial V_{\rm eff}(h=v)/\partial h=0$ at the broken phase:
\begin{equation}\label{eq:ex_condition}
\frac{\partial V_{\text{eff}}(h=v)}{\partial h} = m_{\text{loc}}^2 v - 3E_{\rm eff}T v^2 + \lambda_H v^3 = 0,
\end{equation}
together with the local minimum condition at the broken phase:
\begin{equation}
    \frac{\partial^2 V_{\text{eff}}(h=v)}{\partial h^2}>0 \Rightarrow v(T,s)>\frac{3 E_{\rm eff} T}{2 \lambda_H}
\end{equation}
we then obtain the VEV at temperature $T$:
\begin{equation}\label{eq:vev_T}
v(T,s)
= \frac{3 E_{\rm eff} T + \sqrt{(3 E_{\rm eff} T)^2 - 4 \lambda_H m_{\rm loc}^2(T,s)}}{2 \lambda_H}.
\end{equation}

A necessary condition for a first-order phase transition is that Eq.~\eqref{eq:ex_condition} has two distinct positive real roots, i.e.\, the discriminant is positive:
\begin{equation}\label{eq:Delta}
\Delta \equiv \big(3 E_{\text{eff}}(s) T\big)^2 - 4\lambda_H\,m_{\text{loc}}^2(T,s) > 0 .
\end{equation}
Inserting the $s$-dependent local mass leads to
\begin{equation}\label{eq:lower_impl}
s^2 \;>\; \frac{2}{\kappa}\left[ \Big(D_{\text{eff}}(s) - \frac{9 E_{\text{eff}}^2(s)}{4\lambda_H}\Big)T^2 - \mu^2 \right].
\end{equation}
Recall that the barrier grows as $E_{\text{eff}}\propto 1/Z(s)^{3/2}$,
while the thermal mass $D_{\text{eff}}(s)$ grows only as $1/Z(s)$. For sufficiently large $s$, we have $Z(s)\to Z_{\infty}\ll1$. 
The right-hand side of Eq.~\eqref{eq:lower_impl} then tends to a large negative number, making the inequality $s^2 > \text{negative constant}$ automatically satisfied for all physical $s^2 > 0$.
In other words, the lower bound completely disappears in the large-$s$ regime.  The condition $\Delta > 0$ does not restrict the spectator field amplitude from above.

Furthermore, to ensure that the symmetric phase \(h=0\) is a minimum (not a maximum), we require 
\begin{equation}\label{eq:mlocal}
    \frac{\partial^2 V_{\rm eff}(h=0)}{\partial h^2}>0 \Longrightarrow m_{\rm loc}^2 > 0.
\end{equation}
If \(m_{\rm loc}^2 \le 0\), the potential barrier disappears and the transition becomes second-order or a smooth crossover.
In the small field limit, i.e.\, $s\ll \Lambda$, we can expand
\begin{equation}\label{eq:Zappro}
    \frac{1}{Z(s)}\approx 1 + (1-Z_{\infty})\frac{s^2}{\Lambda^2} + \mathcal{O}(s^4/\Lambda^4).
\end{equation}
Substituting Eq.~\eqref{eq:Deff} and rearranging yields
\begin{equation}
s^2 \left[ 1 - \frac{2 D_{\text{gauge}}(1-Z_{\infty}) T^2}{\kappa\,\Lambda^2} \right] \;<\; \frac{2}{\kappa}\left(D_{\text{SM}} T^2 - \mu^2\right).
\end{equation}
Define the dimensionless parameter
\begin{equation}\label{eq:varsigma}
\varsigma  \equiv \frac{2 D_{\text{gauge}}(1-Z_{\infty}) T^2}{\kappa\,\Lambda^2}.
\end{equation}
\begin{itemize}
\item If $\varsigma  \ge 1$, the left-hand side has a non-positive coefficient, and the inequality holds automatically for any $s^2$, provided the right-hand side is positive.  In this regime, the upper bound on $s$ disappears entirely.
\item If $\varsigma  < 1$, we obtain the explicit upper limit
      \begin{equation}\label{eq:upper}
      s^2 \;<\; \frac{\frac{2}{\kappa}(D_{\text{SM}} T^2 - \mu^2)}{1-\varsigma }.
      \end{equation}
\end{itemize}
In the typical parameter space at the electroweak scale, $T\sim 100$~GeV, the numerator $2 D_{\text{gauge}} T^2 \approx 1.7\times 10^3\,\text{GeV}^2$.
The condition $\varsigma \gtrsim 1$ gives $\kappa\Lambda^2 \;\lesssim\; 1.7\times 10^3\;\text{GeV}^2$.
For $\Lambda \sim 10^{10}\,\text{GeV}$, this translates into $\kappa \lesssim 1.7\times 10^{-17}$, which is precisely the range of $\kappa$ relevant for realising a large first-order volume fraction.
Thus, for the vast majority of the parameter space, the upper bound is trivially evaded.

The higher-dimensional kinetic operator modifies the finite-temperature coefficients in such a way that the original first-order phase transition constraint in the SM becomes less restrictive.
The ``window'' of $s$ values capable of driving a first-order phase transition is open towards large $s$.  Any non-Gaussian tail that produces sufficiently 
large spectator fluctuations will naturally fall into the allowed region, and the transition will be correspondingly stronger.
This confirms that the mechanism remains fully self-consistent, and the field-dependent gauge coupling operator can simultaneously enhance the phase transition 
strength while preserving the stability of the symmetric phase.

The critical temperature \(T_c\) is defined as the temperature at which the free-energy densities of the symmetric and broken phases are equal, i.e.\, the pressures in the two phases coincide:
\begin{equation}
V_{\rm eff}(0, T_c, s)
= V_{\rm eff}\big(v_c(T_c,s), T_c, s\big),
\end{equation}
where \(v_c \equiv h_+(T_c)\) is the field value in the broken phase at \(T_c\). Substituting the potential~\eqref{eq:loc_Veff} with \(V_{\rm eff}(0)=0\) as the zero of energy, we obtain
\begin{equation}
\frac{1}{2} m_{\rm loc}^2 v_c^2 - E_{\rm eff} T_c v_c^3 + \frac{\lambda_H}{4} v_c^4 = 0 ,
\end{equation}
Combining this with the local minimum condition~\eqref{eq:ex_condition} at \(T_c\), we obtain a general relation:
\begin{equation}\label{eq:vtoT}
    \frac{v_c}{T_c}=\frac{2E_{\rm eff}}{\lambda_H} .
\end{equation}
With this result, we have:
\begin{equation}
T_c(s) = \sqrt{\frac{\mu^2 + \frac{\kappa}{2}s^2}{D_{\rm eff} - \frac{2E_{\rm eff}^2}{\lambda_H}}},
\end{equation}
\begin{equation}
v_c(s) = \frac{2E_{\rm eff}}{\lambda_H} \sqrt{\frac{\mu^2 + \frac{\kappa}{2}s^2}{D_{\rm eff} - \frac{2E_{\rm eff}^2}{\lambda_H}}}.
\end{equation}
We observe that to have real solutions for $T_c$ and $v_c$, the condition
\begin{equation}\label{eq:cond3}
    D_{\rm eff} - \frac{2E_{\rm eff}^2}{\lambda_H}>0
\end{equation}
should be satisfied. We first solve the equation:
\begin{equation}\label{eq:fx0}
    f(y)=D_{\rm SM} + D_{\rm gauge}y - \frac{2E_{\rm SM}^2}{\lambda_H}(1+y)^3=0,
\end{equation}
where $y=s^2/\Lambda^2$. We have adopted the approximation~\eqref{eq:Zappro} and $Z_{\infty}\sim0$ for Eq.~\eqref{eq:fx0}.
The numerical solution is found to be $y_0\simeq 12.39$. Then the condition~\eqref{eq:cond3} gives:
\begin{equation}\label{eq:Tcbound}
    \frac{s^2}{\Lambda^2} < y_0\simeq 12.39 \Rightarrow s(T_c) < 3.52 \Lambda .
\end{equation}
Using Eqs.~\eqref{eq:Eeff} and~\eqref{eq:vtoT}, we find that the field-dependent gauge coupling increases the SM value $r_{\rm SM}\equiv 2E_{\rm SM}/\lambda_H\simeq 0.098$ 
to $r(y)\approx r_{\rm SM}(1+y)^{3/2}$. Consequently, in our framework the ratio can reach $r=v_c/T_c\approx 4.78$, which satisfies the condition for a strongly 
first-order phase transition, i.e.\, $v_c/T_c\gtrsim 1.0$, required to sufficiently suppress the electroweak sphaleron process.

We emphasise that the analytical results above, derived from the cubic potential~\eqref{eq:loc_Veff}, provide important qualitative insight into the mechanism. 
For the quantitative benchmark predictions of Table~\ref{tab:ewpt_parameters}, the full one-loop thermal functions~\eqref{eq:Vfull} (including the daisy-resummed ring term~\eqref{eq:Vdaisy}) 
are used at the level of individual \((s, \Lambda, \kappa)\) points; the large parameter-space scans instead employ the cubic approximation. 
Numerically, for \(\Lambda = 10^8 \, \text{GeV}\), \(Z_\infty = 0.1\), \(\kappa = 10^{-14}\), the full one-loop potential exhibits no first-order transition 
(a crossover) for \(s/\Lambda = 0.5, 1.0, 2.0\), where the cubic approximation spuriously predicts \(v_c/T_c \approx 0.13, 0.24, 0.66\), whereas for 
\(s/\Lambda = 3.0, 5.0\) it gives \(v_c/T_c \approx 1.06, 1.14\), far below the cubic 1.18, 1.98; the cubic prediction overshoots the full one-loop result for 
\(s/\Lambda \gtrsim 3\), where Boltzmann suppression of the gauge-boson modes weakens the thermal barrier relative to the high-\(T\) cubic form. 
The core mechanism survives. The non-Gaussian tail produces regions with \(Z(s)\ll 1\), enhancing the gauge coupling and inducing a first-order barrier. 
The condition \(v_c/T_c>1\) for the full one-loop resummed potential holds only for \(s/\Lambda\gtrsim 3\), and the cubic approximation strongly overestimates it.
For a representative benchmark point (\(H_{\text{inf}} = 10^{12} \, \text{GeV}\), \(\kappa = 1.14 \times 10^{-16}\), \(\Lambda = 10^8 \, \text{GeV}\), \(Z_\infty = 0.15\)), 
the daisy-resummed calculation gives \(T_n \simeq 157 \, \text{GeV}\), \(\alpha \simeq 2.8 \times 10^{-3}\), \(\beta/H_n \simeq 5.7 \times 10^3\), \(v_c/T_c \simeq 1.05\) (cf. Table~\ref{tab:ewpt_parameters}). 
The reduced \(\alpha\) compared to the cubic prediction (\(\alpha \sim 0.07\)) implies a weaker GW signal, consistent with the benchmark range quoted in the table.

We stress that Eq.~\eqref{eq:Tcbound} is a discriminant derived within the \emph{cubic} high-$T$ effective potential (Eq.~\eqref{eq:fx0}) in the limit $Z_\infty\to0$, and is evaluated at the critical temperature $T_c$. Its upper bound $s/\Lambda<3.52$ is \emph{highly dependent on the value of $Z_\infty$}, because the effective gauge coupling $g_{\rm eff}=g/\sqrt{Z_\infty}$ grows as $Z_\infty$ is reduced and the barrier condition accordingly tightens or relaxes substantially away from the $Z_\infty\to0$ limit. The bound therefore applies \emph{only} to that cubic $Z_\infty\to0$ set-up.

The spectator values $s/\Lambda=3.0$ and $5.0$ used in the cubic-versus-full comparison above are field-space coordinates sampled at fixed $s$; for these points the \emph{full} daisy-resummed potential (with $Z_\infty=0.1$) is evaluated at its own $T_c$ to obtain $v_c/T_c$. Because this employs a different potential \emph{and} a different $Z_\infty$ from the cubic $Z_\infty\to0$ regime underlying Eq.~\eqref{eq:Tcbound}, the bound $s/\Lambda<3.52$ does \emph{not} constrain these points and they must not be read as violating it. In all numerical calculations we have explicitly verified that the bounds~\eqref{eq:Delta}, \eqref{eq:mlocal}, and \eqref{eq:cond3} are satisfied.

For the \emph{full one-loop daisy-resummed} potential of Table~\ref{tab:ewpt_parameters}, by contrast, the FOPT window and the actual nucleation amplitude $s^*/\Lambda$ are obtained numerically and are \emph{not} bounded by Eq.~\eqref{eq:Tcbound}; the resummed treatment shifts the FOPT window and yields nucleation points at $s^*/\Lambda\sim 40$--$80$ (Table~1), fully consistent with the mechanism developed below.

\subsection{Perturbativity bounds}
\label{sec:perturbativity}

The enhancement of the phase transition lives in the tail of the $s$ distribution where $Z(s) \to Z_\infty \ll 1$, so that $g_{\rm eff} = g/\sqrt{Z_\infty}$ becomes 
large. It is essential to determine the range of $Z_\infty$ for which the perturbative treatment remains reliable.

The loop expansion parameter for the gauge sector at finite temperature is
\begin{equation}\label{eq:pertbound}
\epsilon_{\rm gauge} \equiv \frac{g_{\rm eff}^2}{4\pi} = \frac{g^2}{4\pi\,Z_\infty}.
\end{equation}
For the SM value $g \simeq 0.65$:
\begin{itemize}
\item For $Z_\infty = 0.1$ one has $\epsilon_{\rm gauge} \approx 0.34$, so perturbation theory is marginal but still controlled.
\item For $Z_\infty = 0.05$ one has $\epsilon_{\rm gauge} \approx 0.67$, and perturbation theory is unreliable.
\item For $Z_\infty \simeq 0.034$ one has $\epsilon_{\rm gauge} \approx 1.0$, where perturbation theory breaks down.
\end{itemize}
We therefore adopt $Z_\infty \geq 0.1$ as a conservative lower bound throughout this work. This bound is used to constrain the parameter space in all subsequent numerical results.

This bound controls the leading-order thermal barrier and also propagates to higher orders. In our treatment we evaluate the complete daisy-resummed one-loop 
thermal potential, retaining the full bosonic and fermionic thermal functions $J_B$ and $J_F$ rather than a high-temperature cubic truncation. The gauge loop 
expansion is then organised in powers of the effective coupling, with the leading thermal barrier set by the tree-level gauge masses entering $J_B$ and the first 
subleading (NLO) corrections appearing at $\mathcal{O}(g_{\rm eff}^4)$. Because the bound $\epsilon_{\rm gauge}=g_{\rm eff}^2/(4\pi)\lesssim 0.34$ for 
$Z_\infty\geq 0.1$ implies $g_{\rm eff}^4/(4\pi)^2\lesssim 0.12$, the NLO corrections are suppressed by an additional factor of $\epsilon_{\rm gauge}$ relative to 
the leading term; the dominant two-loop ``basketball'' diagram estimated below is indeed at the $\sim 12\%$ level. This is the self-consistency condition under 
which the perturbative treatment of the $Z(s)$-enhanced gauge sector remains reliable for $Z_\infty \geq 0.1$.

The infrared (Linde) problem of the gauge sector also warrants attention. The non-perturbative magnetic mass of the SU(2) gauge boson is 
$m_{\rm mag} \sim g_{\rm eff}^2\,T$, which cannot be computed perturbatively. For $Z_\infty \geq 0.1$, $m_{\rm mag}$ remains a subleading correction to the 
electric (Debye) mass $m_D \sim g_{\rm eff} T$, and the one-loop potential is a reasonable first approximation. For $Z_\infty < 0.1$, a lattice or 
dimensional-reduction treatment would be necessary, which is beyond the scope of this work.

We have estimated the size of the dominant two-loop correction, which arises from the gauge-boson ``basketball'' diagram and scales as 
$\sim g_{\rm eff}^4 T^4/(16\pi^2)$. For $Z_\infty = 0.1$, this is approximately $(0.34)^2 \approx 12\%$ of the one-loop contribution, confirming that perturbation 
theory is still meaningful but with non-negligible higher-order corrections. A full two-loop calculation along the lines of Ref.~\cite{Arnold:1993rc} would be 
desirable for a definitive treatment.

Finally, we comment on the standard dimensional-reduction (DR) framework. In the presence of the field-dependent gauge kinetic function $Z(s)$, the three-dimensional 
effective theory is valid \emph{locally} (for fixed $s$), with matching coefficients that acquire $s$-dependence through $g_{\rm eff}(s)$. The DR framework provides 
a systematic organisation of the resummation~\cite{Laine:1997nq} and would allow a controlled inclusion of higher-order effects. We leave a full DR analysis for 
future work, noting that our daisy-resummed one-loop treatment, together with the perturbativity bound $Z_\infty \geq 0.1$, provides a self-consistent estimate.

%==========================================================================
\section{Scalar perturbation from inflation}\label{sec:scalar_evolution}
\subsection{Evolution of quantum fluctuations}
In this section, we present the evolution of the $s$ field from inflation to the present day.
In the inflationary epoch, the universe undergoes approximately de Sitter exponential expansion with an almost constant Hubble parameter $H_{\text{inf}}$. 
The $s$ field is light, satisfying $m_s \ll H_{\text{inf}}$, and the power spectrum of its quantum fluctuations reads~\cite{Baumann:2009ds}
\begin{equation}
\mathcal{P}_s(k) = \left( \frac{H_{\text{inf}}}{2\pi} \right)^2.
\end{equation}
Each field mode is frozen upon exiting the Hubble horizon, with a characteristic amplitude $H_{\text{inf}}/(2\pi)$. Due to non-Gaussianity ($f_{\text{NL}} \neq 0$), 
the probability distribution of the field value deviates from a Gaussian form, while the typical fluctuation magnitude is still governed by $H_{\text{inf}}$.
During inflation, the $s$ field acquires quantum fluctuations. The initial mean-squared displacement is give by:
\begin{equation}
\langle s^2\rangle = \int_{-\infty}^{\infty} s^2\,\mathcal{P}(s)\,ds = 2\int_0^\infty s^2\,\mathcal{P}(s)\,ds.
\end{equation}
We define the root-mean-square amplitude as
\begin{equation}\label{eq:srms}
s_{\text{rms}} \equiv \sqrt{\langle s^2 \rangle} \approx \frac{H_{\text{inf}}}{2\pi} \sqrt{1 + \frac{3}{4}f_{\text{NL}}^2},
\end{equation}
where the correction corresponds to the non-Gaussian contribution (see Eq.~\eqref{eq:non_Gaussian} below).

The derivation of Eq.~\eqref{eq:srms} proceeds as follows. Using the local non-Gaussian parametrisation $s = \sigma_s\,(g + \frac{f_{\rm NL}}{2}\,g|g|)$ with $g\sim\mathcal{N}(0,1)$ and $\sigma_s = H_{\rm inf}/(2\pi)$, we compute
\begin{equation}
\langle s^2\rangle = \sigma_s^2\left\langle\left(g + \frac{f_{\rm NL}}{2}\,g|g|\right)^{\!2}\right\rangle = \sigma_s^2\left(\langle g^2\rangle + f_{\rm NL}\,\langle g^2|g|\rangle + \frac{f_{\rm NL}^2}{4}\,\langle g^4\rangle\right).
\end{equation}
The cross term vanishes by symmetry, $\langle g^2|g|\rangle = 0$ (odd integrand), and the standard Gaussian moments give $\langle g^2\rangle = 1$ and $\langle g^4\rangle = 3$. Therefore
\begin{equation}
\langle s^2\rangle = \sigma_s^2\!\left(1 + \frac{3\,f_{\rm NL}^2}{4}\right),
\end{equation}
which yields Eq.~\eqref{eq:srms}. We note that this relation is specific to the local-type non-Gaussianity model of Eq.~\eqref{eq:non_Gaussian}; other non-Gaussianity types (equilateral, orthogonal, folded) would modify the quantitative relation between $s_{\rm rms}$ and $f_{\rm NL}$ but not the qualitative mechanism.

After inflation, these super-Hubble perturbations evolve into classical, spatially inhomogeneous field configurations, and their amplitudes remain constant before 
horizon re-entry due to Hubble friction.
For the quadratic potential $V(s)=\tfrac12 m_s^2 s^2$ in a radiation-dominated background, the classical equation of motion for the $s$ field is
\begin{equation}
\ddot{s} + 3H\dot{s} + m_s^2 s = 0,
\end{equation}
where the scalar mass $m_s^2=\mu_s^2-\kappa \langle H \rangle^2$ is treated as a free parameter.
For $H \gg m_s$, the Hubble damping term dominates over the mass term, leading to the approximate solution $s(t) \approx \text{constant} + \mathcal{O}(m_s^2/H^2)$.
The field value is therefore frozen at its initial condition set by inflation. Before the onset of oscillation, the $s$ value at each spatial point remains unchanged, 
and $s_{\text{rms}}$ is conserved. The corresponding energy density $\rho_s = \tfrac12 m_s^2 s^2$ is negligible owing to the tiny bare mass $m_s$, which is far 
subdominant compared to the radiation energy density and does not affect cosmic expansion.

Coherent oscillation is triggered when the Hubble parameter drops to the scale of $m_s$, such that Hubble damping becomes subdominant. 
The $s$ field starts oscillating when $3H(T_{\text{osc}}) = m_s$. Solving for $T_{\text{osc}}$ gives:
\begin{equation}
T_{\text{osc}} = \left(\frac{m_s M_{\text{Pl}}}{3\cdot1.66\sqrt{g_*}}\right)^{1/2}
\simeq 27.2~{\rm GeV}\left(\frac{m_s}{10^{-6}~{\rm eV}} \right)^{1/2}\left(\frac{g_{*}(T_{\rm osc})}{10.75} \right)^{-1/4},
\end{equation}
where $g_*(T)$ is the relativistic degrees of freedom at that epoch. For $T\gg 100\ \text{GeV}$, $g_*\approx 106.75$ and for $T\sim \text{MeV}$, $g_*\approx 10.75$.

Well after oscillation starts, $H \ll m_s$, the mass term dominates and Hubble damping can be safely neglected. The equation of motion reduces to simple harmonic motion:
\begin{equation}
\ddot{s} + m_s^2 s \approx 0,
\end{equation}
with formal solution $s(t) \approx A\cos(m_s t + \phi)$. Nevertheless, cosmic expansion gradually suppresses the oscillation amplitude. 
The covariant field equation takes the form
\begin{equation}
\frac{d}{dt}\big(a^3 \dot{s}\big) + a^3 m_s^2 s = 0.
\end{equation}
By introducing conformal time or employing the WKB approximation, we substitute the ansatz $s(t) = a(t)^{-3/2}\chi(t)$ for $m_s \gg H$, 
where $\chi(t)$ undergoes undamped harmonic oscillation. The solution is:
\begin{equation}
s(t) \approx s_i \left( \frac{a_{\text{osc}}}{a(t)} \right)^{\!\frac32} \cos(m_s t + \phi),
\end{equation}
in which $s_i$ is the initial oscillation amplitude equal to $s_{\text{rms}}$ at $T_{\text{osc}}$. 
The field amplitude evolves independently at each spatial position while obeying the same dilution law, scaling as $a^{-3/2}$. 
The total energy density $\rho_s\approx \frac{1}{2}m_s^2s^2\propto a^{-3}$ consistent with matter-like dilution. 

From entropy conservation in the thermal plasma, the scale factor is related to the temperature by
\begin{equation}
\left( \frac{a(T_{\text{osc}})}{a(T)} \right)^{\!3}
= \frac{g_{*s}(T)}{g_{*s}(T_{\text{osc}})} \left( \frac{T}{T_{\text{osc}}} \right)^{\!3}.
\end{equation}
We therefore can write the scalar fluctuation as:
\begin{equation}\label{eq:s_evolution}
s(t)\approx\left\{\begin{matrix}
s_{\rm rms}  & {\rm for}\quad T>T_{\rm osc}, \\
s_{\rm rms}\left(\frac{g_{*s}(T)}{g_{*s}(T_{\text{osc}})}\right)^{\!1/2} \left( \frac{T}{T_{\text{osc}}} \right)^{\!3/2}  
& {\rm for}\quad T<T_{\rm osc}.
\end{matrix}\right.
\end{equation}

\subsection{Oscillating dark matter}\label{sec:DM}
The initial displacement is generated by non-Gaussian fluctuations during inflation, and it subsequently undergoes coherent oscillations as the universe expands, behaving like cold DM.

 A key point that warrants clarification is the following. The $\mathbb{Z}_2$ symmetry $s\to -s$ enforces a vanishing ensemble average, $\langle s\rangle = 0$. However, the dark matter energy density depends on $s^2$, not $s$:
\begin{equation}
\rho_s(\mathbf{x}) = \frac{1}{2}\,m_s^2\,s(\mathbf{x})^2 \;\geq\; 0 .
\end{equation}
While $\langle s\rangle = 0$, the mean-squared amplitude $\langle s^2\rangle = s_{\rm rms}^2 > 0$ is non-zero. At each spatial point, the field $s(\mathbf{x})$ has a definite value (set by the inflationary fluctuation at that point), and the local energy density $\rho_s(\mathbf{x})$ is positive definite. After the onset of coherent oscillations ($H\lesssim m_s$), the field at each point oscillates with amplitude $s(\mathbf{x})$, and the energy density redshifts as $a^{-3}$, exactly as cold dark matter. The spatially averaged (equivalently, ensemble-averaged) energy density is
\begin{equation}
\bar{\rho}_s = \frac{1}{2}\,m_s^2\,\langle s^2\rangle = \frac{1}{2}\,m_s^2\,s_{\rm rms}^2 ,
\end{equation}
which is the quantity used in Eq.~\eqref{eq:DM_abundance} for the relic abundance.

This mechanism is directly analogous to the axion dark matter scenario, where the axion field $\theta(\mathbf{x})$ may have $\langle\theta\rangle = 0$ in general, but $\langle\theta^2\rangle > 0$ provides the dark matter density~\cite{Preskill:1982cy,Abbott:1982af,Dine:1982ah}. The $\mathbb{Z}_2$ symmetry plays a role analogous to the axion shift symmetry. It guarantees the stability of $s$ (forbidding its decay into SM particles except through the highly suppressed portal coupling $\kappa$), making $s$ a viable DM candidate. We adopt the supersymmetric UV completion of Appendix~\ref{app:UVcompletion} as the preferred framework, in which the smallness of $m_s$ is protected by supersymmetry and is technically natural. In the limit $m_s\to 0$, both the $\mathbb{Z}_2$ symmetry and supersymmetry are restored.

\paragraph{Radiative stability of the spectator mass.}
The correct DM relic-density condition~\eqref{eq:DM_abundance} correlates the spectator mass with the fluctuation amplitude as 
$m_s\propto s_{\rm rms}^{-4}$, so that across the range $s_{\rm rms}\sim10^{10}$--$10^{13}$~GeV probed by our scan the required mass spans roughly 
$m_s\sim10^{6}$--$10^{-6}$~eV. The value $m_s\sim10^{-6}$~eV corresponds to the benchmark region $s_{\rm rms}\sim10^{13}$~GeV, 
where $f_{\rm FOPT}$ is largest. A small $m_s$ is radiatively unstable unless it is protected. Besides the portal coupling $\kappa$, the 
field-dependent kinetic operator itself couples $s$ directly to the electroweak gauge sector. Expanding Eq.~\eqref{eq:Zparam} at $s\ll\Lambda$ gives 
$\mathcal{L}\supset-\tfrac14 Z(s)F_{\mu\nu}F^{\mu\nu}\supset\tfrac{c}{4}\frac{s^{2}}{\Lambda^{2}}F_{\mu\nu}F^{\mu\nu}$, with $c=1-Z_{\infty}=\mathcal{O}(1)$. 
Closing a gauge-boson loop with two insertions of this dimension-six operator yields
\begin{equation}\label{eq:dm_s_gauge}
\delta m_s^{2}\;\sim\;\frac{N}{16\pi^{2}}\,\frac{c}{\Lambda^{2}}\,\Lambda_{\rm UV}^{4}
\;\sim\;\frac{c}{16\pi^{2}}\,\Lambda^{2}
\;\sim\;6\times10^{13}~{\rm GeV}^{2}
\qquad (\Lambda_{\rm UV}\sim\Lambda=10^{8}~{\rm GeV}),
\end{equation}
i.e.\ $\delta m_s\sim10^{7}$~GeV, many orders of magnitude above the values required by the DM fit, though still below $H_{\rm inf}$, so that the inflationary 
fluctuations themselves are unaffected. An unprotected spectator would nevertheless be hopelessly fine-tuned, since its late-time DM phenomenology would be set 
by the radiative correction rather than by the bare mass.

This is precisely why we adopt the globally supersymmetric completion of Appendix~\ref{app:UVcompletion}. There the gauge kinetic function is a holomorphic 
function of the chiral superfield $S$; the quadratic and quartic divergences of Eq.~\eqref{eq:dm_s_gauge} cancel between the gauge-boson loop and its gaugino 
partner, and the non-renormalisation theorem forbids the generation of a supersymmetric mass for $S$. After soft breaking, the residual correction is of order 
the gravitationally mediated soft mass itself, so a small $m_s$ is technically natural rather than tuned, whichever value in the above range is selected. We stress that this protection is a 
property of the UV completion and not of the low-energy effective theory. In the non-supersymmetric completions of Appendix~\ref{app:UVcompletion} the 
estimate~\eqref{eq:dm_s_gauge} applies unchanged, and an additional stabilisation mechanism would be required there.

At the onset of oscillations, the energy density of the $s$ field is dominated by the potential term:
\begin{equation}
\rho_s(T_{\text{osc}}) = \frac{1}{2} m_s^2 s_{\text{rms}}^2.
\end{equation}
After oscillation begins, the energy density of the $s$ field dilutes as matter with cosmic expansion:
\begin{equation}
\rho_s(T_0) = \rho_s(T_{\text{osc}})\left(\frac{a_{\text{osc}}}{a_0}\right)^3 = \rho_s(T_{\text{osc}})\frac{g_{*s}(T_0)}{g_{*s}(T_{\text{osc}})}\left(\frac{T_0}{T_{\text{osc}}}\right)^3,
\end{equation}
where $T_0=2.725\,\text{K}=2.35\times10^{-13}\ \text{GeV}$, and the quantities $g_{*s}(T_0)=3.91$ and $g_{*s}(T_{\text{osc}})\approx g_*(T_{\text{osc}})$ are the 
effective entropy degrees of freedom at the present epoch and at $T_{\text{osc}}$, respectively.
Combining the results, the DM energy relic abundance today is given by:
\begin{equation}\label{eq:DM_abundance}
\Omega_s h^2 = \frac{\rho_s(T_0)}{\rho_c/h^2} = 0.184\left(\frac{m_s}{10^{-6}~{\rm eV}} \right)^{1/2}\left(\frac{s_{\rm rms}}{10^{13}~{\rm GeV}} \right)^2\left(\frac{g_{*}(T_{\rm osc})}{3.91} \right)^{-1/4},
\end{equation}
where we have used the critical density of today $\rho_c/h^2 = 1.05\times10^{-5}\ \text{GeV/cm}^3$. 

%===========================================================
\subsection{$\mathbb{Z}_2$-symmetric non-gaussian distribution}
During inflation, the light spectator field acquires quantum fluctuations that are stretched to cosmological scales. These fluctuations are generally 
non-Gaussian~\cite{Bartolo:2004if,Lyth:2005fi}, which significantly enhances the probability of large-amplitude fluctuations.

To satisfy the symmetry $s \leftrightarrow -s$, the nonlinear term must be an odd function (ensuring symmetric amplification of positive and negative fluctuations). 
The lowest-order $\mathbb{Z}_2$-invariant nonlinear term is $g |g|$, leading to the definition:
\begin{equation}\label{eq:non_Gaussian}
s = \sigma_s \cdot \left[ g + \frac{f_{\text{NL}}}{2} \cdot g |g| \right]~,
\end{equation}
where $g \sim \mathcal{N}(0,1)$ is the dimensionless normalised Gaussian random variable, satisfying $\mathbb{E}[g]=0$, $\mathbb{E}[g^2]=1$. 
This ensures $\mathbb{E}[s] = 0$, automatically satisfying the zero background field required by $\mathbb{Z}_2$ symmetry.
$\sigma_s = H_{\text{inf}}/(2\pi)$ is the root-mean-square amplitude of the scalar field perturbation, representing the characteristic scale of primordial 
fluctuations generated by inflation. $f_{\text{NL}}$ is a dimensionless non-Gaussian parameter, describing the degree of nonlinearity in the fluctuations.
In this work, we fix $f_{\rm NL}=5.0$ in our numerical calculations.
It is easy to check that under the transformation $g \to -g$, the fluctuation changes as $s(-g)=-s(g)$.
That is, $g$ maps to $s$, and $-g$ necessarily maps to $-s$. 

The exact $\mathbb{Z}_2$-symmetric probability density function (PDF) of $s$, derived via change of variables and the Jacobian determinant 
(see Appendix~\ref{app:exact_PDF}), is given by:
\begin{equation}\label{eq:non_Gaussian_distribution}
\mathcal{P}(s) = \frac{1}{\sqrt{2\pi} \sigma_s \cdot \left( 1 + f_{\text{NL}} |g(s)| \right)} \exp\left( -\frac{g(s)^2}{2} \right)~.
\end{equation}
The one-point probability density function \(\mathcal{P}(s)\) is defined such that, at any spatial point \(x\), the probability that \(s\) lies within an infinitesimal 
interval \([s, s+ds]\) is \(\mathcal{P}(s)\,ds\).
Using Eq.~\eqref{eq:non_Gaussian}, one finds that the probability distribution is an even function, $\mathcal{P}(-s)=\mathcal{P}(s)$, 
thereby strictly respecting the $\mathbb{Z}_2$ symmetry.

In the bulk of the distribution where $2f_{\text{NL}} \cdot |s/\sigma_s| \ll 1$, the PDF in Eq.~\eqref{eq:non_Gaussian_distribution} reduces to the Gaussian distribution. 
Non-Gaussianity significantly enhances the probability of moderately large fluctuations, even within the bulk of the distribution. This enhancement is essential for 
producing a substantial volume fraction of the Universe that undergoes a strong first-order phase transition.

\subsection{Isocurvature constraints and the power spectrum}\label{sec:isocurvature}

The spectator field $s$ acquires super-Hubble fluctuations during inflation and, as discussed in Section~\ref{sec:DM}, later constitutes the dark matter. In this respect, $s$ behaves similarly to a curvaton field, and its fluctuations generate isocurvature perturbations that are constrained by CMB observations~\cite{Planck:2018vyg}. In this subsection we discuss the impact of isocurvature bounds on our scenario and specify the assumed shape of the power spectrum.

\paragraph{Isocurvature constraint.}
If $s$ constitutes all of the dark matter, its relative fluctuations contribute to the CMB isocurvature power spectrum
\begin{equation}
\alpha_{\rm iso} = \frac{P_{\mathcal{I}}}{P_{\mathcal{R}} + P_{\mathcal{I}}} < 0.038 \quad (95\%\,{\rm CL},\;\text{Planck}),
\end{equation}
where $P_{\mathcal{R}} \simeq 2.1\times 10^{-9}$ is the curvature perturbation power spectrum and 
\begin{equation}\label{eq:Piso}
P_{\mathcal{I}}(k) = 4\,\frac{P_s(k)}{s_{\rm rms}^2}
\end{equation}
is the dimensionless isocurvature power spectrum, evaluated at CMB scales ($k_{\rm CMB}\sim 0.05$--$0.2$~Mpc$^{-1}$). The factor of $4$ follows from 
$\rho_s\propto s^2$, which gives $\delta\rho_s/\rho_s = 2\,\delta s/s_{\rm rms}$. The Planck bound therefore requires 
$P_s(k_{\rm CMB})\lesssim \tfrac{\alpha_{\rm iso}}{4}\,s_{\rm rms}^2\simeq10^{-2}\,s_{\rm rms}^2$ at CMB scales.

\paragraph{Scale-invariant spectrum.} 
Were the spectrum scale invariant, $P_s(k)=\sigma_s^2\equiv(H_{\rm inf}/2\pi)^2$ at all $k$, the variance would read 
$s_{\rm rms}^2=\int d\ln k\,P_s(k)=\sigma_s^2\,\Delta\ln k$, with $\Delta\ln k$ the logarithmic range of modes contributing to the variance. 
Equation~\eqref{eq:Piso} would then give
\begin{equation}\label{eq:Piso_SI}
P_{\mathcal{I}} = \frac{4\,\sigma_s^2}{s_{\rm rms}^2} = \frac{4}{\Delta\ln k}
=\mathcal{O}(10^{-1})\qquad \bigl(\Delta\ln k=\mathcal{O}(10\text{--}10^{2})\bigr).
\end{equation}
Since this exceeds $P_{\mathcal{R}}\simeq2.1\times10^{-9}$ by many orders of magnitude, the normalised ratio saturates at 
$\alpha_{\rm iso}=P_{\mathcal{I}}/(P_{\mathcal{R}}+P_{\mathcal{I}})\simeq1$, in severe conflict with the Planck bound. 
\emph{A scale-invariant spectrum for $s$ is therefore excluded} as soon as $s$ is required to account for the dark matter.

\paragraph{Peaked power spectrum.}
The resolution is to assume that the power spectrum of $s$ is \emph{not} scale-invariant, but instead has a peak at a comoving scale $k_*$ that re-enters the horizon 
during the electroweak epoch, well below the CMB scales:
\begin{equation}
P_s(k) = \left(\frac{H_{\rm inf}}{2\pi}\right)^{\!2} \times \mathcal{F}(k/k_*),
\end{equation}
where $\mathcal{F}(k/k_*)$ is a peaked profile with $\mathcal{F}\ll 1$ at CMB scales and $\mathcal{F}\sim 1$ near $k\sim k_*$. With this ansatz:
\begin{itemize}
\item At CMB scales, $P_s(k_{\rm CMB}) \ll (H_{\rm inf}/2\pi)^2$, so the isocurvature contribution is suppressed and the Planck bound is evaded.
\item At the electroweak scale $k_* \sim H_{\rm EW}$, the power is enhanced, providing the fluctuations needed to trigger the phase transition.
\end{itemize}
The RMS amplitude is dominated by the peak: $s_{\rm rms} \simeq (H_{\rm inf}/2\pi)\sqrt{\Delta\ln k}$, where $\Delta\ln k$ is the logarithmic width of the peak, 
which we take to be $\gtrsim \mathcal{O}(100)$.

% \emph{Normalisation used in the numerical analysis.} 
For a peaked spectrum one has, strictly, $s_{\rm rms}\simeq\sigma_s\sqrt{\Delta\ln k}$ with 
$\sigma_s=H_{\rm inf}/(2\pi)$, whereas throughout our numerical analysis we employ the simpler scale-invariant normalisation 
$\sigma_s=H_{\rm inf}/(2\pi)$ of Section~\ref{sec:scalar_evolution}. This is not an additional assumption. It amounts to absorbing the 
$\sqrt{\Delta\ln k}$ enhancement into the inflationary scale $H_{\rm inf}$, which is a free parameter scanned over several decades. All 
phase-transition observables depend on $s$ only through the dimensionless combinations $s/\Lambda$ and $Z(s)$, so retaining the explicit 
$\sqrt{\Delta\ln k}$ factor for $\Delta\ln k\sim\mathcal{O}(100)$ would merely shift the preferred $H_{\rm inf}$ by an $\mathcal{O}(1\text{--}10)$ factor 
along the $H_{\rm inf}$ axis, without changing the mechanism or the qualitative picture. We therefore keep the scale-invariant normalisation in our numerical scan.

\paragraph{Inflationary origin of the peak.}
A peaked power spectrum for a spectator field is a well-motivated possibility in inflationary cosmology. Several mechanisms can produce it:
\begin{enumerate}
\item \emph{Transient non-slow-roll phase:} A brief departure from slow-roll in the inflaton potential (e.g.\, an inflection point or a near-flat region) amplifies 
fluctuations at the corresponding comoving scale~\cite{Garcia-Bellido:2017mdg,Ezquiaga:2017qlvi}.
\item \emph{Tachyonic instability in the spectator sector~\cite{Bassett:1997az,Dufaux:2006ee,Markkanen:2015xuw,Kitajima:2023pby}:} If the effective mass of $s$ briefly 
becomes tachyonic ($m_{\rm eff}^2 < 0$) during inflation, its fluctuations are exponentially amplified at a specific scale.
\item \emph{Multi-field dynamics~\cite{Stewart:1997wg,Drees:2011hb,Alabidi:2012ex,Silk:1986vc,GarciaBellido:1996mdl}:} In multi-field inflation, a brief period of 
negative effective mass squared for $s$ can arise from cross-couplings with other spectator fields.
\end{enumerate}
These mechanisms have been extensively studied in the context of primordial black hole production and are known to produce spectra with 
$\mathcal{F}(k_{\rm CMB})/\mathcal{F}(k_*) \ll 10^{-3}$, sufficient to evade the isocurvature bound.

\paragraph{Perturbativity condition.}
The non-Gaussian expansion requires $f_{\rm NL}^2\,P_{\delta s}(k) \ll 1$ at each scale $k$, where $P_{\delta s}(k) = P_s(k)/s_{\rm rms}^2$ is the power spectrum 
of the relative fluctuation. At the peak one has $P_s(k_*)\simeq\sigma_s^2$ while $s_{\rm rms}^2\simeq\sigma_s^2\,\Delta\ln k$, so that 
$P_{\delta s}(k_*)\simeq1/\Delta\ln k$ and
\begin{equation}\label{eq:pert_nongauss}
f_{\rm NL}^2\,P_{\delta s}(k_*)\;\simeq\;\frac{f_{\rm NL}^2}{\Delta\ln k}
\;\simeq\;\frac{25}{\Delta\ln k}
\;\lesssim\;\mathcal{O}(10^{-1})
\qquad \bigl(f_{\rm NL}=5,\ \Delta\ln k\gtrsim\mathcal{O}(100)\bigr),
\end{equation}
so the expansion is perturbative. Equivalently, in field space the non-Gaussian term is suppressed relative to the Gaussian one by 
$\tfrac12 f_{\rm NL}|s|/\sigma_s$, and we require $f_{\rm NL}|s|/\sigma_s\lesssim\mathcal{O}(1)$; this is the criterion adopted in our numerical 
calculations, so that the leading-order non-Gaussian treatment is self-consistently controlled by construction.

%===========================================================

\section{Bubble nucleation}
\label{sec:nucleation}

\subsection{Thermal nucleation rate}
The rate of thermal nucleation of true vacuum bubbles per unit time per unit volume is~\cite{Apreda:2001us,Espinosa:2008kw,Caprini:2015zlo}:
\begin{equation}\label{eq:loc_nucleation_rate}
\Gamma(T,s) = A(T) \exp\left(-\frac{S_3(T,s)}{T}\right),
\end{equation}
where $A(T) \approx T^4$ is the dynamical prefactor and $S_3(T,s)$ is the Euclidean action of the critical bubble.

In the thin-wall approximation~\cite{Coleman:1977py,Cline:2017qpe}, the 3D Euclidean action \(S_3\) of the critical bubble is related to the surface tension \(\sigma\) 
and vacuum energy difference \(\Delta V\) by
\begin{equation}
S_3 = \frac{16\pi}{3} \frac{\sigma^3}{(\Delta V)^2},
\end{equation}
where the surface tension \(\sigma\), the energy per unit area of the bubble wall, is defined as
\begin{equation}
\sigma (T,s)= \int_0^{v(T,s)} \sqrt{2 V_{\rm eff}(h,T,s)}\, dh,
\end{equation}
and the vacuum energy difference between the symmetric and broken phases is:
\begin{equation}\label{eq:vacuum_energy_difference}
\Delta V(T,s) = V_{\text{eff}}(0,T,s) - V_{\text{eff}}(v(T,s),T,s) .
\end{equation}

\subsection{Global average nucleation rate under saddle point approximation}

The effective global nucleation rate, which describes the average number of bubble nucleations per unit time per unit volume in the entire universe, 
is obtained by statistical averaging over all possible fluctuation amplitudes:
\begin{equation}
\Gamma_{\text{avg}}(T) = \int_{-\infty}^{\infty} \Gamma(T, s) \cdot \mathcal{P}(s) \,ds = 2 \int_{0}^{\infty} \Gamma(T, s) \cdot \mathcal{P}(s) \, ds~,
\end{equation}
where we have used $\Gamma(T, -s) = \Gamma(T, s)$ and $\mathcal{P}(-s) = \mathcal{P}(s)$ due to the $\mathbb{Z}_2$ symmetry.

When the integrand has an exponential form and the exponential function has a sharp minimum, the integral is dominated by a small region near the minimum. 
We can use the saddle point approximation to evaluate the integral, which is highly accurate for exponentially peaked functions.
Assuming $F(s)$ attains its minimum at the optimal fluctuation amplitude $s = s_*$, the saddle-point condition is:
\begin{equation}
\left. \frac{dF}{ds} \right|_{s = s_*} = 0.
\end{equation}
Expanding $F(s)$ in a Taylor series around the saddle point $s_*$, keeping terms up to second order:
\begin{equation}
F(s) \approx F(s_*) + \frac{1}{2} F''(s_*) (s - s_*)^2~,
\end{equation}
where $F''(s_*)$ is the second derivative of $F(s)$ at the saddle point.

Substituting the Taylor expansion into the integral, the integration limits can be extended to infinity (since the exponential function decays rapidly far from 
the saddle point):
\begin{equation}
\int_{0}^{s_{\text{max}}} e^{-F(s)} ds \approx e^{-F(s_*)} \int_{-\infty}^{\infty} e^{-\frac{1}{2} F''(s_*) (s - s_*)^2} ds~.
\end{equation}
Evaluating this standard Gaussian integral, we then obtain:
\begin{equation}
\int_{0}^{s_{\text{max}}} e^{-F(s)} ds \approx \sqrt{\frac{2\pi}{|F''(s_*)|}} \cdot e^{-F(s_*)}~.
\end{equation}
Substituting into the expression for the global average nucleation rate, we obtain:
\begin{equation}\label{eq:Gamma_avg}
\Gamma_{\text{avg}}(T) \approx 2T^4 \cdot \sqrt{\frac{2\pi}{|F''(s_*)|}} \cdot \exp\left[-F(s_*)\right]~,
\end{equation}
where the exponent is given by:
\begin{equation}
    F(s) = \frac{S_3(T,s)}{T} - \ln\mathcal{P}(s).
\end{equation}

%=======================================================
\begin{figure}
    \centering
    \includegraphics[width=1.0\linewidth]{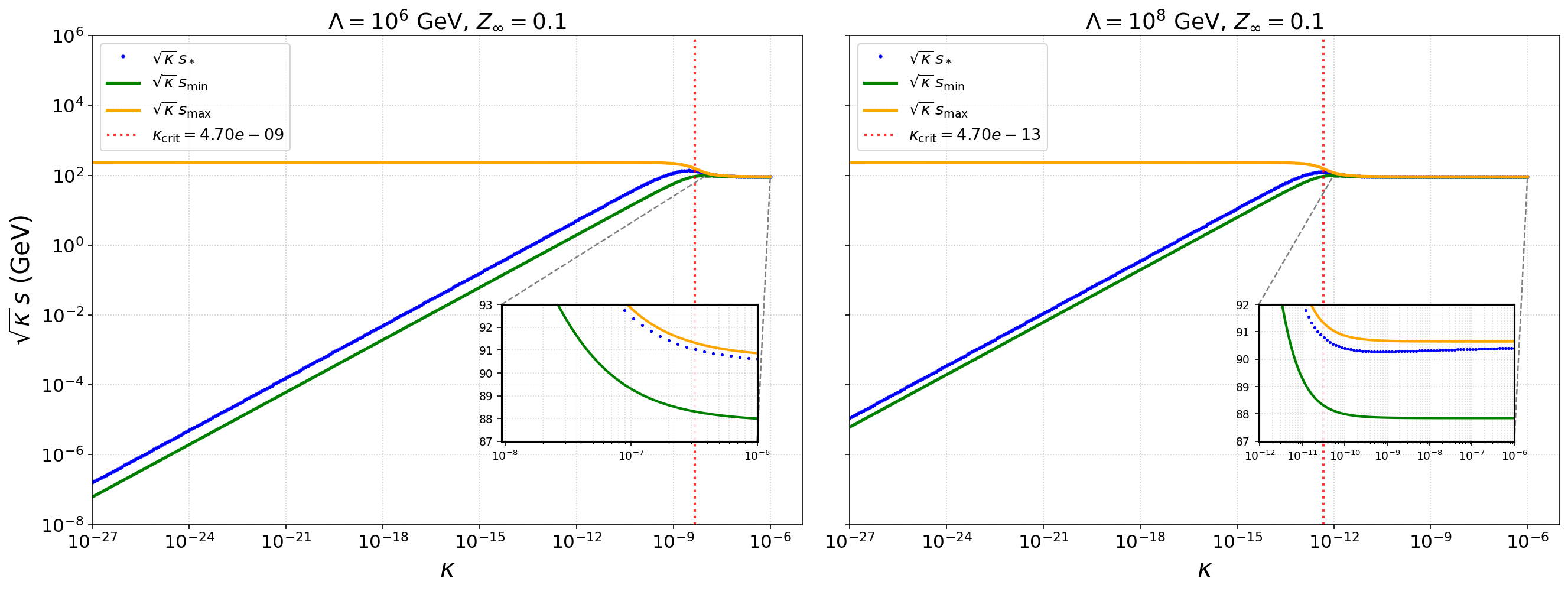}
    \caption{The blue points represent the saddle point $\sqrt{\kappa}s_*$ as a function of the coupling $\kappa$. The orange and green lines represent the upper 
    and lower bounds from conditions $m_{\rm loc}^2>0$ and $\Delta>0$, respectively. We take ($\Lambda=10^6$~GeV, $Z_{\infty}=0.1$) and ($\Lambda=10^8$~GeV, 
    $Z_{\infty}=0.1$) for the left and right panels, respectively. The temperature is fixed at $T = 150$~GeV.}
    \label{fig:saddle}
\end{figure}
Figure~\ref{fig:saddle} shows the saddle point (blue points) of the rescaled spectator field amplitude $\sqrt{\kappa}\,s$ as a function of the portal 
coupling $\kappa$ for two different values of the new physics scale, $\Lambda = 10^6$~GeV and $10^8$~GeV. The temperature is fixed at $T = 150$~GeV. 
The green and orange lines represent the lower bound ($\Delta>0$) and the upper bound ($m_{\rm loc}^2>0$) on the scalar field, respectively. In this plot, 
the saddle point is determined using the exact distribution~\eqref{eq:non_Gaussian_distribution}. We observe that the saddle points always lie within the lower 
and upper bounds for the first-order phase transition. The vertical dotted line denotes the critical value of $\kappa$, determined by the condition:
\begin{equation}
    \varsigma=1\Rightarrow \kappa_c=\frac{2D_{\rm gauge}(1-Z_{\infty})T^2}{\Lambda^2} .
\end{equation}
As stated in the previous section, for $\kappa<\kappa_c$ (i.e.\, $\varsigma>1$), the higher-dimensional scalar operators can effectively modify the gauge coupling 
coefficients, thereby widening the parameter window for a first-order phase transition. In contrast, when $\kappa>\kappa_c$ (i.e.\, $\varsigma<1$), the influence of 
higher-dimensional scalar operators on the gauge couplings diminishes, the model gradually approaches the SM parameters, and the first-order phase 
transition becomes weaker.

On the other hand, by setting the limit $\Lambda \to \infty$ (or $s\to 0$) we return to the scenario without the higher‑dimensional gauge‑kinetic operator, 
then the effective potential coefficients reduce to the constant SM values $D_{\rm eff}=D_{\rm SM}\approx 0.40$ and $E_{\rm eff}=E_{\rm SM}\approx 0.0063$.
In that case, the lower and upper bounds from the requirements $\Delta>0$ and $m_{\rm loc}^2>0$ on the scalar field amplitude is given by:
\begin{equation}\label{eq:SMbound}
\frac{2}{\kappa} \left[ \left( D_{\rm SM} - \frac{9E_{\rm SM}^2}{4\lambda_h} \right) T^2 - \mu^2 \right] < s^2 < \frac{2}{\kappa} \left( D_{\rm SM} T^2 - \mu^2 \right).
\end{equation}
For a temperature $T=150$~GeV, the bounds from Eq.~\eqref{eq:SMbound} is found to be around 90~GeV, as indicated by the ``plateau'' at $\kappa>\kappa_c$ in Fig.~\ref{fig:saddle}.
Therefore, the introduction of a finite $\Lambda$ through the gauge‑kinetic operator relaxes the bound~\eqref{eq:SMbound}, allowing the spectator fluctuations to 
reach much larger amplitudes and thereby enhancing the strength of the first‑order phase transition.

%=======================================================

\subsection{Nucleation temperature}
The local nucleation probability in the Universe is estimated by the globally averaged nucleation rate. The nucleation of true vacuum bubbles occurs at the 
temperature at which the local nucleation rate equals the fourth power of the Hubble expansion rate~\cite{Espinosa:2008kw}
\begin{equation}\label{eq:nucleation_condition}
\Gamma(T_n,s) \simeq H^4(T_n)~,
\end{equation}
where $T_n$ is the nucleation temperature. 
During the radiation-dominated epoch of the universe, the Hubble rate is given by:
\begin{equation}
H(T) = \sqrt{\frac{8\pi^3{g_*(T)}}{90}} \cdot \frac{T^2}{M_p} = 1.66 \sqrt{g_*} \cdot \frac{T^2}{M_p},
\end{equation}
where $M_p = 1.22 \times 10^{19}$~GeV is the Planck mass.
Taking the natural logarithm on both sides of the nucleation condition~\eqref{eq:nucleation_condition}, we have $\ln\Gamma(T_n,s)\simeq \ln H^4(T_n) \approx 120$ 
at the electroweak scale, i.e.\, $T \sim 100$~GeV for the Hubble expansion rate.
When condition~\eqref{eq:nucleation_condition} is satisfied, at least one critical bubble is produced per Hubble volume per Hubble time. 
After nucleation, the bubbles begin to expand and collide in large numbers, and the phase transition then commences.

%====================================================

\subsection{First-order phase transition space fraction}
 Because \(s(x)\) is statistically homogeneous and the observable Universe contains a huge number of independent Hubble volumes, the field 
is \emph{self-averaging}. In a sufficiently large spatial volume $V$ containing many statistically independent sub-volumes, the spatial average of any 
statistical quantity converges to its ensemble average:
\begin{equation}
\frac{1}{V}\int_V \mathcal{O}[s(\mathbf{x})]\,d^3x \;\xrightarrow{V\to\infty}\; \langle \mathcal{O}[s] \rangle,
\end{equation}
for any functional $\mathcal{O}[s]$. This is a consequence of the ergodic theorem applied to a statistically homogeneous and isotropic random field. The observable Universe at the electroweak epoch contains approximately
\begin{equation}
N_H \sim \left(\frac{H_{\rm inf}}{H_{\rm EW}}\right)^{\!3} \sim \left(\frac{10^{13}\;\text{GeV}}{10^{-15}\;\text{GeV}}\right)^{\!3} \sim 10^{84}
\end{equation}
independent Hubble volumes, so the relative fluctuations of the spatial average around the ensemble average are of order $N_H^{-1/2}\sim 10^{-42}$, making the self-averaging approximation essentially exact.

Under the assumption of ergodicity, an ensemble average over realisations of the field is equivalent to a spatial average over a single, sufficiently large 
comoving volume \(V\). The volume fraction of the region in which the field falls in the interval \(s_1<s<s_2\) is therefore given by the probability density,
\begin{equation}
    \frac{dV}{V} \;=\; \mathbb{P}\bigl(s_1<s<s_2\bigr) \;=\; \mathcal{P}(s)\,ds~,
\end{equation}
Therefore, the fraction of the universe that undergoes a first-order phase transition is given by the integration of the PDF over scalar field value.

To determine the fraction of the Universe that undergoes a first-order EWPT, \(f_{\mathrm{FOPT}}\), we must identify all initial fluctuation 
amplitudes \(s_0\) (defined at the end of inflation) that, after evolving through the thermal history, eventually satisfy the nucleation condition 
\(\Gamma(T,s) \gtrsim H^4(T)\). This is achieved by the following numerical procedure, 
which is performed independently for each portal coupling \(\kappa\) and new physics scale \( \Lambda \).

\begin{figure}
    \centering
    \includegraphics[width=1.0\linewidth]{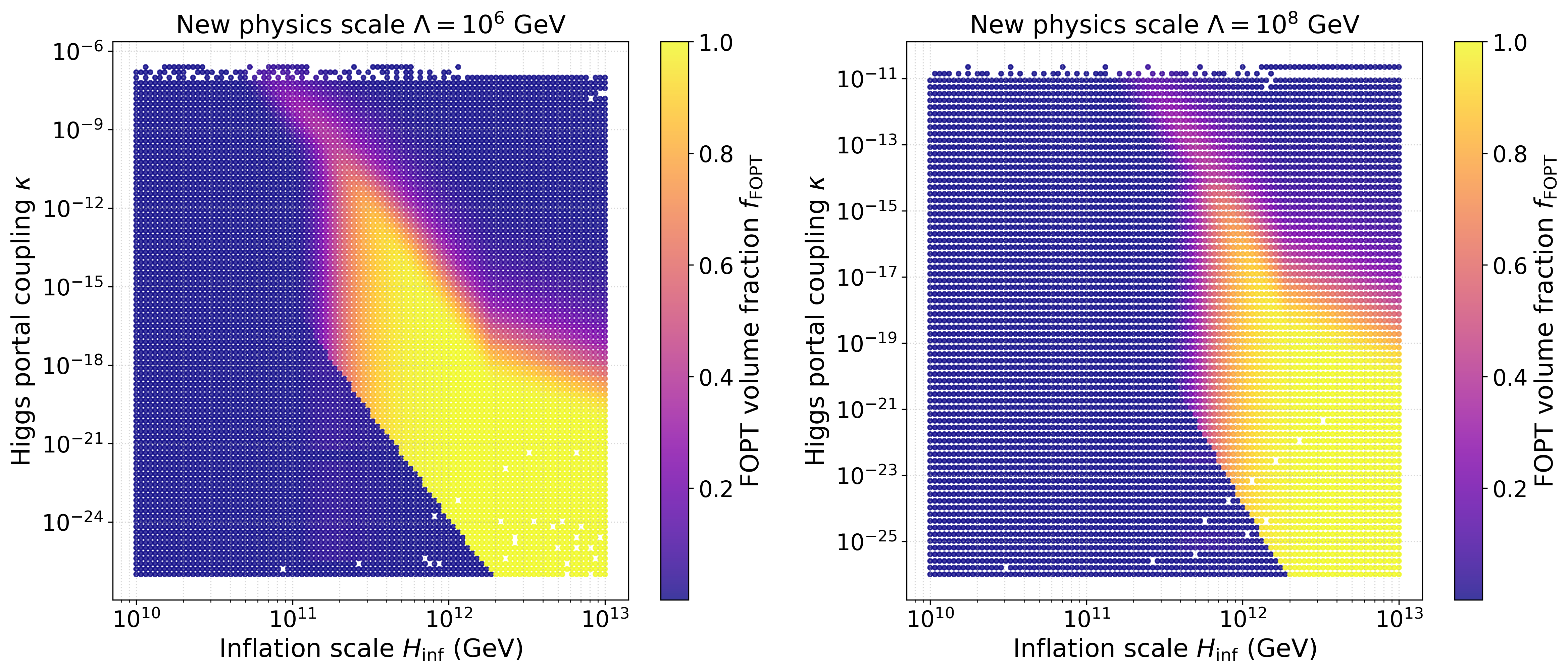}
    \caption{The fraction of the Universe that undergoes a first-order phase transition, $f_{\rm FOPT}$, in $H_{\rm inf}-\kappa$ space, with cubic approximation. 
    We fix the non-Gaussian parameter $f_{\rm NL}=5$ and the asymptotic value $Z_{\infty}=0.1$. The new physics scale $\Lambda$ is taken as $10^6$~GeV and $10^8$~GeV 
    for the left and right panels, respectively. The scan is performed in the high-temperature cubic approximation (see Sec.~\ref{sec:model}); the accurate full 
    one-loop benchmark values are collected in Table~\ref{tab:ewpt_parameters}.}
    \label{fig:fFOPT}
\end{figure}

\begin{enumerate}
\item \textbf{Temperature and field scanning.}\\
We consider a dense grid of temperatures \(T \in [T_{\min}, T_{\max}]\) that covers the expected electroweak epoch. For each \(T\) (i.e.\, cosmic time \(t\)), 
we first determine the allowed range of the instantaneous spectator field \(s\) for which the effective potential exhibits a first-order barrier and the symmetric 
minimum remains stable, i.e.\  conditions \(\Delta>0\) and \(m_{\mathrm{loc}}^2>0\) (see Eqs.~\eqref{eq:Delta} and~\eqref{eq:mlocal}).
To achieve phase transition from the symmetric phase to the broken phase, we additionally confirm that the free energy difference satisfies \(\Delta V(T,s)>0\).
Within the range \([s_{\rm min}(T),s_{\rm max}(T)]\), a fine scan of \(s\) is performed. At each point we evaluate the local thermal nucleation rate \(\Gamma(T,s)\) 
(given by Eq.~\eqref{eq:loc_nucleation_rate}) and compare it with the Hubble expansion rate \(H(T)\). A value of \(s\) is considered to trigger a first‑order 
transition if the logarithm of the nucleation rate per unit volume exceeds that of \(H^4\).
Contiguous intervals of \(s\) satisfying this condition are recorded for that temperature.

\item \textbf{Back‑mapping to the initial fluctuation.}\\
The field \(s\) evolves with cosmic expansion. As derived in Section~\ref{sec:scalar_evolution}, for temperatures above the oscillation temperature \(T_{\mathrm{osc}}\), the amplitude of the spectator field is frozen; below \(T_{\mathrm{osc}}\) it dilutes as \(s \propto T^{3/2}\). 
The evolution of the amplitude of the spectator field is estimated by Eq.~\eqref{eq:s_evolution}.
Applying this mapping, the physical intervals of \(s\) that satisfy the nucleation criterion are translated into initial \(s_0\) intervals, all defined at the common reference time immediately after inflation.

\item \textbf{Merging of intervals.}\\
The procedure above yields a large set of initial‑condition intervals, one for each temperature at which successful nucleation is possible. 
Because the observable Universe is a single realisation of the random field, a spatial point enters the first‑order channel if its initial amplitude \(s_0\) 
lies in any of these intervals. We therefore sort all obtained \(s_0\) intervals and merge overlapping or contiguous ones, producing a minimal set of disjoint 
intervals \(\{[s_{0,i}^{\mathrm{low}}, s_{0,i}^{\mathrm{high}}]\}\).

\item \textbf{Volume fraction integration.}\\
By ergodicity, the spatial volume fraction occupied by fluctuations that eventually undergo a first‑order transition equals the probability for \(s_0\) to fall 
inside the merged intervals. Using the exact \(\mathbb{Z}_2\)-symmetric probability density \(\mathcal{P}(s_0)\) derived in Eq.~\eqref{eq:non_Gaussian_distribution}, 
and taking into account that both positive and negative \(s_0\) give identical contributions, we have
\begin{equation}
   f_{\mathrm{FOPT}} = 2 \sum_i \int_{s_{0,i}^{\mathrm{low}}}^{s_{0,i}^{\mathrm{high}}} \mathcal{P}(s_0)\, ds_0~. 
\end{equation}
In practice, the integration is performed numerically for each merged interval, and the factor of \(2\) accounts for the symmetry \(s \leftrightarrow -s\). 
The sum of these integrals directly gives \(f_{\mathrm{FOPT}}\) for the chosen \(\kappa\) and \(\Lambda\).
\end{enumerate}
In Fig.~\ref{fig:fFOPT}, we show the volume fraction of the Universe that undergoes a first-order phase transition. We observe that for a wide range of $\kappa$ 
and $H_{\rm inf}\lesssim 10^{12}$~GeV, the volume fraction satisfies $f_{\rm FOPT}\sim 1$. The value of $\kappa$ required to achieve $f_{\rm FOPT}\sim 1$ decreases 
with increasing inflationary Hubble scale $H_{\rm inf}$. The bound from the condition $\Delta > 0$ becomes stronger as the new physics scale $\Lambda$ increases; 
consequently, the parameter space for $f_{\rm FOPT}\sim 1$ shrinks as $\Lambda$ rises from $10^6$~GeV to $10^8$~GeV.

\subsection{Percolation condition}\label{sec:percolation}
A first-order phase transition is cosmologically viable only if the true-vacuum bubbles \emph{percolate}, i.e.\, the bubble network occupies a sufficient volume 
fraction to complete the transition. In three spatial dimensions, percolation theory gives a critical volume fraction $f_{\rm perc}\approx 0.29$ for site percolation 
on a simple cubic lattice~\cite{Stauffer:1994}; a more conservative cosmological criterion, accounting for the requirement that the transition completes within a 
Hubble time, uses $f_{\rm true}\gtrsim 0.5$~\cite{Gleiser:1998}.

Two distinct quantities must be kept separate here. The first is $f_{\rm FOPT}$, the fraction of the \emph{Universe's volume} in which the local value of $s$ is 
such that the transition proceeds by bubble nucleation rather than by a smooth crossover; this is the quantity computed numerically (Eq.~\eqref{eq:Gamma_avg} and the 
procedure above) and displayed in Figure~\ref{fig:fFOPT}. The second is $f_{\rm true}(t)$, the fraction of volume \emph{inside} a first-order patch that has been 
converted to the true vacuum by expanding bubbles at time $t$; percolation requires $f_{\rm true}$ to exceed $f_{\rm perc}\approx0.29$, or conservatively $0.5$. 
A large $f_{\rm FOPT}$ does not by itself guarantee percolation, because a patch can undergo a first-order transition and still nucleate too slowly for its bubbles to connect.

Percolation within the first-order patches is nonetheless ensured by the large inverse duration of the transition. For exponential nucleation, 
$\Gamma\propto e^{\beta t}$, the false-vacuum fraction falls as $f_{\rm false}(t)=\exp[-e^{\beta(t-t_0)}]$, so the interval between the nucleation temperature 
$T_n$ (one bubble per Hubble volume, $\Gamma\simeq H_n^{4}$) and the percolation temperature $T_p$ (defined by $f_{\rm true}=1-f_{\rm false}=0.29$) is 
$\Delta t\sim\mathcal{O}(1)/\beta$. Normalised to the Hubble time this gives
\begin{equation}
\frac{\Delta t}{H_n^{-1}}\sim\frac{H_n}{\beta}=\left(\frac{\beta}{H_n}\right)^{-1}\sim2\times10^{-4},
\qquad\text{so that}\qquad
\frac{T_n-T_p}{T_n}\sim\left(\frac{\beta}{H_n}\right)^{-1}\sim2\times10^{-4},
\end{equation}
where we have used $\overline{\beta/H_n}\simeq(5.5\text{--}5.9)\times10^{3}$ from Table~\ref{tab:ewpt_parameters}. Percolation therefore occurs essentially 
simultaneously with nucleation, $T_p\simeq T_n$, and the true-vacuum fraction reaches unity well within a Hubble time; the criterion 
$f_{\rm true}>f_{\rm perc}$ is satisfied in every first-order patch of the viable parameter space.
For the case $f_{\rm FOPT} \lesssim 1$, where a fraction of the Universe undergoes a smooth crossover while the remainder undergoes a first-order transition. 
In this mixed scenario, percolation may still occur within the first-order-transition region, but the global completion of the transition is delayed. 
This regime can potentially generate inhomogeneities in the baryon-to-entropy ratio, which is not the focus of the present work.

\subsection{Validity of the local approximation}\label{sec:local_approx}
The use of a purely local effective potential $V_{\rm eff}(h, T, s(\mathbf{x}))$ and a local Euclidean bounce action $S_3(T, s_*)$ is justified by a strong hierarchy 
between the bubble scale and the scale of spectator-field variation.

\paragraph{Scale hierarchy.}
The spectator field $s(\mathbf{x})$ varies on the scale of the modes that re-enter the horizon during the electroweak epoch. For the peaked power spectrum discussed 
in Section~\ref{sec:isocurvature}, the relevant scale is
\begin{equation}
L_s \sim k_*^{-1} \sim H_{\rm EW}^{-1} \sim \frac{M_{\rm Pl}}{T_{\rm EW}^2} \sim 10^{15}\,\text{GeV}^{-1}.
\end{equation}
The bubble wall thickness is set by the inverse gauge boson mass in the broken phase,
\begin{equation}
R_{\rm wall} \sim \frac{1}{m_W^{\rm eff}} \sim \frac{1}{g_{\rm eff}\,v/2} \sim 10^{-2}\,\text{GeV}^{-1},
\end{equation}
and the bubble radius at nucleation is
\begin{equation}
R_{\rm bubble} \sim \frac{v_w}{\beta} = \frac{v_w}{H_n\,(\beta/H_n)} \sim \frac{0.1\text{--}0.6}{10^{-12}\,\text{GeV}} \sim 10^{11}\text{--}10^{12}\,\text{GeV}^{-1},
\end{equation}
where $H_n\sim 10^{-15}\,\text{GeV}$ is the Hubble rate at $T_n\sim 100$ GeV and we have used $\beta/H_n\sim 10^{3}$ from Table~\ref{tab:ewpt_parameters}. 
The hierarchy $R_{\rm bubble} \ll L_s$ (by more than three orders of magnitude) ensures that $s$ is \emph{effectively constant} across a bubble and its immediate surroundings.

\paragraph{Gradient corrections.}
The leading gradient correction to the effective potential is of order
\begin{equation}
\frac{\delta V_{\rm grad}}{V_{\rm eff}} \sim \left(\frac{R_{\rm bubble}}{L_s}\right)^{2}\sim \left(\frac{v_w}{\beta/H_n}\right)^2 \lesssim 10^{-6},
\end{equation}
utterly negligible. The Euclidean bounce solution is computed in the local potential $V_{\rm eff}(h, T, s)$ and the global nucleation rate is then obtained by 
averaging over $s$. This two-step procedure (local bounce plus global average) is justified by the same scale separation, because the bubble dynamics are 
local, while the statistical distribution of $s$ is global.

% \paragraph{Limits of the approximation.}
The hierarchy is lost when the spectator re-enters the horizon well before the bubbles nucleate.
This can happen in a strongly supercooled transition, where the completion of the transition is
delayed until $T_n\ll T_c$. The spectator modes have then already re-entered at $T\sim T_c$, the
inverse duration has dropped to $\beta/H_n\lesssim \mathcal{O}(1)$, and the bubbles reach radii
$R_{\rm bubble}\gtrsim L_s$, large enough for $s$ to vary appreciably across a single bubble.
The gradient terms are then no longer suppressed, and the local treatment has to give way to one
that resolves the profile of $s$ over the bubble.

In our work, the spectator re-enters the horizon at the electroweak scale
(Section~\ref{sec:isocurvature}), the transition stays far from the supercooled regime, and
$\beta/H_n\sim\mathcal{O}(10^{3})$ as in Table~\ref{tab:ewpt_parameters}, so that
$R_{\rm bubble}/L_s\lesssim10^{-3}$. The conclusion is unchanged if the spectator re-enters the horizon
anywhere between the electroweak and the CMB epochs, with an even larger $L_s$.

%=================================================================
\subsection{Three-dimensional effective-theory criterion}
\label{sec:xsm_comparison}

The electroweak phase transition in the real-singlet extension of the SM has been studied extensively with dimensional reduction, 
two-loop effective potentials, and lattice simulations~\cite{Gould:2019qek,Niemi:2021qvp,Niemi:2024vzw,Niemi:2024axp}. 
In that setup the $\mathbb{Z}_2$ symmetry is broken, the singlet $S$ acquires a VEV $v_s$, and the barrier arises at tree level from the portal coupling. 
The transition typically proceeds along a two-step path $(0,0)\to(0,v_s)\to(v,0)$, and the portion of parameter space that yields a strong 
first-order transition is tightly constrained by the need to complete nucleation together with collider bounds on the singlet-Higgs mixing. 
None of those constraints applies to the mechanism studied here. Our spectator field preserves $\mathbb{Z}_2$ at all temperatures, $\langle s\rangle=0$, 
and the barrier is generated thermally by the gauge ring term whose coefficient is enhanced through $g_{\rm eff}(s)=g/\sqrt{Z(s)}$. 
In our scenario, the phase transition proceeds in a single step rather than through a two-step process, because no singlet phase transition occurs.

It is instructive to place our mechanism within the same three-dimensional effective-field-theory language. The dimensional-reduction program 
is the systematic, all-orders resummation of thermal effects. In the dimensionally reduced theory the daisy and ring resummations are 
reproduced automatically by the Debye masses of the temporal gauge bosons, and the order of the transition is then read from the 
three-dimensional effective theory constructed to two-loop accuracy. Because $s$ carries no VEV, no self-coupling, and couples to the 
Higgs only through the portal $\kappa\sim10^{-16}$ (whose thermal mass contribution $\kappa T^2/16\sim10^{-17}D_{\rm eff}T^2$ is negligible), 
$s$ does not participate in the infrared Higgs dynamics. It enters the dimensionally reduced theory as a fixed background parameter, a spurion, 
that rescales the gauge coupling. We therefore carry out the standard two-loop dimensional reduction, applying the gauge-coupling replacement 
$g\to g_{\rm eff}(s)$ of Eq.~\eqref{eq:geff} to SM matching relations. 
In the general singlet- and triplet-extended reduction of~\cite{Niemi:2018asa,Schicho:2022wty} the result is parameterised by three integers $(N_d,N_t,N_f)$, 
where $N_d$ is the number of Higgs doublets, $N_t$ is the number of real scalar triplets, and $N_f$ is the number of fermion families. 
Our spectator $s$ is a gauge singlet, it enters the effective theory only as a static spurion that rescales the gauge coupling, and is removed 
at the superheavy scale. Therefore, the reduced theory in our scenario is SM-like with $(N_d,N_t,N_f)=(1,0,3)$. 
% The complete two-loop matching relations for the SM were first derived in~\cite{Kajantie:1995dw} and are reproduced 
% in ~\cite{Niemi:2018asa} with a real triplet extension of the SM.

At the matching scale $\mu=4\pi T e^{-\gamma_E}$ the Debye masses (for which a one-loop matching is sufficient~\cite{Niemi:2018asa,Schicho:2022wty}) become
\begin{equation}
m_D^2(s)=\frac{11}{6}\,g_{\rm eff}^2(s)\,T^2=\frac{11}{6}\frac{g^2}{Z(s)}\,T^2,\qquad
m_{D'}^2(s)=\frac{11}{6}\frac{g'^2}{Z(s)}\,T^2,
\end{equation}
so that the electric screening scale is enhanced by $1/Z(s)$ relative to the SM. For the benchmark points of Table~\ref{tab:ewpt_parameters}, 
with $Z_\infty\in[0.12,0.15]$ and $s^*/\Lambda\sim40$--$80$ (so that $Z(s^*)\simeq Z_\infty$), this gives $m_D^2/T^2\in[5.13,6.42]$, 
a factor of $6.7$ above the SM value $m_D^2/T^2=11g^2/6\simeq0.77$ ($Z=1$). The two-loop dimensionally reduced gauge and scalar 
couplings are~\cite{Kajantie:1995dw,Niemi:2018asa,Schicho:2022wty}
\begin{equation}
\begin{aligned}
\bar g_3^2(s)&=g_{\rm eff}^2(s)\,T\Bigl[1+\frac{g_{\rm eff}^2(s)}{(4\pi)^2}\Bigl(\frac{2}{3}-16\ln2\Bigr)\Bigr]
=\frac{g^2T}{Z(s)}\Bigl[1+\frac{g^2}{(4\pi)^2Z(s)}\Bigl(\frac{2}{3}-16\ln2\Bigr)\Bigr],\\
\bar\lambda_3(s)&=T\Bigl\{\lambda_H+\frac{1}{(4\pi)^2}\Bigl[\tfrac18\bigl(3g_{\rm eff}^4(s)+g_{\rm eff}^{\prime 4}(s)+2g_{\rm eff}^2(s)g_{\rm eff}^{\prime 2}(s)\bigr)
+3L_f\bigl(y_t^4-2\lambda_H y_t^2\bigr)\Bigr]\Bigr\},
\end{aligned}
\end{equation}
where $L_f=4\ln2$ and we have used $L_b=0$ at the matching scale $\mu=4\pi T e^{-\gamma_E}$ (see Ref.~\cite{Niemi:2018asa}). 
Both relations follow from the two-loop SM matching of~\cite{Niemi:2018asa,Schicho:2022wty} under the replacement~\eqref{eq:geff}. 
The three-dimensional gauge coupling scales as $\bar g_3^2(s)\propto1/Z(s)$ (with a two-loop correction enhanced by $1/Z(s)^2$), 
while the quartic $\bar\lambda_3(s)$ is dominated by its tree-level $\lambda_HT$ piece, whose two-loop gauge corrections are enhanced by $1/Z(s)^2$. 
The dimensionless ratio $x\equiv\bar\lambda_3/\bar g_3^2$ that governs the order of the transition, with the critical value $x_c\simeq0.11$ separating a first-order transition from a crossover~\cite{Rummukainen:1998as}, therefore scales, to leading order, as
\begin{equation}
x_{\rm match}(s)\equiv \frac{\bar\lambda_3(s;T_c)}{\bar g_3^2(s;T_c)} \simeq x_{{\rm SM},{\rm match}}\,Z(s).
\label{eq:x_of_s}
\end{equation}
where $x_{{\rm SM},{\rm match}}\simeq0.38$ is the two-loop matching-scale value of the Standard Model, obtained by evaluating the relations 
above at $Z=1$. Evaluating the full two-loop relations numerically, the ratio at the matching scale (in the $\overline{\rm MS}$ scheme) 
is $x_{\rm match}(s^*)\in[0.055,0.065]$ for the seven benchmark points, a factor of $6$ to $7$ below the SM reference. 
This already shows that the $1/Z(s)$ enhancement drives the benchmark points toward the first-order side of the phase diagram in the perturbative description.

The comparison with the lattice-measured phase boundary requires a conversion to the lattice renormalisation scheme. 
In that scheme the SM sits at $x_{\rm SM}\simeq0.17$, just above the critical value $x_c\simeq0.11$ (the crossover region)~\cite{Rummukainen:1998as,Kajantie:1996mn}, 
and the benchmark points are described by $x(s^*)\simeq0.17\,Z_\infty\in[0.020,0.026]$. The two estimates differ only by the $\overline{\rm MS}$-to-lattice 
scheme conversion, since the three-dimensional running of $x$ is small to one-loop order. Both therefore agree that the benchmark points lie far below 
the critical value $x_c=0.11$, confirming that the tail regions of the $s$ distribution undergo a first-order transition. This provides an independent 
cross-check of the four-dimensional daisy-resummed analysis of Section~\ref{sec:nucleation}.

Renormalisation-group improvement of the input couplings does not affect this conclusion. The three-dimensional running from the matching scale 
to the critical point is already absorbed in the lattice-calibrated anchor $x_{\rm SM}\simeq0.17$, which is the outcome of the full three-dimensional 
RG evolution matched to the non-perturbative phase boundary. What remains is the four-dimensional running of the couplings from the electroweak 
scale to the matching scale $\mu=4\pi T e^{-\gamma_E}\sim\mathcal O({\rm TeV})$. Over the range $M_Z$ to $1\,{\rm TeV}$ the gauge, Yukawa, 
and Higgs couplings vary only by a few percent. This is completely dominated by the $1/Z(s)$ enhancement, which rescales $g_{\rm eff}^2$ by 
a factor $1/Z(s)$ of order $6.7$ at $Z_\infty=0.15$ and $10$ at $Z_\infty=0.10$. Since $x(s)$ lies a factor of $4$ to $5$ below 
the critical value $x_c=0.11$ (and $6$ to $8$ below the SM reference $x_{\rm SM}\simeq0.17$), the few-percent four-dimensional 
running cannot change the order of the transition.

We emphasise the division of labour between the two determinations. The 3D effective theory fixes the order of the transition,
and its lattice-calibrated critical ratio $x_c\simeq0.11$ shows that the tail regions with $x(s)\ll x_c$ undergo a first-order transition. 
Thermodynamic quantities such as the latent heat, $\bar\alpha$, and $\overline{\beta/H}$ can also equally be computed within the 3D effective theory, 
both perturbatively and on the lattice~\cite{Niemi:2021qvp,Niemi:2024vzw}. In the present work these quantities are instead evaluated in 
the daisy-resummed four-dimensional potential of Section~\ref{sec:nucleation}. 
This four-dimensional treatment is a theoretical approximation: the one-loop daisy-resummed potential 
incorrectly predicts a weak first-order transition rather than the crossover at the SM point~\cite{Rummukainen:1998as,Kajantie:1996mn}, 
and it does not capture the full higher-loop accuracy of the 3D effective theory. For this reason the present values of $f_{\rm FOPT}$, 
$\bar\alpha$, $\overline{\beta/H}$, and the resulting gravitational-wave spectra should be regarded as approximate results. The first-order 
conclusion itself is robust, because in the regime $x(s)\ll x_c$ the qualitative picture is expected to hold. Moreover, $x(s)<x_c$ is 
satisfied for every field value $s$ that enters $f_{\rm FOPT}$ and the Section~\ref{sec:Teff} averages, not only at the saddle point $s^*$: 
since $x(s)\propto Z(s)$ decreases monotonically into the non-Gaussian tail~[Eq.~\eqref{eq:x_of_s}], the largest ratio $x(s)/x_c$ within the 
contributing set $\mathcal S$ is attained at its lower edge, where $|s|$ is smallest. We have verified on the same logarithmic $s$-grid used 
for the nucleation scan that $x(s)<x_c$ holds throughout $\mathcal S$.

We note that the quantitative strength obtained from the one-loop-with-daisy 4D potential is a first estimate. Two-loop computations
within the dimensionally reduced 3D effective theory can shift the critical temperature and the transition strength by $20\%$--$50\%$ 
\cite{Niemi:2021qvp,Niemi:2024vzw}; these references perform the two-loop calculation inside the 3D effective theory rather than in a 
four-dimensional potential, and they scan its parameter space. The broader two-loop 3D EFT program for the electroweak phase transition 
\cite{Niemi:2018asa,Gould:2019qek,Schicho:2022wty,Patel:2011th} supports the same assessment. A transition that is strong at one loop 
may weaken at two loops. This theoretical uncertainty does not affect the first-order conclusion. The effective-theory ratio 
$x(s^*)$ lies a factor of $4$--$5$ below the critical value $x_c$ at two-loop level, so the order of the transition is robust to the size of the missing higher-loop corrections.
% A complete two-loop dimensional reduction with lattice validation is left for future work.

Whether the added scalar acts as a dynamical field (case (i)) or as a static background (case (ii)) is the relevant distinction for the effective theory. A dynamical singlet 
acquires a VEV and participates in the infrared Higgs dynamics, whereas a static background only rescales the gauge coupling. Our spectator $s$ is of 
the static-background type (case (ii)). The three-dimensional effective theory is therefore SM-like, with $s$ entering only through its effect on the matching. The crucial 
difference from a uniform singlet-extended model is that $x(s)$ is not a fixed parameter but a stochastic variable set by the inflationary fluctuations 
of $s$. Its distribution straddles the lattice critical value $x_c=0.11$, with the bulk of the volume at $Z(s)\simeq1$ (crossover) and the non-Gaussian 
tail at $Z(s)\ll1$ (first order). The volume fraction $f_{\rm FOPT}\in(0,1)$ is then a consequence of this distribution rather than an assumption.

Finally, we address the thermal correction from the portal $\kappa$. The tree-level background term $-\kappa s^2/2$ is retained in the local mass 
parameter $m_{\rm loc}^2$ of Eq.~\eqref{eq:mlocal} and is not dropped. At the benchmark saddle points $s^*/\Lambda\sim40$--$80$ it contributes 
between $12\%$ and $47\%$ of $\mu^2$ for the representative point of Table~\ref{tab:ewpt_parameters} (and a smaller fraction for the points with 
$\Lambda=10^7$~GeV), so its inclusion is already accounted for in the quoted values of $T_n$, $\bar\alpha$, and $\beta/H_n$. What is genuinely negligible 
is the loop-induced thermal mass $\propto\kappa T^2/16\sim10^{-13}\,\text{GeV}^2$, which is smaller than $D_{\rm eff}T^2$ by a factor of $1.8\times10^{-17}$. 
The statement that $\kappa$ corrections are negligible therefore refers to this loop thermal mass only.

A caveat remains. For $Z_\infty\lesssim0.1$ the loop expansion parameter $\epsilon_{\rm gauge}=g_{\rm eff}^2/(4\pi)$ grows and the SM-like three-dimensional 
description may require higher-dimensional operators. We adopt $Z_\infty\ge0.1$ as a conservative bound throughout, as discussed in Section~\ref{sec:conditions}. 
The residual uncertainty in the intermediate regime $s/\Lambda\sim\mathcal{O}(1)$ affects the normalisation of $f_{\rm FOPT}$ but not the qualitative conclusion 
that a first-order transition is driven by the non-Gaussian tail.

\section{Stochastic gravitational wave background}
\label{sec:GW}

A first-order EWPT proceeds via the nucleation, expansion, and collision of true vacuum bubbles. This violent process in the early 
universe generates a SGWB that could be detected by space-based interferometers~\cite{No:2011fi,Caprini:2019egz}. In this section, we compute the characteristic 
parameters of the phase transition and derive the predicted GW energy spectrum.

\subsection{Phase transition parameters}
\label{subsec:GWparams}

The GW signal from a first-order phase transition is controlled by two primary dimensionless parameters, namely the phase transition 
strength $\alpha$ and the inverse time duration $\beta/H$.

\paragraph{Phase transition strength $\alpha$.}
The parameter $\alpha$ quantifies the latent heat released during the transition relative to the radiation energy density of the plasma. It is defined as
\begin{equation}
\label{eq:alpha}
\alpha \equiv \frac{1}{\rho_{\text{rad}}(T_n)} \left[ \Delta V(T_n, s) - \frac{T_n}{4} \frac{\partial \Delta V(T, s)}{\partial T} \bigg|_{T=T_n} \right],
\end{equation}
where $\Delta V(T,s) = V_{\text{eff}}(0,T,s) - V_{\text{eff}}(v(T,s),T,s)$ is the vacuum energy difference, and the radiation energy density 
is $\rho_{\text{rad}}(T) = \frac{\pi^2}{30} g_* T^4$. The second term accounts for the entropy released during the transition.
Using the effective potential at finite temperature, we obtain:
\begin{equation}\label{eq:analytic_alpha}
\alpha(T_n, s) = \frac{1}{\rho_{\mathrm{rad}}(T_n)} 
\Biggl[ 
\frac{1}{4} D_{\mathrm{eff}}(s)\, T_n^2 v_n^2 
- \frac{3}{4} E_{\mathrm{eff}}(s)\, T_n v_n^3 
+ \frac{\lambda_H}{4} v_n^4 
\Biggr],
\end{equation}
where $v_n\equiv v(T_n,s)$.
The analytic result~\eqref{eq:analytic_alpha} neglects the temperature dependence of $s$. 

\paragraph{Inverse time duration $\beta/H$.}
The nucleation rate per unit volume is given by $\Gamma(T,s)$. Expanding the action around the nucleation temperature $T_n$ gives
\begin{equation}
\label{eq:beta}
\frac{\beta}{H_n} \equiv -T_n \frac{d \ln \Gamma (T,s)}{d T} \bigg|_{T=T_n}.
\end{equation}
A smaller value of $\beta/H_n$ corresponds to a stronger and longer-lasting phase transition, leading to a larger GW amplitude.

\paragraph{Bubble wall velocity $v_w$.}

For sufficiently strong transitions with $\alpha\gtrsim1/3$, a steady detonation can be realised, and its wall velocity is well 
approximated by the Chapman--Jouguet value, the minimum velocity of a steady supersonic detonation~\cite{Steinhardt:1981ct,Espinosa:2010hh},
\begin{equation}\label{eq:Jouguet}
v_{\rm CJ}(\alpha) = \frac{\sqrt{\alpha(2+3\alpha)} + 1}{\sqrt{3}\,(1+\alpha)}\,.
\end{equation}
While a strong first-order transition can arise in the cubic approximation, the full finite-temperature one-loop potential lowers the typical strength 
to $\bar\alpha\sim 3\times10^{-3}$ (Table~\ref{tab:ewpt_parameters}), rendering the transition considerably weaker.

For the weak transitions studied here the system lies firmly in the deflagration regime; rather than fixing a single Chapman--Jouguet value, 
we bracket $v_w$ between the model-independent LTE upper bound of Refs.~\cite{Ramsey-Musolf:2024ykk,Ai:2023jcap} and the ballistic 
lower bound of Ref.~\cite{Ai:2024btx}. The analytic LTE fit (derived in Ref.~\cite{Ai:2023jcap} and employed in 
Ref.~\cite{Ramsey-Musolf:2024ykk}) yields
\begin{equation}\label{eq:vw_LTE}
v_w^{\rm LTE} = \Bigl[\Bigl|\frac{3\alpha+\Psi-1}{2(2-3\Psi+\Psi^3)}\Bigr|^{c/2}
+ \Bigl|v_{\rm CJ}\Bigl(1-a\frac{(1-\Psi)^b}{\alpha}\Bigr)\Bigr|^c\Bigr]^{1/c},
\end{equation}
with $a=0.2233$, $b=1.704$, $c=-3.433$, and $\Psi=\omega_t/\omega_f$ the enthalpy ratio. In the near-equilibrium limit $\Psi\simeq1$ 
the fit approaches the Chapman--Jouguet velocity $v_{\rm CJ}$ of Eq.~\eqref{eq:Jouguet}, which at the benchmark strength $\bar\alpha\simeq3\times10^{-3}$ evaluates to $v_{\rm CJ}\simeq0.62$, only marginally above the sound speed $c_s=1/\sqrt3\simeq0.58$ that it reduces to as $\alpha\to0$.

For the ballistic lower bound we adopt the analysis of Ref.~\cite{Ai:2024btx}, whose $\Delta V=\mathcal{P}^{\rm b}_{\rm friction}$ matching condition, 
applied to SM-like weak transitions, gives $v_w^{\rm b}\sim0.1$--$0.4$ at $\alpha\sim0.01$. Our seven benchmarks in Table~\ref{tab:ewpt_parameters} 
share the same SM-like plasma but have $\bar\alpha\sim3\times10^{-3}$, placing the bound near the lower end of this interval. Although a precise 
per-benchmark value requires the full ballistic pressure integral evaluated for our particle content, the seven points have nearly identical $\bar\alpha$, 
so the variation of their lower bounds lies well within the quoted spread. We therefore adopt the conservative bracket
\begin{equation}\label{eq:vw_bracket}
 v_w\in[v_w^{\rm b},v_{\rm CJ}]\simeq[0.1,0.62]\;,
\end{equation}
which spans the ballistic and near-equilibrium LTE limits.
Since the sound-wave gravitational-wave amplitude scales approximately linearly with $v_w$, even this sixfold variation leaves the signal many orders 
of magnitude below projected detector sensitivities.

%=================================================================
\subsection{Average phase transition parameters}
\label{sec:Teff}

A unique feature of our inhomogeneous scenario is that different spatial regions undergo the first‑order phase transition at different temperatures, depending on the local initial value of the spectator field $s_0$.
To characterise the global properties of the transition we define a volume‑weighted effective nucleation temperature $\overline{T}_n$ and similarly averaged values of the strength parameter $\alpha$ and the inverse duration $\beta/H_n$.
These are obtained through the following steps.

\begin{enumerate}
    \item For each $s_0$ on a dense logarithmic grid we first determine its critical temperature $T_c(s_0)$, defined as the temperature at which the free energies of the symmetric and broken phases become equal,
    \begin{equation}
        \Delta V\bigl(T,\,s(T)\bigr) \equiv V_{\rm eff}(0,T,s(T)) - V_{\rm eff}\bigl(v(T),T,s(T)\bigr) = 0,
        \label{eq:Tc_def}
    \end{equation}
    where $s(T)$ follows the cosmological evolution~\eqref{eq:s_evolution}.
    Equation~\eqref{eq:Tc_def} is solved numerically by bisection on the interval $[T_{\rm min},T_{\rm max}]$.
    If no solution exists, the corresponding $s_0$ is discarded because it can never host a first‑order transition.

    \item Starting from $T_c(s_0)$ and moving downwards in temperature along the discrete grid $\{T_i\}$, we search for the highest temperature $T_n(s_0) \le T_c(s_0)$ that simultaneously satisfies
    \begin{align}
        \Delta V\bigl(T,\,s(T)\bigr) &> 0, \label{eq:DeltaV_pos} \\
        \Gamma\bigl(T,\,s(T)\bigr)   &\ge H^4(T). \label{eq:Gamma_condition}
    \end{align}
    Condition~\eqref{eq:DeltaV_pos} guarantees that the broken phase is thermodynamically favoured, while condition~\eqref{eq:Gamma_condition} ensures that thermal 
    bubble nucleation is efficient enough to complete the transition within a Hubble time.
    The first temperature (counting from $T_c$) that fulfills both criteria is identified as the nucleation temperature of that spatial volume.
    If no temperature down to $T_{\rm min}$ satisfies both conditions, the volume is considered to undergo a smooth crossover and is excluded from the averages.

    \item For each initial field value $s_0$ that successfully nucleates a first-order transition, we record its nucleation temperature $T_n(s_0)$.
    We then compute the global averages of the phase-transition strength and the inverse time duration at $T_n$ as approximate quantities.
    The local phase-transition strength $\alpha\bigl(T_n(s_0)\bigr)$ is evaluated at the saddle‑point field value of the global average nucleation rate at that temperature.
    The nucleation rate prefactor $\beta/H_n$ is obtained from the global average nucleation rate $\Gamma_{\rm avg}(T)$ through
    \begin{equation}
        \frac{\beta}{H_n}(T) = -T\,\frac{d\ln\Gamma_{\rm avg}(T)}{dT},
        \label{eq:beta_def}
    \end{equation}
    which is evaluated by a central finite difference with step $\Delta T = 4$~GeV around the temperature of interest.

    \item Finally, the volume‑weighted averages are computed using the exact probability density $\mathcal{P}(s_0)$ of initial field fluctuations:
    \begin{align}
        \overline{T}_n   &= \frac{1}{N}\displaystyle \int_{s_0\in\mathcal{S}} T_n(s_0)\,\mathcal{P}(s_0)\,ds_0, \label{eq:Teff_avg} \\[8pt]
        \bar{\alpha}     &= \frac{1}{N}\displaystyle \int_{s_0\in\mathcal{S}} \alpha\bigl(T_n(s_0)\bigr)\,\mathcal{P}(s_0)\,ds_0, \label{eq:alpha_avg} \\[8pt]
        \overline{\beta/H_n} &= \frac{1}{N}
            \left(\displaystyle \int_{s_0\in\mathcal{S}} \bigl[\beta/H_n\bigl(T_n(s_0)\bigr)\bigr]^{-1}\,\mathcal{P}(s_0)\,ds_0\right)^{\!-1}. \label{eq:beta_avg}
    \end{align}
    where $N=\displaystyle \int_{s_0\in\mathcal{S}} \mathcal{P}(s_0)\,ds_0$.
    The integrals run over the set $\mathcal{S}$ of all initial field values that successfully nucleate a first‑order phase transition, i.e.\  the same set that determines $f_{\rm FOPT}$.
    Numerically the integrals are evaluated on the same logarithmic $s_0$ grid, with weights $w_i = \mathcal{P}(s_{0,i})\,\Delta s_{0,i}$.
    The inverse‑variance weighting for $\beta/H_n$ is adopted because the gravitational‑wave amplitude is approximately proportional to $(\beta/H_n)^{-1}$; 
    it gives a conservative estimate of the spectral amplitude when the nucleation temperatures are broadly distributed.
\end{enumerate}

The quantities $\overline{T}_n$, $\bar{\alpha}$ and $\overline{\beta/H_n}$ defined above serve as effective global parameters that characterise the spatially 
averaged phase‑transition dynamics. They are the ones used in the following to compute the SGWB, supplemented by the multi‑temperature superposition described there.

%====================================================
\begin{figure}
    \centering
    \includegraphics[width=1.0\linewidth]{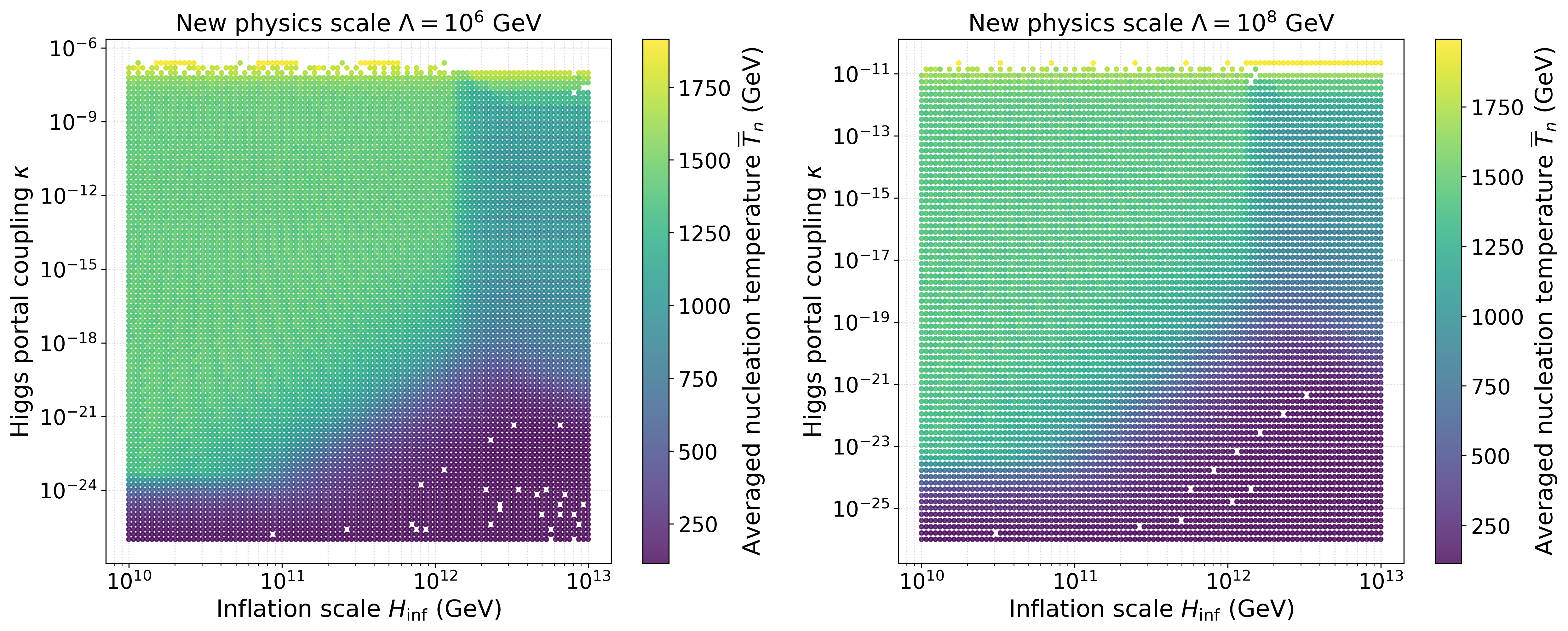}
    \caption{Scatter points distribution in $H_{\rm inf}-\kappa$ plane, with cubic approximation. The colorbar denotes the values of the nucleation temperature $T_n$. We take the 
    new physics scale $\Lambda=10^{6}$~GeV and $10^8$~GeV for the left and right panels, respectively.
    The scalar mass $m_s$ is determined by requiring the correct DM relic abundance $\Omega_s h^2 = 0.12$.}
    \label{fig:Average_Tn}
\end{figure}
In Fig.~\ref{fig:Average_Tn}, we present the scan results for the averaged nucleation temperature $\overline{T}_n$ in the $H_{\rm inf}$-$\kappa$ parameter space using 
the truncated cubic approximation. 
For a given inflationary Hubble scale $H_{\rm inf}$, we determine the scalar mass $m_s$ by requiring the correct DM relic abundance $\Omega_s h^2 = 0.12$ 
according to Planck observations. We observe that $\overline{T}_n$ decreases monotonically as $H_{\rm inf}$ increases and spans the electroweak window 
$T_n\sim100$--$300$~GeV in the most efficient region, with $\sim 1300$ parameter points reaching the EWPT range $T_n\in[100,200]$~GeV. The upper tail extends to 
$\sim 1$--$2$~TeV (the limit of our temperature scan) for smaller $H_{\rm inf}$ or sub-optimal $\kappa$. 

%====================================================
\begin{figure}
    \centering
    \includegraphics[width=1.0\linewidth]{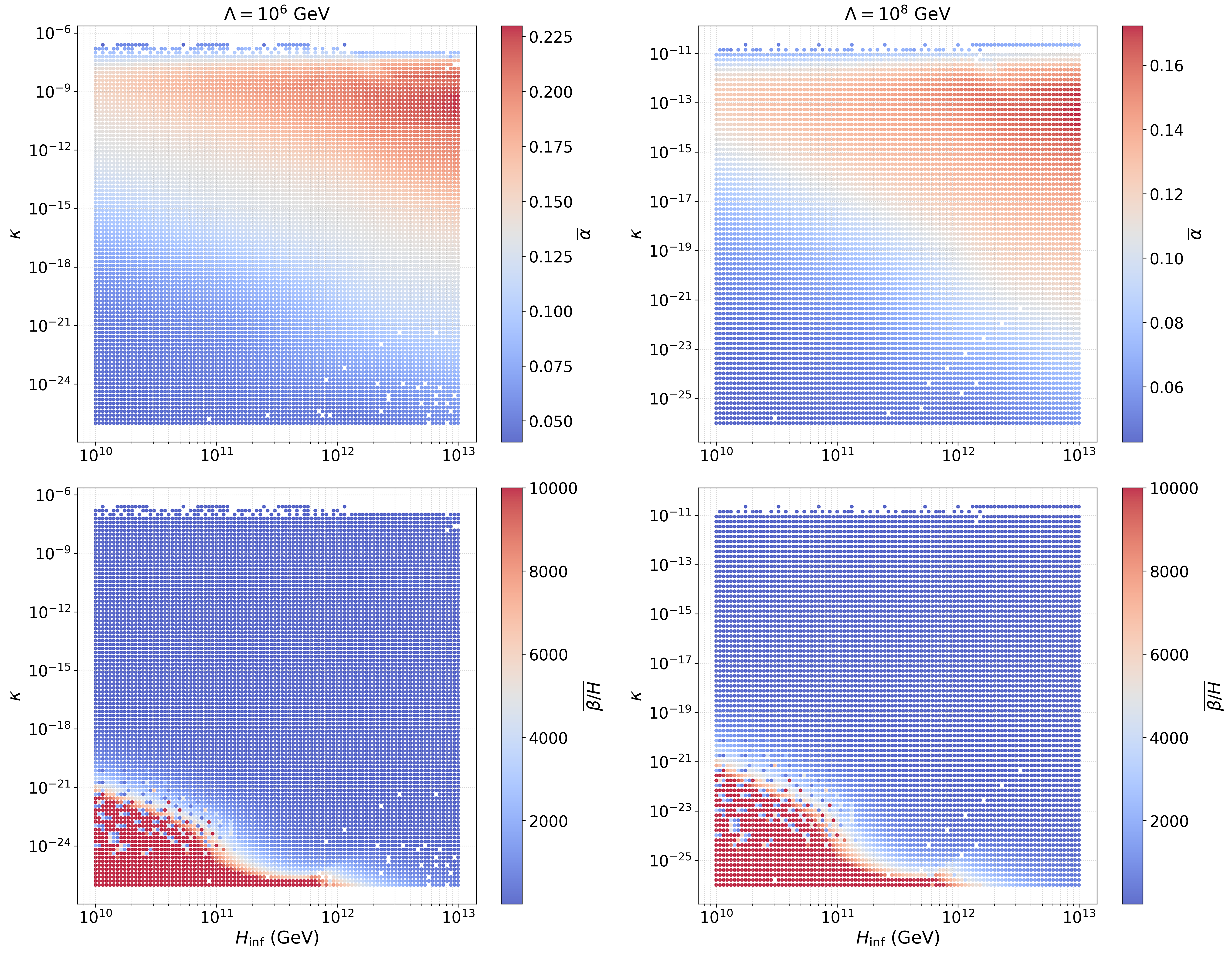}
    \caption{Scatter plot in the $H_{\rm inf}$-$\kappa$ plane, with cubic approximation. \emph{Upper panels}: colorbar is the averaged latent heat $\bar{\alpha}$. 
\emph{Lower panels}: colorbar is the averaged inverse time duration $\overline{\beta/H}$ (computed with the inverse-variance weighting appropriate 
for the gravitational-wave spectrum, Eq.~\eqref{eq:beta_avg}). The left and right panels correspond to new physics scales $\Lambda = 10^6$~GeV and $10^8$~GeV, respectively.
The scalar mass $m_s$ is determined by requiring the correct DM relic abundance $\Omega_s h^2 = 0.12$.}
    \label{fig:ab}
\end{figure}
In Fig.~\ref{fig:ab}, we present numerical calculations of the averaged phase transition parameters $\overline{\alpha}$ and $\overline{\beta/H}$ in the 
$H_{\rm inf}$-$\kappa$ parameter space with the truncated cubic approximation. The transition strength lies in the range $\bar\alpha\sim 0.04$--$0.23$, 
i.e.\ $\bar\alpha=\mathcal{O}(0.1)$ over most of the parameter space. 
We observe that for a given $\Lambda$, the transition is strengthened as $\kappa$ and $H_{\rm inf}$ increase, with $\overline{\alpha}$ rising from 
about $0.04$ to $\sim 0.23$. 
We also observe from the figure that the two phase transition parameters $\overline\alpha$ and $\overline{\beta/H}$ are anti-correlated.
This is because a weaker transition releases less latent heat and completes more quickly.
% a weaker transition releases 
% less latent heat and completes more quickly, so $\alpha\lesssim 10^{-2}$ is typically accompanied by $\beta/H\sim 10^{3}$--$10^{4}$, 
% whereas a strongly supercooled transition with $\alpha\sim 1$ has $\beta/H\sim\mathcal{O}(1)$. 
% Equivalently, a \emph{smaller} $\beta/H$ signals a slower and more delayed transition. In the moderately strong regime relevant here ($\bar\alpha\sim 0.1$) one 
% therefore expects $\beta/H\sim 10^{2}$--$10^{3}$, which is indeed what we find for the points in the electroweak window ($T_n\in[100,200]$~GeV), where 
% $\overline{\beta/H}\sim 400$--$500$. 
% Throughout this range the transitions are of the \emph{moderately} strong type: genuine strong supercooling requires $\alpha\gtrsim1$, 
% which is reached neither in the cubic scan nor in the full one-loop results of Table~\ref{tab:ewpt_parameters} (where $\overline{\alpha}\sim3\times10^{-3}$).
% The tail reaching $\mathcal{O}(10^{4}$--$10^{5})$ visible in the lower panels corresponds instead to the \emph{weakest} transitions (smallest $\bar\alpha$) of the scan.
Since a larger $H_{\rm inf}$ leads to stronger scalar fluctuations, $\overline{\alpha}$ increases with the inflationary scale $H_{\rm inf}$. Furthermore, comparing 
the left and right panels shows that a lower new physics scale $\Lambda$ yields a stronger phase transition, i.e.\ a larger $\bar\alpha$ and therefore, by the 
anti-correlation above, a smaller $\overline{\beta/H}$. Equivalently, a higher $\Lambda$ gives a larger $\overline{\beta/H}$, i.e.\ a faster and less delayed 
transition.

We note that the cubic approximation overestimates the strength of the phase transition. Using the full one-loop thermal functions with daisy resummation, the averaged nucleation 
temperature is $\overline{T}_n\simeq 154$--$166$~GeV, the transition strength is $\overline{\alpha}\sim (2.7$--$3.3)\times10^{-3}$ and the inverse duration is $\overline{\beta/H}\sim(5.5$--$5.9)\times10^{3}$ 
for the viable points (Table~\ref{tab:ewpt_parameters}), whereas the cubic scan restricted to the same electroweak window ($T_n\in[100,200]$~GeV) gives 
$\overline{\alpha}\sim 0.05$--$0.12$ and $\overline{\beta/H}\sim 400$--$500$. The truncated cubic approximation therefore overestimates $\overline{\alpha}$ by a factor of $\sim 15$--$45$ 
and gives a $\overline{\beta/H}$ smaller by a factor of $\sim 10$ than the full one-loop result. The strong-first-order criterion $v_c/T_c\gtrsim 1$ remains satisfied in both 
treatments, but the latent-heat parameter $\alpha$ is more than an order of magnitude smaller than in the cubic approximation.

%====================================================

\subsection{Gravitational waves}
\label{subsec:GWspectrum}

The total GW energy density spectrum $\Omega_{\text{GW}} h^2$ receives contributions from three primary sources:
\begin{enumerate}
    \item \textbf{Bubble collisions} – the kinetic energy of the scalar field gradient across the colliding bubble walls.
    \item \textbf{Sound waves} – compression waves in the plasma generated by the expanding bubbles.
    \item \textbf{Magnetohydrodynamic (MHD) turbulence} – chaotic motion of the plasma after the transition completes.
\end{enumerate}
The full spectrum can be approximated by the sum
\begin{equation}
\label{eq:GWsum}
\Omega_{\text{GW}} h^2 \approx \Omega_{\text{col}} h^2 + \Omega_{\text{sw}} h^2 + \Omega_{\text{turb}} h^2.
\end{equation}
The total GW spectrum is also scaled by a factor $f_{\rm FOPT}$ to take into account the fraction of first-order phase transition.
We adopt the state-of-the-art parameterisations of the three individual contributions. The envelope approximation for bubble collisions follows Ref.~\cite{Huber:2008hg}. The sound-wave source follows Refs.~\cite{Hindmarsh:2015qta,Hindmarsh:2017gnf}, with the efficiency factor $\kappa_v$ taken from Ref.~\cite{Espinosa:2010hh}. The MHD turbulence spectrum follows Ref.~\cite{Caprini:2009yp}. The overall phenomenological framework and the signal-to-noise-ratio method are reviewed in Refs.~\cite{Caprini:2015zlo,Caprini:2019egz}.
The formulae of these modelling are provided in Appendix~\ref{app:GW_spectrum}.

\begin{table}[htbp]
  \centering
  \small
  
  \caption{Benchmark points for a first-order EWPT in the one-loop daisy-resummed finite-temperature potential.  
  }
  \label{tab:ewpt_parameters}
  \begin{tabular}{|c|ccc|cccc|c|}
    \toprule
    $Z_\infty$ & $\Lambda$ [GeV] & $H_{\rm inf}$ [GeV] & $\kappa$
      & $f_{\rm FOPT}$ & $\bar\alpha$ & $v_c/T_c$ & $\overline{\beta/H_n}$ & $\bar T_n$ [GeV] \\
    \midrule
    0.15 & $10^9$          & $2\times10^{12}$ & $1.14\times10^{-18}$ & 0.510 & 0.00284 & 1.053 & 5603 & 156 \\
    0.15 & $10^8$          & $10^{12}$       & $1.14\times10^{-16}$ & 0.614 & 0.00284 & 1.053 & 5659 & 157 \\
    0.13 & $10^8$          & $10^{12}$       & $7.39\times10^{-17}$ & 0.657 & 0.00276 & 1.078 & 5844 & 164 \\
    0.15 & $3\times10^7$   & $5\times10^{11}$ & $1.69\times10^{-15}$ & 0.711 & 0.00313 & 1.032 & 5755 & 160 \\
    0.13 & $3\times10^7$   & $5\times10^{11}$ & $8.21\times10^{-16}$ & 0.724 & 0.00276 & 1.078 & 5853 & 164 \\
    0.14 & $10^7$          & $5\times10^{11}$ & $5.01\times10^{-16}$ & 0.821 & 0.00327 & 1.170 & 5527 & 154 \\
    0.12 & $10^7$          & $5\times10^{11}$ & $5.01\times10^{-16}$ & 0.819 & 0.00270 & 1.155 & 5886 & 166 \\
    \bottomrule
  \end{tabular}
  
\end{table}

These seven points are selected from an extended daisy-resummed parameter scan over $\Lambda\in[10^7,\,10^9]$~GeV, $H_{\rm inf}\in[5\times10^{11},\,2\times10^{12}]$~GeV 
and $\kappa$ spanning several decades, with $Z_\infty\in[0.10,\,0.15]$. In every case the transition is first order with $v_c/T_c\simeq 1.03$--$1.17$, 
and the seven benchmark points above keep $f_{\rm FOPT}>0.5$ and $v_c/T_c>1$ with $\overline{\alpha}\sim (2.7$--$3.3)\times10^{-3}$.

EFT-validity remark: the seven benchmark points with one-loop effective potential require a saddle-point spectator amplitude $s^*/\Lambda$ in the range $\sim 40$--$80$, i.e.\ well above 
the naive cutoff scale $\Lambda$ of the effective theory. We have assumed that the $Z_2$-symmetric non-Gaussian spectator potential remains a valid effective 
description in this regime; the only relevant deformation is the field-dependent gauge coupling $g_{\rm eff}(s)=g/\sqrt{Z(s)}$, and perturbativity is guaranteed by 
the bound $Z_\infty\ge 0.1$. A fully UV-complete treatment of the spectator sector, including resummation of higher-dimensional operators suppressed by $\Lambda$, 
is beyond the scope of current work. Our qualitative conclusion, that a first-order EWPT is driven by the non-Gaussian tail, does not depend on the precise value of $s^*/\Lambda$ 
provided $Z(s)\ll 1$.

\begin{figure}
    \centering
    \includegraphics[width=0.7\linewidth]{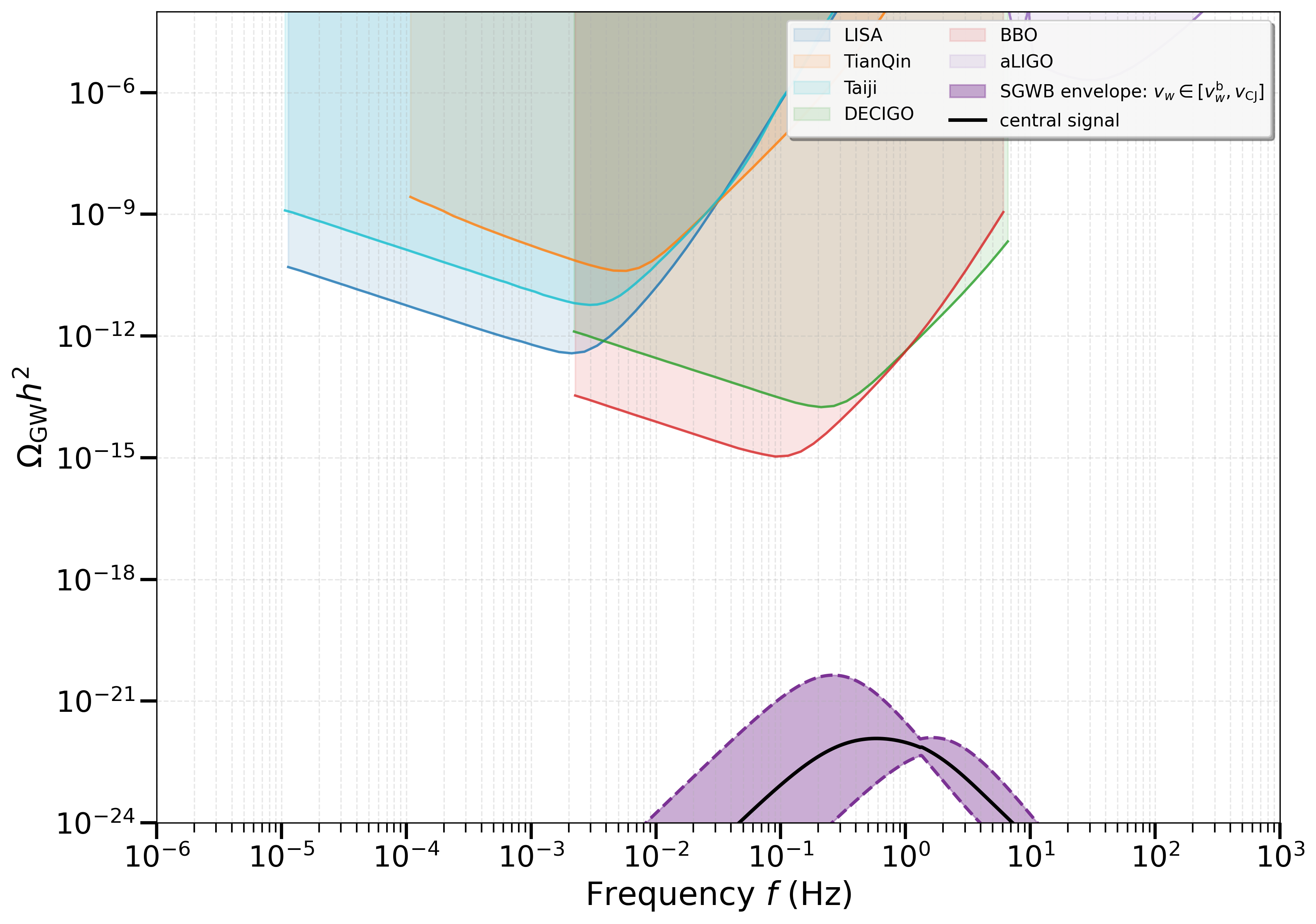}
    \caption{Banded SGWB envelope from the seven Table~\ref{tab:ewpt_parameters} benchmarks as a function of frequency $f$. The purple band is obtained by evaluating each benchmark 
    at the wall-velocity bracket $v_w\in[v_w^{\rm b},v_{\rm CJ}]\simeq[0.1,0.62]$. The black solid curve shows the geometric 
    mean of the envelope, a representative central signal.}
    \label{fig:gw_sepcttrum}
\end{figure}
In Fig.~\ref{fig:gw_sepcttrum} we present the banded SGWB envelope from the seven Table~\ref{tab:ewpt_parameters} benchmarks. 
The purple band reflects the model-independent wall-velocity uncertainty, $v_w\in[v_w^{\rm b},v_{\rm CJ}]\simeq[0.1,0.62]$.
For the points in Table~\ref{tab:ewpt_parameters} the daisy-resummed phase transition has strength $\overline{\alpha}\sim (2.7$--$3.3)\times10^{-3}$ and a 
distribution-averaged inverse duration $\overline{\beta/H_n}\sim (5.5$--$5.9)\times10^{3}$. Since the sound-wave GW amplitude scales as $\alpha^2$ 
and is additionally suppressed by the large $\beta/H$ (the spectral peak shifts to higher frequency as $\beta/H$ grows while the envelope falls 
roughly as $(\beta/H)^{-1}$), the resummed treatment yields a signal suppressed by several orders of magnitude relative to the cubic prediction. 
Even the upper edge of the band lies far below the projected LISA/Taiji/TianQin/DECIGO/BBO sensitivities, so our conclusion is robust against the 
wall-velocity uncertainty.

%=================================================================

\section{Conclusions and discussions}
\label{sec:conclusion}

In this work, we have proposed and systematically investigated a novel mechanism for triggering a cosmological first-order EWPT, 
utilising non-Gaussian primordial fluctuations of a $\mathbb{Z}_2$-symmetric spectator scalar field. Unlike conventional scenarios in which the phase transition 
proceeds uniformly throughout the Universe, our mechanism naturally gives rise to a spatially inhomogeneous transition, where different regions experience either 
bubble nucleation or a smooth crossover, depending on the local amplitude of the spectator field fluctuation.
The spectator field can couple to the SM and significantly modify the SM couplings at high energy scales. In this work, we show that the effective Higgs mass parameter and the gauge couplings are altered by these interactions. The portal coupling shifts the former, 
while the $s$-dependent gauge kinetic function rescales the latter, $g_{\rm eff}=g/\sqrt{Z(s)}$. This substantially modifies the thermal Higgs mass and enhances the cubic barrier 
term in the finite-temperature effective potential, thereby enabling a first-order phase transition. (The physical Higgs-boson mass is held at its measured value 
$m_h=125$~GeV throughout; it is the effective mass parameter and the thermal corrections that are $s$-dependent.)

We have derived the exact $\mathbb{Z}_2$-symmetric probability density function for the spectator field fluctuations, which exhibits significantly enhanced tails 
compared to a purely Gaussian distribution. Using the saddle-point approximation, we have computed the global average nucleation rate by statistically averaging 
over all possible fluctuation amplitudes. Our numerical results demonstrate that the volume fraction of the Universe undergoing a first-order phase transition 
$f_{\text{FOPT}}$ can reach unity for a wide range of portal couplings $\kappa$ and inflationary scales $H_{\text{inf}} \lesssim 10^{13}$ GeV within the cubic approximation, 
whereas the daisy-resummed benchmark points give the more modest values $f_{\rm FOPT}\simeq 0.51$--$0.82$ (Table~\ref{tab:ewpt_parameters}). 
CMB isocurvature constraints preclude a scale-invariant power spectrum for $s$ and instead require a peaked spectrum at electroweak scales. Such a 
feature is naturally realised via transient non-slow-roll dynamics during inflation. The local approximation adopted in this work, wherein $s$ is treated as constant 
across the bubble, is well justified by the $\sim 10^{3}$ hierarchy that separates the bubble size from the spectator-field variation scale.

A distinctive feature of our scenario is that different spatial regions undergo the phase transition at different temperatures, determined by their local initial 
spectator field amplitude. We have developed a volume-weighted averaging procedure to compute the effective global phase transition parameters, including the average 
nucleation temperature $\bar{T}_n$, average transition strength $\bar{\alpha}$, and average inverse duration $\overline{\beta/H_n}$. 
Our daisy-resummed calculations at the benchmark points give an average nucleation temperature $\bar{T}_n\simeq 154$--$166$~GeV. 
The average transition strength is $\bar{\alpha}\sim (2.7$--$3.3)\times10^{-3}$; this is smaller than the cubic prediction 
($\overline{\alpha}_{\rm cubic}\sim 0.04$--$0.23$ over the full scan, and $\sim0.05$--$0.12$ once restricted to the same electroweak window) by more than an order of magnitude, 
up to a factor of $\sim45$, while the transition remains first order with $v_c/T_c\simeq 1.03$--$1.17$, 
and the average inverse duration is $\overline{\beta/H_n}\sim (5.5$--$5.9)\times10^{3}$.

The spectator field in our model also naturally serves as a cold DM candidate. We have shown that the observed DM relic abundance can be reproduced 
over a wide range of parameters, thereby linking the origin of electroweak symmetry breaking to the nature of DM.
We have also calculated the SGWB from the daisy-resummed full one-loop parameters; the resulting signal lies far below the projected sensitivities of all planned detectors (LISA, TianQin, Taiji, DECIGO, BBO, aLIGO).

The core idea of using non-Gaussian primordial fluctuations to trigger cosmological phase transitions via portal couplings is a general paradigm that can be applied to 
other early-Universe phase transitions. For example, it could be used to trigger a dark sector phase transition or a Peccei-Quinn phase transition, where constraints 
from the SM thermal bath are less stringent. In these contexts, even stronger first-order transitions with louder GW signals might be realised, opening 
up new windows into the physics of the early Universe.

\section*{Acknowledgements}
BQL is supported in part by the National Natural Science Foundation of China under Grant No.~12405058 and by the Zhejiang 
Provincial Natural Science Foundation of China under Grant No.~LQ23A050002.

%==========================================================================
\appendix

\section{UV completions for the kinetic function}\label{app:UVcompletion}
In this appendix we present renormalisable UV models that generates the effective gauge kinetic function $Z(s)$ used in Sec.~\ref{sec:gauge_coupling}.

%===============================================================
\subsection{UV completion with heavy scalar}
First, let us consider the extension of the SM (SM) electroweak sector by a heavy modulus $\Phi$ that is a gauge singlet and even under the $\mathbb{Z}_{2}$ symmetry $s\to -s$ ($\Phi\to\Phi$). The Lagrangian reads
\begin{equation}\label{eq:UV}
\begin{aligned}
\mathcal{L}_{\mathrm{UV}} = &\; \frac{1}{2}(\partial_{\mu}\Phi)^{2} - \frac{1}{2}M^{2}\Phi^{2} - g\,\Phi s^{2} - \frac{1}{2}y^{2}\Phi^{2}s^{2} \\
& -\frac{1}{4}\Bigl(1 + \frac{\Phi}{\Lambda_{0}}\Bigr)\Bigl(W_{\mu\nu}^{a}W^{a\mu\nu} + B_{\mu\nu}B^{\mu\nu}\Bigr) ,
\end{aligned}
\end{equation}
where $M$, $g$, $y$ and $\Lambda_{0}$ are positive constants with $\Lambda_{0}$ a high energy scale.
All interactions respect the $\mathbb{Z}_{2}$ symmetry. 
% $s^{2}$ and $\Phi$ are even, and the term linear in $\Phi$ is allowed because $\Phi$ itself is even.
The pure mass term $\frac{1}{2}M^{2}\Phi^{2}$ ensures that in the absence of $s$ the field $\Phi$ has a minimum at $\Phi=0$, so that the tree‑level gauge kinetic term reduces to the canonical SM form.

At energies far below $M$, the heavy scalar is nearly static and the kinetic energy can be ignored.
The heavy scalar $\Phi$ is frozen to its $s$‑dependent classical background value, which is obtained by minimising the effective potential
\begin{equation}
V_{\rm eff}(\Phi,s) = \frac{1}{2}M^{2}\Phi^{2} + g\,\Phi s^{2} + \frac{1}{2}y^{2}\Phi^{2}s^{2}.
\end{equation}
The minimum condition $\partial V_{\rm eff}/\partial\Phi = 0$ gives
\begin{equation}
M^{2}\Phi + g s^{2} + y^{2}s^{2}\Phi = 0 \quad\Longrightarrow\quad
\Phi(s) = -\,\frac{g\,s^{2}}{M^{2}+y^{2}s^{2}} .
\label{eq:Phi}
\end{equation}
Substituting this background into the gauge kinetic term of~\eqref{eq:UV} yields the effective kinetic function
\begin{equation}
Z(s) = 1 + \frac{\Phi(s)}{\Lambda_{0}}
      = 1 - \frac{g}{\Lambda_{0}}\,\frac{s^{2}}{M^{2}+y^{2}s^{2}} .
\label{eq:Zinter}
\end{equation}
It is convenient to define the two parameters
\begin{equation}
\Lambda^{2} \equiv \frac{M^{2}}{y^{2}} , \qquad
c \equiv \frac{g}{y^{2}\Lambda_{0}} = \frac{g\Lambda^{2}}{M^{2}\Lambda_{0}} .
\end{equation}
In terms of these equation~\eqref{eq:Zinter} becomes
\begin{equation}
Z(s) = 1 - c\,\frac{s^{2}/\Lambda^{2}}{1 + s^{2}/\Lambda^{2}} = 1 - c + \frac{c}{1 + s^{2}/\Lambda^{2}}.
\label{eq:Zc}
\end{equation}
Finally, introducing the asymptotic value
\begin{equation}
Z_{\infty} \equiv \lim_{s\to\infty} Z(s) = 1 - c ,
\end{equation}
we obtain exactly the form used in the main text,
\begin{equation}
Z(s) = Z_{\infty} + \frac{1-Z_{\infty}}{1 + s^{2}/\Lambda^{2}} .
\label{eq:Zfinal}
\end{equation}
From this derivation we can check the three required properties:
\begin{enumerate}
\item $Z(0)=1$, recovering the SM at small $s$;
\item $Z(\infty)=Z_{\infty}$, so that the gauge coupling enhancement saturates at a finite value $g_{\rm eff}(\infty)=g/\sqrt{Z_{\infty}}$;
\item $Z(s)>0$ for all $s$ as long as $c<1$, i.e.\ $g<y^{2}\Lambda_{0}$, which can always be satisfied by a suitable choice of parameters.
\end{enumerate}
The parameter $Z_{\infty}$ is therefore identified with the fraction of the tree‑level kinetic term that survives when the heavy scalar is pushed far away from its vacuum by a large $s$ fluctuation.
Since $c>0$ by construction, $Z_{\infty}<1$ follows naturally, corresponding to a finite wave‑function renormalisation of the gauge fields induced by the frozen heavy modulus.
In the cosmological context, $s$ is large only in rare regions produced by inflationary non‑Gaussian tails; after the EWPT $s$ redshifts away as DM, so that today $Z=1$ and the SM is exactly recovered.

We note that the scalar $s$ could receive a mass at the loop level:
\begin{equation}
\delta m_s^2 \sim \frac{g^2}{16\pi^2} \ln\left(\frac{\Lambda_{\text{UV}}^2}{M^2}\right) \sim \frac{(10^8\ \text{GeV})^2}{16\pi^2} \times \mathcal{O}(1) \approx 10^{14}\ \text{GeV}^2,
\end{equation}
where we have assumed a typical value $g\sim 0.9M\sim 10^8$~GeV, which leads to a scalar mass correction $\delta m_s\sim 10^7$~GeV. 
Since $\delta m_s\lesssim H_{\rm inf}$, this mass correction does not significantly suppress the quantum fluctuation from inflation.

However, such a large mass correction is incompatible with our scenario. For one thing, since $T_{\rm osc}\propto m_s^{1/2}$, a heavier $s$ mass would significantly 
raise the oscillation temperature, causing the scalar to decay as $s\propto a^{-3/2}$ at a much earlier stage. This would strongly suppress its effects on the gauge 
coupling and Higgs portal interactions, unless we adopt a much lower new physics scale $\Lambda$ and a much larger value of $\kappa$, which may be subject to various 
experimental constraints. For another thing, a heavy scalar mass would also considerably increase the DM abundance, as indicated by Eq.~\eqref{eq:DM_abundance}. 
Consequently, reproducing the observed DM abundance would require a very low inflation scale, which appears infeasible within our theoretical framework.
% \rev{Like the vector-like-fermion and supersymmetric completions below, the heavy-scalar framework can also furnish the required $\mathbb{Z}_2$-even CP-violating 
% source through an analogous $s^{2}$-enhanced coupling carrying a physical phase, so that all three UV completions are viable electroweak-baryogenesis settings; 
% a quantitative analysis is left for future work.}

%===============================================================
\subsection{UV completion with heavy vector‑like fermions}

Let us consider a minimal fermionic completion that reproduces~\eqref{eq:Zparam} at low energies.
We extend the SM by a Dirac fermion $\Psi$ transforming as a doublet under $SU(2)_{L}$ and carrying hypercharge $Y=1/2$ (the quantum numbers are chosen for definiteness; other representations work similarly).
The fermion is odd under the $\mathbb{Z}_{2}$ symmetry, $\Psi\to -\Psi$, which forbids mixing with the SM fermions.
Its mass is generated through a Yukawa coupling to the spectator field $s$:
\begin{equation}\label{eq:fermion_mass}
\mathcal{L}\supset -M(s)\,\overline{\Psi}\Psi,\qquad
M(s)=M_{0}+\frac{y}{\Lambda}\,s^{2}.
\end{equation}
Here $M_{0}$ is a large bare mass ($M_{0}\gg T_{\rm EW}$), $y$ a Yukawa coupling, and $\Lambda$ the same high scale that appears in the kinetic function.
The term proportional to $s^{2}$ respects the $\mathbb{Z}_{2}$ symmetry ($s\to -s$, $\Psi\to -\Psi$) and vanishes at $s=0$, guaranteeing that the fermion mass reduces to $M_{0}$ in the SM vacuum.
The electroweak gauge interactions of $\Psi$ are contained in the standard covariant derivative
\begin{equation}
D_{\mu}\Psi = \partial_{\mu}\Psi - i g \frac{\sigma^{a}}{2} W_{\mu}^{a}\Psi - i g' \frac{Y}{2} B_{\mu}\Psi .
\end{equation}

% \subsection*{A.2 \ \ Integrating out the heavy fermion}

At energies far below $M(s)$, the fermion $\Psi$ can be integrated out.
Its one‑loop contribution to the gauge‑field two‑point function generates a field‑dependent wave‑function renormalisation.
Using dimensional regularisation with $\overline{\rm MS}$ subtraction at a renormalisation scale $\mu$, the effective gauge kinetic term becomes
\begin{equation}\label{eq:eff_kin_fermion}
\mathcal{L}_{\rm eff}^{\rm gauge} = -\frac{1}{4}\Bigl[1+\Delta Z(s)\Bigr]F_{\mu\nu}F^{\mu\nu},
\end{equation}
where $\Delta Z(s)$ is obtained from the well‑known fermion‑loop vacuum polarisation:
\begin{equation}
\Delta Z(s) = -\frac{g^{2}N}{16\pi^{2}}\,\ln\!\Bigl(\frac{M(s)^{2}}{\mu^{2}}\Bigr) .
\end{equation}
The group‑theory factor $N$ depends on the representation; for an $SU(2)$ doublet $N=1/2$, and a similar factor appears for the $U(1)_{Y}$ contribution.
The overall sign is negative because the fermion loop gives a diamagnetic contribution, analogous to QED vacuum polarisation.

We impose the renormalisation condition that at $s=0$ the kinetic term is canonical, i.e. $Z(0)=1$.
This fixes the constant piece and leads to
\begin{equation}
Z(s) = 1 + \Delta Z(s) - \Delta Z(0)
      = 1 - \frac{g^{2}N}{16\pi^{2}}\,\ln\!\Bigl(\frac{M(s)^{2}}{M_{0}^{2}}\Bigr).
\label{eq:Z_log}
\end{equation}
Substituting the mass function~\eqref{eq:fermion_mass} gives
\begin{equation}
Z(s) = 1 - \frac{g^{2}N}{16\pi^{2}}\,\ln\!\Bigl(1+\frac{y s^{2}}{M_{0}\Lambda}\Bigr).
\label{eq:Z_log_explicit}
\end{equation}
Equation~\eqref{eq:Z_log_explicit} already exhibits the two essential properties required for our mechanism:
(i) $Z(0)=1$,
(ii) $Z(s)<1$ for $s>0$ (since the logarithm is positive),
leading to an enhanced gauge coupling.
Moreover, as $s\to\infty$ the logarithm grows slowly (logarithmically), so the enhancement never diverges.

For practical calculations, and to facilitate a direct comparison with Eq.~\eqref{eq:Zparam}, it is convenient to approximate the logarithmic function by a rational form that captures the same low‑energy expansion and saturates at large $s$.
Expanding~\eqref{eq:Z_log_explicit} for $s^{2}\ll M_{0}\Lambda/y$,
\begin{equation}
Z(s) \approx 1 - \frac{g^{2}N}{16\pi^{2}}\,\frac{y s^{2}}{M_{0}\Lambda}
      = 1 - c\,\frac{s^{2}}{\Lambda^{2}},
\qquad 
c \equiv \frac{g^{2}N}{16\pi^{2}}\frac{y\Lambda}{M_{0}} .
\label{eq:low_expansion}
\end{equation}
This has exactly the form $Z(s)=1- \text{constant}\times s^{2}/\Lambda^{2}$ that appears in the low‑energy effective field theory.
To obtain a UV‑safe expression that saturates, we replace the logarithm by the rational function
\begin{equation}
Z(s) \equiv Z_{\infty} + \frac{1-Z_{\infty}}{1+s^{2}/\Lambda^{2}},
\qquad
Z_{\infty} \equiv 1 - c = 1 - \frac{g^{2}N}{16\pi^{2}}\frac{y\Lambda}{M_{0}} .
\label{eq:Z_rational}
\end{equation}
This choice reproduces the exact low‑$s$ expansion~\eqref{eq:low_expansion} up to $\mathcal O(s^{4}/\Lambda^{4})$ and ensures that $Z(s)\to Z_{\infty}$ for $s\gg\Lambda$.
The parameter $Z_{\infty}$ can be made arbitrarily small by choosing $c$ close to unity, e.g.\ $c=0.9$ yields $Z_{\infty}=0.1$.
Since $c>0$ by construction, one automatically has $Z_{\infty}<1$, as required.

\paragraph*{\bfseries CP-violating source enhanced by the spectator fluctuation.}
The mass function \eqref{eq:fermion_mass}, $M(s)=M_{0}+y s^{2}/\Lambda$, is engineered so that integrating out $\Psi$ produces the kinetic function $Z(s)$ in 
Eq.~\eqref{eq:Z_log_explicit}; it is \emph{real} (or can be taken real) and therefore carries no CP violation. The spectator field $s$ does \emph{not} acquire a 
VEV during the EWPT ($\langle s\rangle=0$ throughout, by the $\mathbb{Z}_{2}$ symmetry), so it cannot itself supply the \emph{spatial variation} of a CP source — that 
role is played exclusively by the Higgs profile $h(\mathbf{x})$, which interpolates between the symmetric phase ($h=0$) outside the bubble and the broken 
phase ($h=v_{c}$) inside. 

A crucial symmetry constraint must be observed. Since the $\mathbb{Z}_{2}$ symmetry $s\to -s$ (with $\Psi\to -\Psi$) protects $\langle s\rangle=0$, the CP-violating 
operator must respect $\mathbb{Z}_{2}$ and is therefore taken to be quadratic (even) in $s$. A na\"ive Higgs-dependent CP mass insertion 
$M_{0}+|y|e^{i\theta}h^{2}/\Lambda'/2$ is far too small: for $\Lambda'\sim 10^{8}$~GeV and $\langle h\rangle\sim 10^{2}$~GeV the CP-odd component is only 
$\sim 10^{-4}$~GeV. Instead, we exploit the large fluctuation $s/\Lambda\sim\mathcal{O}(1)$ in the FOPT-active regions to boost the source through the 
$\mathbb{Z}_{2}$-even coupling. The gauge-invariant parent of this $h$--$\Psi$ coupling is the dimension-five operator 
$-\frac{y_{\rm CP}(s)}{\sqrt{2}\,v}\,\overline{\Psi}\Psi\,H^{\dagger}H$ ($H$ the SM Higgs doublet, $v=246$~GeV); in unitary gauge $H^{\dagger}H=(v+h)^{2}/2$, 
whose linear part yields the effective mass term below.
\begin{equation}\label{eq:fermion_mass_CP}
\mathcal{L}\supset -\mathcal{M}(z;s)\,\overline{\Psi}\Psi,\qquad
\mathcal{M}(z;s) = \frac{y_{\rm CP}(s)}{\sqrt{2}}\,h(z),\qquad
y_{\rm CP}(s)\equiv y_{0} + |y_{s}|\,e^{i\theta}\,\frac{s^{2}}{\Lambda^{2}},
\end{equation}
where $h(z)$ is the Higgs profile (a function of the wall coordinate $z$) and $\theta$ is a physical CP-violating phase. Because $y_{\rm CP}(s)$ depends on $s^{2}$, 
it is the \emph{same} for the $+s$ and $-s$ tails, guaranteeing that both fluctuation tails reinforce the same sign of the baryon asymmetry. Because the spectator 
$s$ is spatially constant inside a given bubble, the spatial dependence of $\mathcal{M}$ comes entirely from $h(z)$; the CP-violating source 
(a $\gamma_{5}$-scalar / wall-gradient type) reads
\begin{equation}\label{eq:CP_source}
S_{\rm CP}(z;s)\;\propto\;\operatorname{Im}\!\bigl[m_{\rm tot}^{*}\partial_{z}m_{\rm tot}\bigr]
\;\propto\;M(s)\,\operatorname{Im}\!\bigl[y_{\rm CP}(s)\bigr]\,\partial_{z}h(z)
\;\propto\;M(s)\,|y_{s}|\sin\theta\,\frac{s^{2}}{\Lambda^{2}}\,\partial_{z}h(z),
\end{equation}
(where $m_{\rm tot}(z;s)=M(s)+\mathcal{M}(z;s)$) which is non-zero only across the wall and, crucially, has its \emph{overall magnitude} enhanced by 
$M(s)\,\operatorname{Im}[y_{\rm CP}(s)]\simeq M(s)\,|y_{s}|\sin\theta\,(s^{2}/\Lambda^{2})$. In the high-$s$ fluctuation regions that actually drive a first-order 
EWPT, $s/\Lambda\sim\mathcal{O}(1)$ and the source reaches order-unity strength, whereas a region with $s=0$ (the SM vacuum) has no enhanced CP source. 
This ties the two EWBG ingredients together. The \emph{same} large non-Gaussian fluctuation of $s$ both strengthens the phase transition (via $Z(s)$) and amplifies 
the CP-violating source (via $y_{\rm CP}(s)$). The two operators are physically distinct. The mass $M(s)$ in Eq.~\eqref{eq:fermion_mass} yields the spectator-dependent kinetic 
function $Z(s)$, while the $s^{2}$-enhanced Yukawa $y_{\rm CP}(s)$ in Eq.~\eqref{eq:fermion_mass_CP} yields the spatially-varying, fluctuation-boosted, 
$\mathbb{Z}_{2}$-even CP source.

However, similar to the extension with a heavy scalar, scalar $s$ also receives mass correction from loop contributions.
The leading quantum corrections to $m_{s}^{2}$ arise from a closed fermion loop with two insertions of the Yukawa vertex $y s^{2}\overline{\Psi}\Psi/\Lambda$.
The one-loop diagram gives a mass contribution
\begin{equation}
\delta m_{s}^{2} = -\frac{y}{8\pi^{2}}\,\frac{M_0^{3}}{\Lambda}\ln\left(\frac{M_0^2}{\mu^2}-1\right) ,
\end{equation}
For typical parameters $M_{0}\sim\Lambda\sim 10^{8}$~GeV and $y\sim\mathcal O(1)$, one finds $\delta m_{s}\sim 10^{8}$~GeV.

%===============================================================
\subsection{Globally supersymmetric UV completion}

In this appendix we present a fully self‑consistent, globally supersymmetric UV completion that generates the required gauge kinetic function $Z(s)$ of 
Eq.~\eqref{eq:Zparam} while naturally protecting the spectator field $s$ from large quantum corrections.
The key ingredients are:
(i)~a renormalisable superpotential without a bare $S^{2}$ mass term;
(ii)~the non‑renormalisation theorem, which forbids the generation of a supersymmetric mass for $s$;
(iii)~exact cancellation of quadratic divergences between bosonic and fermionic loops;
(iv)~a sequestered SUSY‑breaking sector that provides an ultra‑small soft mass for $s$ via gravitational mediation.

\subsection*{A.3.1 \ \ Field content and superpotential}

We work in global $N=1$ supersymmetry.
The spectator field $s$ is the scalar component of a chiral superfield $S$, which is odd under a $\mathbb{Z}_{2}$ symmetry $S\to -S$.
In addition to the MSSM Higgs doublets $H_{u},H_{d}$, we introduce:
\begin{itemize}
\item A heavy gauge‑singlet chiral superfield $\Phi$, even under $\mathbb{Z}_{2}$.
\item A vector‑like pair of chiral superfields $\Psi$ and $\widetilde{\Psi}$, transforming as an $SU(2)_{L}$ doublet and carrying hypercharge $Y=1/2$ (the precise quantum numbers are inessential).
      They are even under $\mathbb{Z}_{2}$.
\end{itemize}
The K\"ahler potential is taken to be canonical,
\begin{equation}
K = S^{\dagger}S + \Phi^{\dagger}\Phi + \Psi^{\dagger}\Psi + \widetilde{\Psi}^{\dagger}\widetilde{\Psi} + H_{u}^{\dagger}H_{u} + H_{d}^{\dagger}H_{d},
\end{equation}
and the tree‑level superpotential reads
\begin{equation}\label{eq:Wfinal_full}
W = \frac{1}{2} M_{\Phi} \Phi^{2} + \lambda \Phi S^{2} + M_{\Psi} \widetilde{\Psi} \Psi + y \Phi \widetilde{\Psi} \Psi + W_{\rm MSSM},
\end{equation}
where $M_{\Phi},M_{\Psi}\gg T_{\rm EW}$ are large supersymmetric masses, and $\lambda$, $y$ are dimensionless Yukawa couplings.
All terms are manifestly renormalisable and respect the $\mathbb{Z}_{2}$ symmetry.
No supersymmetric mass term $M_{S}S^{2}$ is present in~\eqref{eq:Wfinal_full}.
The tree‑level gauge kinetic function is the canonical one, $f_{a}=1/g_{a,0}^{2}$ ($a=1,2$).

\subsection*{A.3.2 \ \ Integrating out the heavy fields: logarithmic $Z(s)$}

Below the scale $M_{\Phi}$, the heavy field $\Phi$ is integrated out.
Its F‑term equation of motion,
\begin{equation}
\frac{\partial W}{\partial\Phi}= M_{\Phi}\Phi + \lambda S^{2} + y \widetilde{\Psi}\Psi = 0,
\end{equation}
yields the non‑propagating solution
\begin{equation}
\Phi = -\,\frac{1}{M_{\Phi}}\bigl( \lambda S^{2} + y \widetilde{\Psi}\Psi \bigr).
\end{equation}
Substituting this result back into the superpotential generates an effective operator coupling $S^{2}$ to the vector‑like pair,
\begin{equation}
W_{\rm eff} \supset -\,\frac{\lambda y}{M_{\Phi}}\, S^{2}\,\widetilde{\Psi}\Psi .
\end{equation}

At the next step we integrate out $\Psi,\widetilde{\Psi}$ in the $s$‑dependent background.
Their effective mass matrix is
\begin{equation}
M(s) = M_{\Psi} - \frac{\lambda y}{M_{\Phi}}\, s^{2}
      \equiv M_{\Psi}\!\left( 1 - \frac{s^{2}}{\Lambda^{2}} \right),
\qquad
\Lambda^{2} \equiv \frac{M_{\Psi} M_{\Phi}}{\lambda y}.
\end{equation}
The one‑loop contribution of the vector‑like pair to the gauge kinetic function is universal and well‑known.
Using the holomorphic renormalisation scheme ($\overline{\rm DR}$), we obtain
\begin{equation}\label{eq:flog}
f_{a}(s) = \frac{1}{g_{a,0}^{2}} - \frac{b_{a}}{16\pi^{2}}\,
\ln\!\left( \frac{M(s)^{2}}{\mu^{2}} \right),
\end{equation}
where $b_{a}$ is the one‑loop $\beta$‑function coefficient of the corresponding gauge group ($b_{SU(2)}=3$, $b_{U(1)}=1/5$ in standard normalisation).
Imposing the renormalisation condition $f_{a}(0)=1/g_{a,0}^{2}$ eliminates the dependence on the arbitrary scale $\mu$:
\begin{equation}
f_{a}(s) = \frac{1}{g_{a,0}^{2}} \left[ 1 - \frac{b_{a}\,g_{a,0}^{2}}{16\pi^{2}}\,
\ln\!\left( 1 - \frac{s^{2}}{\Lambda^{2}} \right)^{\!2}\, \right].
\end{equation}
Taking the real part and expanding for $s^{2}\ll\Lambda^{2}$ gives
\begin{equation}\label{eq:Z_low}
Z(s) \approx 1 - \frac{2c}{\Lambda^{2}}\,s^{2} + \mathcal{O}(s^{4}),
\qquad
c \equiv \frac{b_{a}\,g_{a,0}^{2}}{16\pi^{2}} > 0 .
\end{equation}
This is precisely the low‑energy form $Z(s)\approx 1 - s^{2}/\widetilde{\Lambda}^{2}$ required for the phase‑transition mechanism, with an effective scale $\widetilde{\Lambda}^{2}=\Lambda^{2}/(2c)$.
To ensure an explicit UV saturation, we can adopt 
\begin{equation}
    Z_{\infty} \equiv 1-2c
\end{equation}
for the parametrisation~\eqref{eq:Zparam}.
The coefficient $c$ is proportional to the one-loop $\beta$-function coefficient of the heavy states that have been integrated out, and hence to their multiplicity. 
With $N$ vector-like pairs carrying the same quantum numbers one simply has $c\to Nc$. A single pair gives a small $c$ and therefore $Z_\infty$ close to unity, but by 
taking $N$ large enough the value $Z_\infty\simeq0.1$ required by the phase-transition analysis is attained while the loop expansion remains parametrically 
controlled, $Nc\lesssim\mathcal{O}(0.1)$. The multiplicity needed is model dependent and we do not attempt to minimise it here; we only note that the $Z_\infty$ 
adopted in the main text is reachable within this perturbative regime.

\subsection*{A.3.3 \ \ Radiative stability}

The spectator field $s$ is massless in the exact SUSY limit, secured by three independent mechanisms.

\paragraph{Non‑renormalisation of the superpotential.}
The tree‑level superpotential~\eqref{eq:Wfinal_full} contains no term proportional to $S^{2}$.
The celebrated non‑renormalisation theorem of global supersymmetry states that the superpotential receives no perturbative quantum corrections beyond wave‑function renormalisation.
Consequently, no holomorphic mass term $M_{S}S^{2}$ can be generated at any loop order.

\paragraph{Exact cancellation of quadratic divergences.}
The coupling $\lambda\Phi S^{2}$ generates, for the self‑energy of $s$, potentially dangerous diagrams, namely a scalar loop with the scalar component $\phi$ of $\Phi$, and a fermion loop with the fermionic partners $\psi_{\Phi}$ of $\Phi$ and $\psi_{S}$ of $S$.
In the unbroken SUSY limit the sum of these diagrams vanishes identically, because the scalar and fermion couplings are precisely related by the supersymmetric Ward identities.
After SUSY breaking, the residual correction is proportional to the soft breaking scale $m_{\rm soft}$ and is always loop‑suppressed.

\paragraph{K\"ahler potential corrections.}
Radiative corrections can generate non‑holomorphic operators in the K\"ahler potential, e.g.\
\begin{equation}
\Delta K \sim \frac{c}{\Lambda_{\rm UV}^{2}} (S^{\dagger}S)(\Phi^{\dagger}\Phi).
\end{equation}
After SUSY breaking, they contribute a soft mass term of order
\begin{equation}
\delta m_{s}^{2} \sim \frac{m_{\rm soft}^{2}}{16\pi^{2}} \ln\frac{\Lambda_{\rm UV}}{M_{\Phi}}.
\end{equation}
Because $s$ is a gauge singlet, its soft mass can be sequestered from the main SUSY‑breaking source, making $m_{\rm soft}$ extremely small compared to $H_{\rm inf}$.

\subsection*{A.3.4 \ \ Sequestered supersymmetry breaking and the Higgs portal}

Supersymmetry breaking is assumed to take place in a hidden sector described by a chiral superfield $X$ with an F‑term VEV $\langle F_{X}\rangle\neq 0$.
The spectator sector communicates with the hidden sector only through gravitational‑strength operators in the K\"ahler potential,
\begin{equation}\label{eq:K_break}
\Delta K = \frac{X^{\dagger}X}{M_{\rm Pl}^{2}}\, S^{\dagger}S .
\end{equation}
When $X$ acquires its F‑term, this operator generates a soft scalar mass for $s$:
\begin{equation}
m_{s}^{2} \sim \frac{|\langle F_{X}\rangle|^{2}}{M_{\rm Pl}^{2}} .
\end{equation}
For an intermediate SUSY‑breaking scale $\sqrt{F_{X}}\sim 10^{8}$~GeV, one obtains $m_{s}\sim 10^{-3}$~GeV. Since $m_{s}\simeq|\langle F_{X}\rangle|/M_{\rm Pl}$ 
grows quadratically with the SUSY‑breaking scale, the range $m_{s}\sim10^{-6}$--$10^{6}$~eV required by Eq.~\eqref{eq:DM_abundance} across the scan is obtained for 
$\sqrt{F_{X}}\sim10^{2}$--$10^{8}$~GeV. No particular value of $m_{s}$ is singled out by the mechanism. It is fixed only once the fluctuation amplitude $s_{\rm rms}$ 
is specified.

\subsection*{A.3.6 \ \ Summary}

The globally supersymmetric UV completion defined by the renormalisable superpotential \eqref{eq:Wfinal_full} and the minimal K\"ahler potential automatically 
generates, upon integrating out the heavy fields, a logarithmic gauge kinetic function that reproduces the required $Z(s)$ in the low‑energy regime.
The supersymmetric non‑renormalisation theorem, combined with exact boson‑fermion cancellations, guarantees that the spectator field $s$ receives no large quantum corrections to its mass.
The tiny mass and the Higgs‑portal coupling $\kappa$ are introduced as soft SUSY‑breaking terms arising from a sequestered hidden sector through gravitational mediation, and are technically natural.

%===============================================================

\section{Exact probability density function}\label{app:exact_PDF}

For a strictly monotonic random variable mapping $Y = f(X)$, if the probability density of $X$ is $\mathcal{P}_X(x)$, then the probability density of $Y$ is:
\begin{equation}
\mathcal{P}_Y(y) = \mathcal{P}_X\left(f^{-1}(y)\right) \cdot \left| \frac{df^{-1}(y)}{dy} \right|~,
\end{equation}
where $f^{-1}(y)$ is the inverse function, and $\left| \frac{df^{-1}(y)}{dy} \right|$ is the Jacobian determinant.
The mapping is:
\begin{equation}
y = f(x) = x + \frac{f_{\text{NL}}}{2} x |x|
\end{equation}
One can check that the derivative $f'(x)>0$ for both $x>0$ and $x<0$. Therefore, $f(x)$ is strictly monotonically increasing over the entire real axis, and the 
inverse function exists and is unique.
The solution of $x + \frac{f_{\text{NL}}}{2} x |x|-y=0$ can be obtained in positive ($x>0$) and negative ($x<0$) regions, respectively.
We obtain the inverse function for the normalised variable:
\begin{equation}\label{eq:xy}
x(y) = \frac{\sqrt{1 + 2f_{\text{NL}}|y|} - 1}{f_{\text{NL}}} \cdot \text{sign}(y)~.
\end{equation}
Substituting $x=g$ and $y = s/\sigma_s$ (normalising $s$) into Eq.~\eqref{eq:xy}, we obtain:
\begin{equation}\label{eq:inverse_variable}
g(s) = \frac{\sqrt{1 + 2f_{\text{NL}} \cdot \left| \frac{s}{\sigma_s} \right|} - 1}{f_{\text{NL}}} \cdot \text{sign}(s)~.
\end{equation}
The Jacobian determinant is given by:
\begin{equation}
\left| \frac{dg}{ds} \right| = \frac{1}{\sigma_s \cdot \left( 1 + f_{\text{NL}} |g(s)| \right)}~.
\end{equation}

The probability density of the normalised Gaussian variable $g$ is given by the Gaussian distribution.
Substituting the inverse function $g(s)$ and the Jacobian determinant into the Gaussian distribution, we obtain the exact PDF for dimensional $s$:
\begin{equation}
\mathcal{P}(s) = \frac{1}{\sqrt{2\pi} \sigma_s \cdot \left( 1 + f_{\text{NL}} |g(s)| \right)} \exp\left( -\frac{g(s)^2}{2} \right)~.
\end{equation}

Let us verify the $\mathbb{Z}_2$ symmetry of the PDF. 
Under the transformation $s \to -s$ we have $g(-s) =-g(s)$ with Eq.~\eqref{eq:inverse_variable}.
Therefore $|g(-s)| = |g(s)|$, $g(-s)^2 = g(s)^2$. Substituting into the PDF, we find $\mathcal{P}(-s) = \mathcal{P}(s)$.
Therefore, the PDF satisfies the $\mathbb{Z}_2$ symmetry.

\section{Gravitational wave spectrum}\label{app:GW_spectrum}
\paragraph{Bubble collision contribution.}
In the envelope approximation and for relativistic walls, the spectrum is~\cite{Huber:2008hg}
\begin{equation}
\label{eq:GWcol}
\Omega_{\text{col}} h^2 = 1.67 \times 10^{-5} \left( \frac{H_n}{\beta} \right)^2 \left( \frac{\kappa_{\phi} \alpha}{1+\alpha} \right)^2 \left( \frac{100}{g_*} \right)^{1/3} \frac{0.11 v_w^3}{0.42 + v_w^2} S_{\text{col}}(f),
\end{equation}
where $\kappa_{\phi}$ is the fraction of the vacuum energy converted into scalar field kinetic energy. For non-runaway walls, $\kappa_{\phi} \approx 0$. For strong transitions, $\kappa_{\phi} \simeq \alpha / (0.73 + 0.083\sqrt{\alpha} + \alpha)$. The spectral shape is
\begin{equation}
\label{eq:Scol}
S_{\text{col}}(f) = \frac{3.8 (f/f_{\text{col}})^{2.8}}{1 + 2.8 (f/f_{\text{col}})^{3.8}},
\end{equation}
with the peak frequency today
\begin{equation}
\label{eq:fcol}
f_{\text{col}} = 16.5 \times 10^{-6} \text{ Hz} \left( \frac{0.62}{1.8 - 0.1 v_w + v_w^2} \right) \left( \frac{\beta}{H_n} \right) \left( \frac{T_n}{100 \text{ GeV}} \right) \left( \frac{g_*}{100} \right)^{1/6}.
\end{equation}

\paragraph{Sound wave contribution.}
The bulk motion of the plasma is the dominant source for most non-runaway phase transitions. The spectrum is~\cite{Hindmarsh:2015qta,Hindmarsh:2017gnf}
\begin{equation}
\label{eq:GWsw}
\Omega_{\text{sw}} h^2 = 2.65 \times 10^{-6}\Upsilon(\tau_{sw}) \left( \frac{H_n}{\beta} \right) \left( \frac{\kappa_v \alpha}{1+\alpha} \right)^2 \left( \frac{100}{g_*} \right)^{1/3} v_w \, S_{\text{sw}}(f),
\end{equation}
where $\kappa_v$ is the efficiency factor for converting latent heat into bulk fluid motion. For our parameter space, $\kappa_v \approx \alpha/(0.73 + 0.083\sqrt{\alpha} + \alpha)$ is a good approximation~\cite{Espinosa:2010hh}. 
The factor
\begin{equation}
    \Upsilon(\tau_{\rm sw}) = 1 - \frac{1}{\sqrt{1 + 2\tau_{\rm sw}H_*}}
\end{equation}
accounts for the decay of the sound wave. Following Ref.~\cite{Hindmarsh:2017gnf}, we approximate the lifetime of the sound wave as $\tau_{\rm sw}\simeq R_n/\overline{U}_f$, 
where the mean bubble separation is given by $R_n = (8\pi)^{1/3} v_w \beta^{-1}$ and the root-mean-squared fluid velocity is $\overline{U}_f = \sqrt{3 \kappa_{sw} \alpha / 4}$.
The spectral shape function is 
\begin{equation}
\label{eq:Ssw}
S_{\text{sw}}(f) = (f/f_{\text{sw}})^3 \left( \frac{7}{4 + 3 (f/f_{\text{sw}})^2} \right)^{7/2},
\end{equation}
with the redshifted peak frequency
\begin{equation}
\label{eq:fsw}
f_{\text{sw}} = 1.9 \times 10^{-5} \text{ Hz} \left( \frac{1}{v_w} \right) \left( \frac{\beta}{H_n} \right) \left( \frac{T_n}{100 \text{ GeV}} \right) \left( \frac{g_*}{100} \right)^{1/6}.
\end{equation}

\paragraph{Turbulence contribution.}
MHD turbulence in the plasma provides a subleading contribution~\cite{Caprini:2009yp}:
\begin{equation}
\label{eq:GWturb}
\Omega_{\text{turb}} h^2 = 3.35 \times 10^{-4} \left( \frac{H_n}{\beta} \right) \left( \frac{\kappa_{\text{turb}} \alpha}{1+\alpha} \right)^{3/2} \left( \frac{100}{g_*} \right)^{1/3} v_w \, S_{\text{turb}}(f),
\end{equation}
where typically $\kappa_{\text{turb}} \approx 0.05 \kappa_v$. The spectral shape is
\begin{equation}
\label{eq:Sturb}
S_{\text{turb}}(f) = \frac{(f/f_{\text{turb}})^3}{\big[1 + (f/f_{\text{turb}})\big]^{11/3} \big(1 + 8\pi f / h_* \big)},
\end{equation}
with the Hubble rate at $T_n$ giving $h_* = 16.5 \times 10^{-6} \text{ Hz} (T_n/100 \text{ GeV})(g_*/100)^{1/6}$. The peak frequency is
\begin{equation}
\label{eq:fturb}
f_{\text{turb}} = 2.7 \times 10^{-5} \text{ Hz} \left( \frac{1}{v_w} \right) \left( \frac{\beta}{H_n} \right) \left( \frac{T_n}{100 \text{ GeV}} \right) \left( \frac{g_*}{100} \right)^{1/6}.
\end{equation}

% \normalem
\bibliographystyle{JHEP}
\bibliography{reference}

\end{document}